\documentclass[aps,prl,twocolumn,groupedaddress, floatfix,longbibliography,nofootinbib,byrevtex]{revtex4}
\usepackage{graphicx}% Include figure files
\usepackage{dcolumn}% Align table columns on decimal point
\usepackage{bm}% bold math
\usepackage[utf8]{inputenc}
\usepackage{amsmath,amsfonts,calc}
\usepackage{comment}
\usepackage[normalem]{ulem}
\usepackage[usenames,dvipsnames]{color}
\usepackage{soul}
\usepackage{graphicx}
\usepackage{scalerel}
\usepackage{mathtools}
\usepackage{stackengine,wasysym}
\usepackage[free-standing-units=true]{siunitx}
\makeatletter
\newsavebox\myboxA
\newsavebox\myboxB
\newlength\mylenA
\makeatother

\newcommand{\ket}[1]{\left|#1\right>}

\newcommand{\calC}{\mathcal{C}}

\newcommand{\catpm}{\ket{\calC_{\alpha}^\pm}}

\newcommand{\op}[1]{\hat{#1}}%{\bm{#1}}
\newcommand{\ns}{\textrm{ns}}
\newcommand{\us}{\textrm{\textmu s}}

\newcommand{\omegaa}{\omega_a} % SNAIL-transmon frequency
\newcommand{\omegap}{\omega_d} % stabilization drive frequency
\newcommand{\omegar}{\omega_b} % readout cavity frequency
\newcommand{\kappar}{\kappa_b} % readout cavity linewidth
\newcommand{\kapparValue}{0.40\,\textrm{MHz}}
\newcommand{\cqrTime}{4.44~\us} % fig 2, Time of a single CQR
\newcommand{\cqrTco}{(482 \pm 4)~\us} % fig 2, coherent state lifetime
\newcommand{\cqrTcocqr}{(485 \pm 1)~\us} % fig 2, coherent state lifetime during repeated CQR
\newcommand{\cqrnbar}{10.7} % fig 2, photon number of cat, catPump = 0.3
\newcommand{\omegaspec}{\omega_{pr}} % fig 3, spectroscopy frequency
\newcommand{\kerr}{320~\textrm{kHz}} % Kerr at bias point. 0.63857 / 2 # Kerr in MHz, full costate Kerr refocusing 1566ns = 2pi/2K => 2K/2pi 0.6385696,
\newcommand{\TkerrGate}{784 ~\ns}
\newcommand{\FockTtwostar}{(2.17 \pm 0.5)~\us}

\newcommand{\detunedsize}{15} %From alpha^2 from eps_2/K-\Delta/2K. (CP = 0.25) eps_2/K ~ 8, Delta/2K ~ 7 (Delta -4.5MHz). See /KerrCat_2021/Figs/timeRabi_vs_RabiPhase.py
\newcommand{\beTauXp}{(101 \pm 4) \,\us} % breakeven X coherence
\newcommand{\beTauXm}{(103  \pm 4) \,\us} % breakeven X coherence
\newcommand{\beTauYp}{(5.9\pm0.2) \,\us} % breakeven Y coherence, 9 (5.9 +/- 0.2), 
\newcommand{\beTauYm}{(6.5\pm0.2) \,\us} % breakeven Y coherence, 10 (6.5 +/- 0.2)
\newcommand{\beTauZp}{(6.1\pm0.1) \,\us} % breakeven Z coherence, 8 (6.2 +/- 0.3), 
\newcommand{\beTauZm}{(6.2\pm0.4) \,\us} % breakeven Z coherence,  11 (6.1 +/- 0.3)
\newcommand{\befactor}{2.8 }

\newcommand{\benbar}{1.85} % breakeven nbar, 20211013exp8,9,10,11: catPump=0.05
\newcommand{\msTauX}{(1.102  \pm 0.008)~\textrm{ms}} % 1 ms X coherence
\newcommand{\msTcat}{(0.76\pm0.08)~\textrm{\textmu s}} %measured by fitting Rabi proof of cat 20211023_0. Fig 5f

\newcommand{\msepsOK}{8.9} % catpump= 0.25 => eps2/K = 8.88
\newcommand{\msDelta}{-4.5 \,\mathrm{MHz}} % Detuning for 1 ms cat.

\newcommand{\figwidth}{87mm} %3.375in or \columnwidth,
\newcommand{\figwidthWide}{183mm}

\begin{document}
\title{The squeezed Kerr oscillator: spectral kissing and phase-flip robustness}
%experimental realization of the squeezed Kerr-Hamiltonian
%Spectral kissing in a driven Kerr oscillator
%Spectral kissing and step-wise enhancement of the logical lifetime  \\in a driven Kerr oscillator \textcolor{blue}{ beyond bifurcation}}

%Spectroscopy of pairwise level kissing in a driven Kerr oscillator}

%\\Pairwise spectral kissing causes a stepped lifetime increase\\in a squeezed Kerr oscillator}

%\\Pairwise spectral kissing leads to stepped lifetime growth\\in a squeezed Kerr oscillator}
%Spectroscopy of the Kerr-cat metapotential}

%Pairwise level kissing in the crossover from a Kerr oscillator to the Schr\"odinger cat regime
%Pairwise spectral kissing in a squeezed Kerr oscillator}
% \title{Period doubling of the Kerr-cat eigenspectrum}
%\title{Pairwise exponential level kissing in the Kerr-cat eigenspectrum}
\author{Nicholas E. Frattini}
\thanks{These authors contributed equally}
\email{nicholas.frattini@colorado.edu,\\ rodrigo.cortinas@yale.edu.}
\affiliation{Department of Applied Physics and Physics, Yale University, New Haven, CT 06520, USA}
\author{Rodrigo G. Corti\~nas}
\thanks{These authors contributed equally}
\email{nicholas.frattini@colorado.edu,\\ rodrigo.cortinas@yale.edu.}
\affiliation{Department of Applied Physics and Physics, Yale University, New Haven, CT 06520, USA}
\author{Jayameenakshi Venkatraman}
\affiliation{Department of Applied Physics and Physics, Yale University, New Haven, CT 06520, USA}
\author{Xu Xiao}
\affiliation{Department of Applied Physics and Physics, Yale University, New Haven, CT 06520, USA}
\author{Qile Su}
\affiliation{Department of Applied Physics and Physics, Yale University, New Haven, CT 06520, USA}
\author{Chan U Lei}
\affiliation{Department of Applied Physics and Physics, Yale University, New Haven, CT 06520, USA}
\author{Benjamin J. Chapman}
\affiliation{Department of Applied Physics and Physics, Yale University, New Haven, CT 06520, USA}
\author{Vidul R. Joshi}
\affiliation{Department of Applied Physics and Physics, Yale University, New Haven, CT 06520, USA}
\author{S. M. Girvin}
\affiliation{Department of Applied Physics and Physics, Yale University, New Haven, CT 06520, USA}
\author{Robert J. Schoelkopf}
\affiliation{Department of Applied Physics and Physics, Yale University, New Haven, CT 06520, USA}
\author{Shruti Puri}
\affiliation{Department of Applied Physics and Physics, Yale University, New Haven, CT 06520, USA}
\author{Michel H. Devoret}
\email{michel.devoret@yale.edu}
\affiliation{Department of Applied Physics and Physics, Yale University, New Haven, CT 06520, USA}
\date{\today}
\begin{abstract}

By applying a microwave drive to a specially designed Josephson circuit, we have realized an elementary quantum optics model, the squeezed Kerr oscillator. This model displays, as the squeezing amplitude is increased, a cross-over from a single ground state regime to a doubly-degenerate ground state regime. In the latter case, the ground state manifold is spanned by Schrödinger-cat states, i.e. quantum superpositions of coherent states with opposite phases. For the first time, having resolved up to the tenth excited state in a spectroscopic experiment, we confirm that the proposed emergent static effective Hamiltonian correctly describes the system, despite its driven character.  We also find that the lifetime of the coherent state components of the cat states increases in steps as a function of the squeezing amplitude. We interpret the staircase pattern as resulting from pairwise level kissing in the excited state spectrum. Considering the Kerr-cat qubit encoded in this ground state manifold, we achieve for the first time quantum nondemolition readout fidelities greater than 99\%, and enhancement of the phase-flip lifetime by more than two orders of magnitude, while retaining universal quantum control. Our experiment illustrates the crucial role of parametric drive Hamiltonian engineering for hardware-efficient quantum computation.

% By driving a superconducting circuit we have experimentally realized an elementary quantum optics model which is equivalent to a double-well potential. We demonstrate pairwise kissing of the quasienergy spectrum produced by the suppression of excited state tunneling  in a Kerr oscillator as it is progressively dressed by squeezed microwave light. This is the spectroscopic fingerprint of a complication-free realization of this model. The pairwise kissing causes the incoherent delocalization timescale of the ground states to increase in steps as a function of the squeezing amplitude. The degenerate ground states span a Kerr-cat qubit which is now protected from thermal activation by the exponential suppression of excited state tunneling. Ultimately, we show that the localization time can be enhanced in this way by more than two orders of magnitude while improving quantum nondemolition readout fidelities up to $>99\%$ and retaining universal quantum control. Additionally, we show an autonomous error correction factor of 2.8 for the Kerr-cat qubit and also modify the double-well potential to achieve an extra factor of two in the enhancement of localization time. Our experiment provides important tools for analog Hamiltonian engineering and hardware-efficient quantum computation.

\end{abstract}
\maketitle 
The cross-over of a dynamical system from the quantum regime to the classical regime involves the competition between nonlinearity and dissipation. To illustrate this rather abstract statement, let us consider the simplest example with these two ingredients, namely the mechanical pendulum. The natural oscillation frequency of the pendulum decreases as a function of its total energy, and in the underdamped regime, for oscillation angles well below 90 degrees, the pendulum can be well-modeled as a simple harmonic oscillator with a so-called Kerr nonlinearity, i.e.\ a quartic dependence of the total energy with respect to the oscillation amplitude.

Besides its small oscillation frequency, the quantum version of the Kerr oscillator is characterized by another frequency: the change in oscillator frequency when one excitation quantum is added to the system, also known as the Kerr frequency. If the damping rate of the oscillator is well below this Kerr frequency, striking quantum effects can be observed: for instance, if the pendulum is displaced from its ground state, it will evolve into the quantum superposition of semiclassical coherent states, i.e. the so-called “cat states” in reference to Schrodinger’s cat \cite{yurke1986, haroche2006, kirchmair2013}. In this regime, the residual dissipation will progressively decohere the superposition and reduce the cat state to a classical mixture. 

The situation becomes more interesting when an oscillating driving force is applied to the pendulum. In the classical case, this introduces two extra parameters in the problem: the amplitude and the frequency of the drive. Richer consequences of nonlinearity can then be observed, as for example in the case of the classical Kapitza pendulum. The state of this pendulum, for large enough drive, bifurcates into two locally stable states and the system is described by a double-well Hamiltonian \cite{landau1976, venkatraman2021}.

In this work, we focus on the quantum regime of such phenomena taking place in a driven nonlinear system.  The Hamiltonian we start from is
\begin{align}\label{eq:nl-osc-H}
\begin{split}
    \op H(t) / \hbar = \omega_o\op a^\dagger \op a &+ \sum_{m=3}^{\infty} \frac{g_m}{m}(\op a + \op a^\dagger)^m \\
    &- i\Omega_d (\op a - \op a^\dagger)\cos\omegap t,
\end{split}
\end{align}
and describes a generic driven nonlinear oscillator.
In this expression, $\hat a$ is the bosonic annihilation operator. The parameters $\omega_o$ and $g_m\ll\omega_o$ are the bare oscillator frequency and the $m$-th order nonlinearity coefficients of the oscillator. The drive is specified by the amplitude $\Omega_d$ and frequency $\omegap$. 

\begin{figure}[t]
    \centering
    \includegraphics[width=\figwidth]{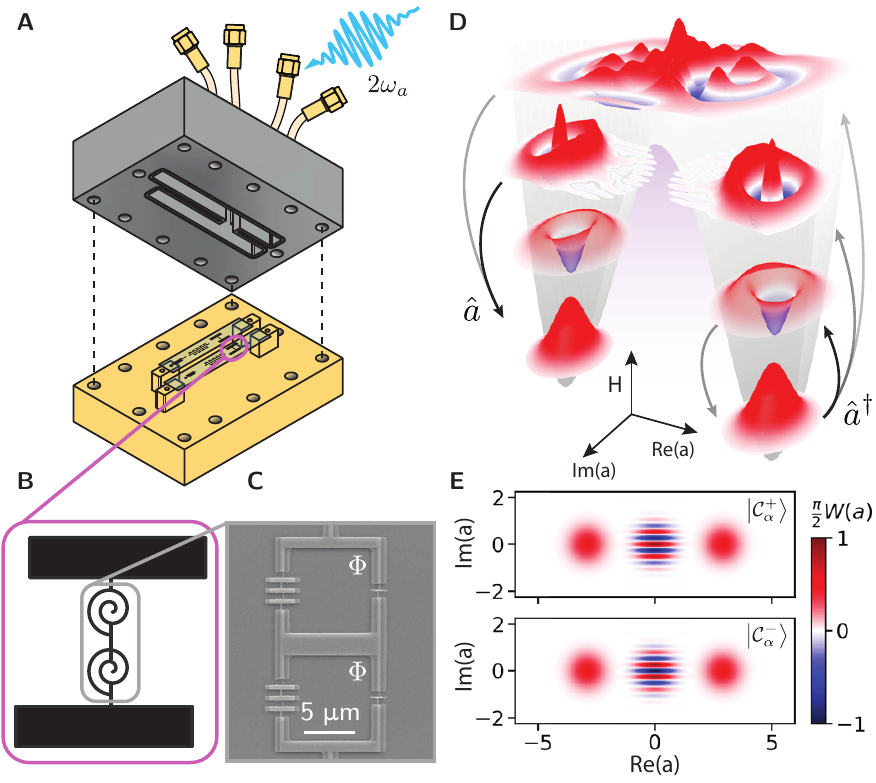}
    \caption{\textbf{Implementation of the squeezed Kerr oscillator.} \textbf{A}, Rendering of the half-aluminum, half-copper sample package containing two sapphire chips, each with a SNAIL-transmon, readout resonator and Purcell filter. Only one chip is used in the present work. Applying a strong microwave drive at $\omegap \approx 2\omegaa$ transforms the SNAIL-transmon Hamiltonian to a squeezed Kerr oscillator. \textbf{B}, Schematic of the SNAIL-transmon: a two-SNAIL array serves as the nonlinear element. The capacitor pads are shifted with respect to the axis of the array to couple it to the readout resonator. \textbf{C}, Scanning electron micrograph of the two-SNAIL array. The SNAIL loops are biased with an external magnetic flux $\Phi/\Phi_0=0.33$, where $\Phi_0$ is the magnetic flux quantum. \textbf{D}, Metapotential (grey) of the squeezed Kerr oscillator static effective Hamiltonian Eq.~(\ref{eq:HKC}) for $\epsilon_2/K = 8.5$. Wigner functions of the first seven eigenstates are shown. The highly nonlinear double-well structure hosts three pairs of degenerate eigenstates. The arrows represent incoherent jumps causing a well-occupation flip from right to left. \textbf{E}, Wigner functions of the even and odd superpositions of the two degenerate coherent states, the Kerr-cat qubit $|\pm Z\rangle$ states. }
    %Phi_0 = 8.25mA, operation point at 5.5mA.
    \label{fig:setup_metapot}
    % Metapotential_drawing_paper_staircase in Qulab_Iphython. 
\end{figure}

This Hamiltonian is implemented experimentally in a superconducting circuit, which is the only engineerable platform for which the competition between nonlinearity and dissipation can be finely controlled in presence of a drive. The central element of the circuit is the Josephson tunnel junction, which is the exact electrical analog of a pendulum. In circuit quantum electrodynamics (cQED) \cite{blais2004,Blais2021}, driven Josephson circuits attain the fully quantum regime. In our experiment, the circuit is based on an array of charge-driven SNAIL  \cite{frattini2017} transmons (see Figure \ref{fig:setup_metapot}, left panels): it is the analog of an asymmetric pendulum with a third-order nonlinearity, or in other words, a system capable of 3-wave mixing. When the drive frequency approaches twice the natural oscillator frequency $\omegap \approx 2\omega_o$, Eq.~(\ref{eq:nl-osc-H}) exhibits a period doubling bifurcation whose ground state, in the quantum regime, leads to the Kerr-cat qubit \cite{grimm2020} which is one representative of \textit{stabilized}  Schr\"odinger cat qubits \cite{mirrahimi2014, leghtas2015, touzard2018, lescanne2020, berdou_one_2022}. After a canonical transformation and keeping terms beyond the rotating wave approximation (RWA) (see Supplementary and \cite{venkatraman2021}), the evolution in the frame rotating at $\omegap/2$ is given by the effective, time-independent squeezed Kerr Hamiltonian
\begin{equation}
\label{eq:HKC}
    \hat{H}_{\textrm{SK}}/\hbar = \epsilon_2(\hat a^{\dagger2} + \hat a^{2}) - K\hat a^{\dagger2} \hat a^2 .
\end{equation}

% We have defined $\omegaa$ as the Lamb-shifted small oscillation frequency and taken the parametric resonant driving condition to be $\omegap=2 \omegaa$ while neglecting the Stark shift.

Here the Kerr nonlinearity is given to leading order  by $K = -\frac{3g_4}{2} +  \frac{10g_3^2}{3\omegaa}$ where $\omegaa$ is the the Lamb- and Stark-shifted small oscillation frequency.
The associated bilinear term conserves solely the parity of the number of excitations and is governed by  the squeezing amplitude $\epsilon_2 = g_3\frac{4\Omega_p}{3\omegap}$. Both the effective Kerr coefficient and the squeezing amplitude  are functions of the bare nonlinearities and the drive parameters.

The Hamiltonian in Eq.~(\ref{eq:HKC}) has received theoretical attention over the previous decades \cite{Harris1969,milburn1991,cochrane1999,wielinga1993,goto2016, zhang2017, puri2017}.
Pioneering experimental proposals involved optical platforms  \cite{milburn1991,Kinsler1991} and trapped ions \cite{cochrane1999}, but were not successfully realized.\footnote{Note that the Paul trap is an example of the emergence of a static effective Hamiltonian from system submitted to an oscillating force and known to have an accessible excited state spectrum; yet it is only weakly nonlinear, and thus an external atomic system is needed to bring it into the quantum regime.}
It is then natural to ask the questions:
Can the static effective Hamiltonian Eq.~(\ref{eq:HKC}) be realized in the cQED platform? And if so, how can we unequivocally demonstrate its properties experimentally? The answers to these questions are of interest to academic and industrial proposals for quantum information processing \cite{Guillaud2019, puri2020, bonilla_ataides_xzzx_2021, darmawan_practical_2021, chamberland_building_2022, puri2017, puri2019,putterman_stabilizing_2022,gautier_combined_2022}.

The cross-over from a nondegenerate spectrum to a pairwise kissing spectrum as a function of the squeezing amplitude is the salient feature of $\hat H_{\textrm{SK}}$ and to observe it we choose the parameters of the circuit of Figure \ref{fig:setup_metapot} adequately.
The array of SNAILs, flux biased at $\Phi/\Phi_0 = 0.33$, provides the target parameters of resonance frequency $\omegaa/2\pi= 6.3~$GHz and Kerr nonlinearity $K/2\pi = \kerr$.
In absence of drive, this SNAIL-transmon has an excitation lifetime of $T_1= (20\pm3)$~\textmu s and the superposition lifetime of its two lowest laying Fock states is $T^*_2= \FockTtwostar$.
We interpret the exponential kissing as the suppression of excited state quantum tunneling across the barrier separating a double-well energy surface and is in excellent agreement with theory predictions without free parameters.
The suppression of excited state dynamical tunneling protects against phase-flips of the Kerr-cat qubit spanned by the two coherent states \cite{puri2017,grimm2020}. This feature is a qualitative improvement over what was achieved in \cite{grimm2020}.
The realization of this Hamiltonian also allows high-fidelity, quantum nondemolition (QND) readout which is an important requirement for applications in quantum information.

% Moreover, we find that the decoherence properties of the ground state manifold are related in a direct, yet nontrivial, way to the eigenspectrum of $\hat H_{\textrm{SK}}$.

In Figure \ref{fig:setup_metapot} \textbf{D}, we show the Wigner phase space representation $H_{\mathrm{SK}}(a,a^*)$ \cite{curtright2013} of $\hat H_{\textrm{SK}}$, which we also refer to as the \emph{metapotential} of the system.
The Wigner functions of the seven lowest lying energy states inhabiting the phase space Hamiltonian surface (grey) are also drawn. The ground state of the system is doubly degenerate and spanned by the even and odd Schr\"odinger cat states $\catpm \propto \ket{+\alpha} \pm \ket{-\alpha}$, where $|\alpha|^2=\epsilon_2/K$.
We refer to these ground states as the Kerr-cat qubit computational states $\ket{\pm Z} = \catpm$ with Wigner functions shown in Figure \ref{fig:setup_metapot} \textbf{E} for  $|\alpha|^2=8.5$.
Their equal weight superpositions $\ket{\pm X} \approx \ket{\pm \alpha}$ correspond, in the lab frame, to two oscillations of the circuit with a relative phase-shift of $180^{\circ}$.

\begin{figure}[t]
\centering
\includegraphics[width=\figwidth]{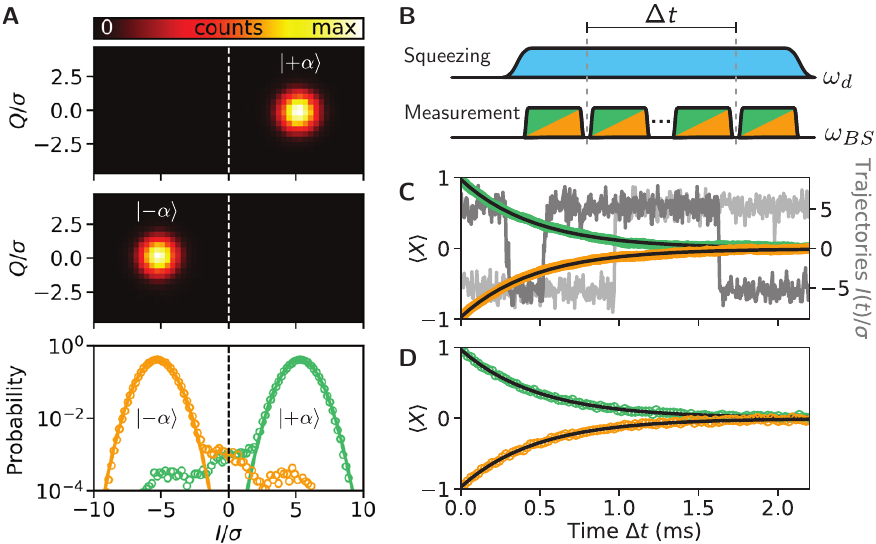}
\caption{\textbf{QND measurement and quantum jumps. A}, Top and middle: histogram of the readout resonator output field when performing $2.5\times 10^8$ measurements after preparation in $\ket{\pm \alpha}$ with a previous, stringently thresholded measurement.
Bottom: corresponding probability distribution along the $I$ quadrature,  and Gaussian fits (solid lines) with standard deviation $\sigma$ used to scale the axes.
Applying a fair threshold represented by the dashed vertical line yields a readout infidelity of 0.46\%. 
\textbf{B}, Pulse sequence to performing repeated measurements, each with a duration of $\cqrTime$.
\textbf{C}, Example quantum jump trajectories (grey) under repeated measurements for $\epsilon_2/K = \cqrnbar $.
Averages of trajectories conditioned on the first measurement of $\ket{\pm \alpha}$ (green/orange) fit with single exponentials (black) with decay time $T^{\textrm{jumps}}_{X} = \cqrTcocqr$.
\textbf{D}, State lifetime for $\ket{\pm \alpha}$ (green/orange) with no intermediate measurements (free decay).
Black lines are single-exponential fits with decay time $T_X = \cqrTco$.}
\label{fig:cqr}
\end{figure}

Measurements of the oscillator state were performed through a separate on-chip readout resonator with frequency $\omegar/2\pi = 8.5\,\mathrm{GHz}$ and coupling rate $\kappar/2\pi = \kapparValue$ to the quantum-limited measurement chain (see other parameters of the setup in Table II in the Supplementary material).
In order to activate a frequency-converting beamsplitter interaction between the squeezed Kerr oscillator and the readout resonator, we apply an additional drive at $\omega_{BS}  = \omegar - \omegap/2$.
This transfers photons from the squeezed Kerr oscillator to the readout resonator, which are subsequently collected by the measurement chain.
The strong squeezing drive ($\epsilon_2> K$) replenishes these radiated photons, thereby maintaining a steady oscillation amplitude. This is a necessary condition for a quantum nondemolition (QND) measurement of the observable $\hat X \approx |+\alpha\rangle\langle +\alpha| - |-\alpha\rangle\langle -\alpha|$ \cite{grimm2020}.
In effect, the readout resonator is displaced conditioned on which of the two metapotential wells is occupied.

%For the ground state of Eq.~(\ref{eq:HKC}) this is a QND measurement when performed in presence of the squeezing drive ($\epsilon_2\neq 0$).
% The measurement is performed by MW activation of a parametric coupling in between the nonlinear oscillator and a linear readout resonator \cite{Touzard2019, grimm2020}. The coupling acts as a frequency converting beam splitter that transfers the state of the nonlinear oscillator to a quantum-limited amplifier chain to which the readout resonator is coupled. The photons fluoresced by the resonator are replenished by the squeezing drive to maintain the self-oscillation. The net result is a collapse of the oscillation into $\sim \pm 180^{\circ}$ performing a ``which-well measurement'' and thus serves as a cat quadrature readout (CQR) scheme \cite{grimm2020}.

% The package allows to measure with high \textcolor{red}{infidelity} the state in the oscillator through use of a separate readout resonator with frequency $\omegar$.
% By MW activation of a parametric coupling between the nonlinear Kerr-cat oscillator and its readout resonator, the readout resonator is displaced conditionally on which-well of the Kerr-cat metapotential \cite{Touzard2019, grimm2020}.

In Figure \ref{fig:cqr} \textbf{A}, we show a histogram of $\hat X$ measurements.
The single shot readout infidelity is 0.46\%. Correlation measurements determined that the QND infidelity in our experiment is 1.5\% (see Supplementary).
These numbers mean that we can continuously monitor our system and reconstruct the trajectories associated with the quantum jumps of the well occupation.

In Figure \ref{fig:cqr} \textbf{B}, we show the experimental protocol for measuring the quantum trajectories.
The squeezing drive is first turned on and a series of measurements is then performed. The sequence of their outcomes constitutes a quantum trajectory record.
Two examples of quantum trajectories are shown in shades of grey in Figure \ref{fig:cqr} \textbf{C}.
The green and orange data points correspond to averages of $5\times10^5$ trajectories, each conditioned on the initial measurement falling on the positive or negative side of a fair threshold defined by the demodulated field quadrature $I=0$.
The decay curve is fitted by a single exponential (black), yielding a (incoherent) delocalization timescale of $T^{\textrm{jumps}}_{\pm X} = \cqrTcocqr$.
We next compare these measurements to the free decay of the coherent states $|\pm \alpha\rangle$. This is obtained by performing only two measurements spaced by a variable idling time, in absence of continuous monitoring; the results are shown in Figure \ref{fig:cqr} \textbf{D}. The decay is also fitted by an exponential whose lifetime is found to be $T_{\pm X} = \cqrTco$, thus showing that continuous monitoring does not significantly modify the coherent state timescale $T_{X}$ in the metapotential.

\begin{figure*}[t!]
    \centering
    \includegraphics[width=\figwidthWide]{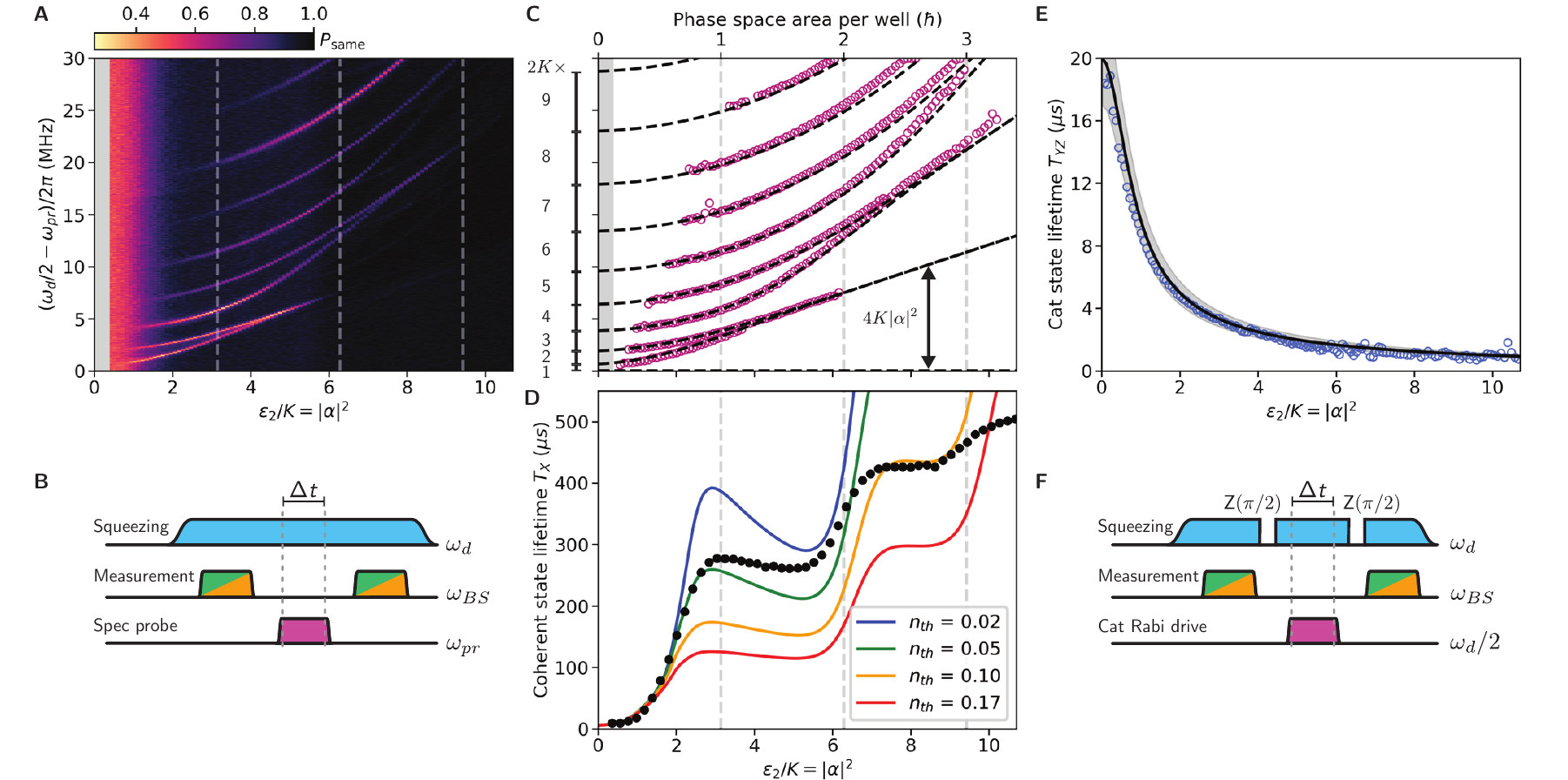}
    \caption{\textbf{Spectroscopic fingerprint of the squeezed Kerr oscillator and coherence of the ground state manifold}.
    \textbf{A}, Spectroscopy data taken with the pulse sequence in \textbf{B}.
    By applying a resonant squeezing drive we perform spectroscopy by scanning $\omegaspec$ as a function of its amplitude $ \epsilon_2/K\; (=|\alpha|^2)$.
    Color denotes probability that both measurements give the same result.
    In the grey region of $|\alpha|^2<0.3$, the measurement is not QND and thus yields a poor preparation. The power of the probe tone is increased to access the upper laying lines.
    \textbf{C}, Purple open dots are the extracted resonances from \textbf{A}. Black-dashed lines are a parameter-free  diagonalization of Eq.~(\ref{eq:HKC}).
    Grey vertical lines indicate the number of levels per Hamiltonian well using quantization calculated by phase space area (see Supplementary).
    \textbf{D}, Coherent state lifetime $T_X$ (black circles) as a function of squeezing amplitude, measured by fitted single-exponential decay timescale from experimental pulse sequences in \textbf{B} (without spectroscopy probe).
    Solid lines are extracted from fits to time-dependent master equation simulations including phenomenological parameters that emulate coupling to a photon bath at rate $\kappa_1$ with nonzero temperature $n_\mathrm{th}$ (colors), quasi-static dephasing noise, and white-noise dephasing $\kappa_\phi = 500 \,\mathrm{s}^{-1}$ (see Supplementary).
    \textbf{E}, Cat state lifetime (blue open dots) as a function its size ($|\alpha|^2$), measured with the Ramsey-like pulse sequence in \textbf{F}.
    The solid line inside the grey band is the prediction with no free parameters.% $T_1/2 \bar{n}$ where $T_1 = \fockTone$ is the bare SNAIL-transmon lifetime and $\bar{n}$ is the average number of photons in the Kerr-cat.
% \textbf{E}, Coherent state (green) and cat state (blue) lifetimes as a function of cat size $|\alpha|^2$, measured by fitted single-exponential decay timescale from experimental pulse sequences in \textbf{A} (without spectroscopy probe) and \textbf{B} respectively with swept delay time $\Delta t$.
% Cat state lifetime follows $T_1/2\bar{n}$ as expected.
% Coherent state lifetime rises exponentially, plateaus, rises again when the first set of excited states become degenerate (dashed grey line), plateaus, and rises again when the second set of excited states become degenerate (dashed grey line).
}
\label{fig:spec}
\end{figure*}

To observe the cross-over to the pairwise degenerate spectrum, we perform spectroscopy of discrete quantum energy levels as a function of the squeezing amplitude. This is achieved by interrupting the idling time, now kept constant, between the readout pulses by a microwave probe tone at frequency $\omegaspec$.
If the frequency of the probe tone coincides with the energy difference between the ground state and an excited state above the metapotential barrier, an interwell transition becomes likely. In Figure \ref{fig:spec} \textbf{A}-\textbf{B} we show the measurement of the survival probability for an initial coherent state as a function of squeezing amplitude $\epsilon_2/K$ and probe frequency $\omega_{pr}$.
In Figure \ref{fig:spec} \textbf{C} we show the fitted location of the spectroscopic lines as open purple dots. Here, in dashed lines, we also show a numerical diagonalization with no free parameters of the static-effective Hamiltonian Eq.~(\ref{eq:HKC}). The agreement between theory and experiment is remarkable given the simplicity of the model.
For $|\alpha|^2\rightarrow 0$, we extrapolate the spectrum to that of the bare SNAIL-transmon exhibiting the expected Kerr anharmonic ladder. As the squeezing amplitude---and therefore $|\alpha|^2$---grows, the spectrum becomes pairwise degenerate with levels of different photon-number parity kissing each other in an exponential fashion. We measure the energy gap separating the ground states from the first pair of degenerate states scales as $-4K|\alpha|^2$, corresponding to two identical, independent wells that stiffen as the squeezing amplitude grows.

The (classical) limiting orbit bound inside the wells is the so-called Bernoulli lemniscate \cite{wielinga1993}. In an approximation along the lines of Bohr's action quantization \cite{BREUER1991249}, its area in units of $\hbar$ counts the number $N$ of quantum states in the wells. Analytically, we find $N=\epsilon_2/\pi K =|\alpha|^2/\pi$. Every time $N$ coincides with an integer value, a new pair of excited states fits within the wells and a spectral kissing takes place (see Supplementary). The vertical dashed lines in Figure \ref{fig:spec} \textbf{A}, \textbf{C} and \textbf{D} correspond to this semiclassical condition. As they sink under the metapotential barrier, the pair of states captured by the wells are coupled only by quantum tunneling. The level splitting is exponentially suppressed with the squeezing amplitude controlling the height of the barrier $\hbar \epsilon_2^2/K$.

% Since this energy gap limits the speed of gate operations, its increase showcases how the effective nonlinearity of the system grows with squeezing strength.
%

% increases quadratically with the drive strength $\epsilon_2$ but its effect is expected to be exponentially suppressed with $|\alpha|^2$ and thus the spectrum is largely insensitive to it (see Suplementary).
% The possibility to reproduce the measured spectrum without including a detuning term accounting for the Stark shift is evidence of the insensitivity to detuning-like errors for $|\alpha|^2\gg1$.

%
%

The most remarkable consequence of the pairwise degenerate spectrum is the staircase-shaped increase of the lifetime ($T_X$) of the coherent ground states as a function of the squeezing amplitude, as shown in Figure \ref{fig:spec} \textbf{D}. This staircase-shaped dependence can be understood, to a first approximation, using Bohr's action quantization.
For $\epsilon_2\ll K$, the measured lifetime corresponds to loss of coherence of the superposition between the ground state and the first excited state of a Kerr oscillator ($T_2^*$) and increases exponentially as each of the two metapotential wells becomes wide enough to host one action quantum each ($\epsilon_2/\pi K\sim1$, first vertical dashed line).
The exponential increase stops when excitations to the first pair of excited states ($\hat a^{\dagger}$-like events) become the limiting factor. An excitation into these states will allow the transition between wells.
The coherent state delocalization time thus plateaus at $\sim$250~\textmu s until Bohr's quantization condition is met again and the first pair of excited states is captured by the metapotential wells ($\epsilon_2/\pi K\sim 2$, second vertical dashed line).
At this point, the tunnel splitting between the first pair of excited states vanishes and the increase of lifetime resumes.
This cycle repeats itself for the next pair of excited states as shown by the third rising slope in lifetime at $\epsilon_2/\pi K\sim 3$ (third vertical dashed line).
We thus interpret the experimental data as a manifestation of the eigenspectrum and eigenfuctions influencing $T_X$ via thermal-like excitations associated with dissipation.

To improve the explanation of the staircase-shaped experimental data, we develop a simple model for the dissipation in our out-of-equilibrium driven quantum system.
This model contains the Hamiltonian dynamics governed by Eq.~(\ref{eq:HKC}) as well as Markovian single-photon gain and loss at the measured coupling rate $\kappa_1$ and with a phenomenological temperature given by a mean photon number in the bath $n_\mathrm{th}$. This corresponds to a Lindbladian description under the RWA for the coupling to the environment.

We also include quasi-static dephasing noise in our model to reproduce the behaviour of our experimental data for $\epsilon_2/K=|\alpha|^2 \lesssim 2$.
%The onset of the first plateau is well-captured by the model;
The phase-flip lifetime $T_X$ in this regime increases exponentially $\propto e^{2|\alpha|^2}$, a dependence that results from the competition between dephasing and nonlinear tunneling (see Supplementary and \cite{puri2017, Frattini2021}). Such dephasing ceases to be the limiting mechanism when the coherent states become nonoverlapping, a condition captured by Bohr's quantization. Once quasi-static frequency errors become nondominant, the first plateau in Figure \ref{fig:spec} \textbf{D} is reached. The insensitivity to low-frequency flux noise as $|\alpha|^2$ increases is one the assets of the Kerr-cat qubit \cite{grimm2020} architecture. 

For values of $|\alpha|^2$ between $\pi$ and $2\pi$, the lifetime of the states remains essentially constant, limited by thermal excitation towards the first pair of excited states that have energies superior to the metapotential barrier. The model including only single photon gain and loss predicts a simple expression for the lifetime saturation value that reads $T_X=(\kappa_1 n_\mathrm{th})^{-1}$.  The slight downward trend in the experimental saturation plateau can be accounted by introducing a phenomenological Markovian detuning-like noise with strength $\kappa_{\phi} = 500~\textrm{s}^{-1}$ into the simple model defined above, keeping $n_{\mathrm{th}}$ as an adjustable parameter. The prediction of this model are shown in the colored curves in Figure \ref{fig:spec} \textbf{D}.% whose effect is efficiently suppressed, for the same reasons as just discussed, and yet increases for larger states as it scales with the state's photon number $\langle\hat a ^{\dagger}\hat a\rangle$; this is, faster than single photon events that scale as $\langle\hat a ^{\dagger}\rangle$ (see Supplementary and \cite{puri2017, Frattini2021}). 

The model correctly predicts that the rising edge after the first plateau occurs at the value of $|\alpha|^2$ where the dissipation rate $\kappa_1$ overcomes the nonlinear tunnel splitting between the first pair of excited states $\delta_1$ (see Supplementary and \cite{gautier_combined_2022}): a thermal excitation populating the first pair of nondegenerate excited states induces a tunnelling from one well to the other at rate $\delta_1$, while single-photon dissipation fights against this effect by bringing the population back to the ground state at rate $\kappa_1$. Thus, if $\delta_1 <\kappa_1 $ the tunneling between wells via the first pair of excited states is suppressed by dissipation. This explains in quantitative terms why the elementary picture based on action quantization works well for our system: the nonlinear splitting of the $n^{\text{th}}$ pair of excited levels $\delta_n$ vanishes as the eigenstates are captured by the lemniscate. At that point, the splitting is exponentially reduced by the tunnel effect and fulfills the condition $\delta_n <\kappa_1 $ almost immediately thereafter. Note that our experiments illustrate how the classical law of thermal activation across an energy barrier, known as the Arrhenius law, is modified by the quantum hypothesis. In the classical limit $\epsilon_2/K\gg1$ (formally equivalent here to the \textit{licentia mathematica} $\hbar\rightarrow 0$), the expected smooth exponential behaviour is recovered in the model (see Supplementary).

%This observation validates the scalablity of the paradigm of Hamiltonian stabilization for cat qubits.

While we have explained the presence and precise location of the plateaus in the measurements of the coherent state lifetime $T_X$, there is no quantitative agreement between the simple model and the data. The data traverse isotherms in Figure \ref{fig:spec} \textbf{D} and seems to suggest that the simple model requires the introduction of extra heating terms. Trivial heating of the attenuators in the microwaves lines is improbable, as checked by changing the duty cycle of the measurement sequence.  We remark that, in our experiment and regardless of this discrepancy, the phase-flips ($|+\alpha\rangle\leftrightarrow|-\alpha\rangle$) of our Kerr-cat qubit are robust to gate drives, readout drives, and flux noise as expected (see Supplementary and \cite{puri2019}). Further work, probably requiring beyond-RWA dissipators \cite{venkatraman2021, petrescu2020}, will be needed to improve the present state of our modeling.

We also measure the lifetime of the Schr\"odinger cat superposition of these coherent states. The results are shown in Figure \ref{fig:spec} \textbf{E}.
As expected, the lifetime of the cat states does not present a staircase structure and is well described by the straightfoward dissipators. This is because the decoherence of cat states is dominated by single-photon-loss events while the system remains in the ground state manifold. The experimental sequence used for this measurement is shown in Figure \ref{fig:spec} \textbf{F}.
The interruption of the stabilization drive for a period of $\TkerrGate \approx \pi/2K$ maps cat states into coherent states (and vice-versa) under free Kerr evolution for preparation and measurement (see Supplementary).
The cat state lifetime ($T_{YZ}$) is measured as the decay time of oscillation between cat states.
This is achieved by applying a drive at $\omegap/2$ while the squeezing drive is on. Under this conditions the coherent states remain locked to the minima of the wells while the phase-space interference fringes of the cat states roll, continuously changing the cat parity from odd to even and producing an oscillatory behaviour in the data (see Supplementary and \cite{touzard2018, grimm2020}).
The black line in Figure \ref{fig:spec} \textbf{E} corresponds to a prediction with no adjustable parameters given by $T_1/2 \langle \bar n\rangle  $ where $ \langle \bar n\rangle  = |\alpha|^2 (1+e^{-4|\alpha|^2}) / (1-e^{-4|\alpha|^2})$ is the time-averaged number of photons in the oscillation (see Supplementary). The gray band accounts for the uncertainty in the independently determined $T_1$.
%Measuring the cat state lifetime in this way, we get a representative number describing the Kerr-cat qubit lifetime, valid for all $|\alpha|^2$.
%
%$\langle \bar n \rangle = |\alpha|^2(r^2+ r^{-2})$  is the time averaged number of photons in the oscillation and $r=\sqrt{\frac{1-e^{-2|\alpha|^2}}{1+e^{-2|\alpha|^2}}}$ 
%
\begin{figure}[t!]
    \centering
    \includegraphics[width=\figwidth]{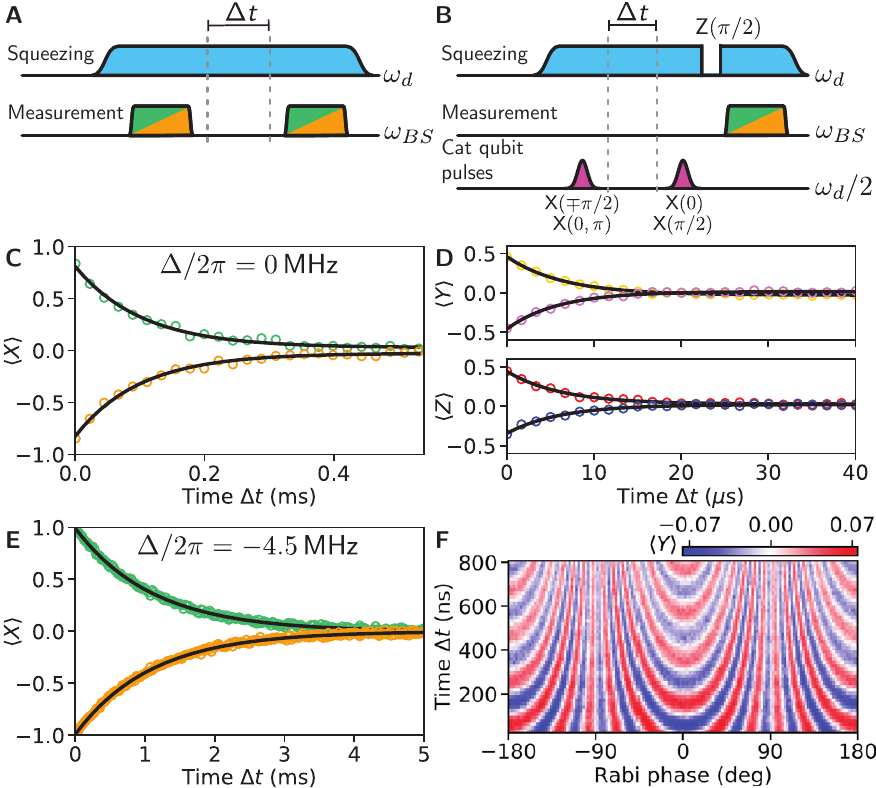}
    \caption{\textbf{Global error protection and large error bias cat-qubits}. \textbf{A}, Pulse sequence for measurement of $T_X$ coherent state lifetime in \textbf{C} and \textbf{E}.
    \textbf{B}, Pulse sequence for measurement of $T_{YZ}$ cat state lifetimes in \textbf{D}.
    \textbf{C--D}, Kerr-cat qubit operating at ($\Delta = 0$ and $\epsilon_2/K = \benbar$) where average coherence surpasses that of the bare system.
    \textbf{C}, Green/orange data for preparation in $\ket{\pm \alpha}$ and black lines are single-exponential fits with decay time $T_{+X} = \beTauXp$ and $T_{-X} = \beTauXm$.
    \textbf{D}, Cat state coherences for the two parityless cats (top yellow/pink) and for the even and odd parity cats (bottom red/blue).
    Black lines are single exponential fits with $T_{+Y} = \beTauYp$, $T_{- Y} = \beTauYm$ and $T_{+ Z} = \beTauZp$, $T_{- Z} = \beTauZm$ respectively.
    \textbf{E--F}, High-biased operation point with drive detuning $\Delta / 2\pi = \msDelta$ and $\epsilon_2/K  = \msepsOK$ implying a mean photon number of $\detunedsize$.
    \textbf{E}, Lifetime measurement for $\ket{\pm \alpha}$ (green/orange). The black line is an exponential fit with timescale $T_{\pm X} = \msTauX$.
    \textbf{F}, Oscillations between cat states as a function of single photon drive time and phase (pulse sequence as in Figure~\ref{fig:spec} \textbf{F}).
    Fits of the line cut at zero phase yield a coherent oscillation with decay time of $T_{YZ} = \msTcat$.}
    \label{fig:1ms}
\end{figure}
%
%\textcolor{blue}{For proof of cat we had  -4.5 MHz detuning on run 20201023exp0. From there we can fir $\alpha^2=\Omega_x/4/\epsilon_x = 9.4$. CP=0.25 which implies $\epsilon_2/K=4.6$ (not 3.8). $\alpha^2$ from $\epsilon_2/K-\Delta/2K =  11.6318225737501$ .  see \url{/KerrCat_2021/Figs/timeRabi_vs_RabiPhase.py} .Tcostate run 20201020exp4 has pumpssb = -44.5 MHz and pumpamp = 0.25}

Our results indicate that, depending on the requirements of a desired quantum information application, different experimental operating points can be selected.
For example, if highest average qubit coherence is desired, we choose $\epsilon_2/K = \benbar$ for which we measure, as described in Figures \ref{fig:1ms} \textbf{A} and \textbf{B}, lifetimes of  $T_X\gtrsim 97$~\textmu s (results shown in Figure \ref{fig:1ms} \textbf{C}) and $T_{YZ} \gtrsim 5.7$~\textmu s (Figure \ref{fig:1ms} \textbf{D}).
At this point, the average decoherence rate over the six cardinal states of the Kerr-cat Bloch sphere is smaller than that of the Fock Bloch sphere in our system by an autonomous error protection gain factor of $\befactor$ (see Supplementary).

If applications instead require a ``large error bias'' ($T_X\gg T_{YZ}>0$) \cite{puri2019, puri2020, bonilla_ataides_xzzx_2021, darmawan_practical_2021}, one could choose to operate the Kerr-cat qubit at $\epsilon_2/K \approx 10$, where we measure $T_X \approx 500$~\textmu s ($T_X/T^*_2\sim 200$) and $T_{YZ}\approx 1$~\textmu s. Note that for quantum applications, ``large error bias'' demands that the short time scale, here $T_{YZ}$, remains long enough to perform the required coherent operations and high fidelity measurements. The Kerr-cat qubit architecture has large margin to accommodate for this. Lately SNAILs transmons with lifetimes of $\sim 100$~\textmu s have been realized and their inclusion in our setup should provide a factor of five increase in our cat-state lifetime. 

At this point of the discussion, note that in addition to the drive amplitude, we can tune the drive frequency to exploit the rich nonlinear dynamics of our driven pendulum equivalent. By pumping the parametric squeezing off-resonantly, a quadratic static effective term arises in Eq.~(\ref{eq:HKC}): $-\Delta \hat a ^{\dagger}\hat a$ where $\Delta = \omegap/2 - \omegaa$ is the drive detuning. Applying Bohr quantization to the separatrix in the new metapotential surface, we find that the number of bound semi-calssical orbits now reads $ N=\frac{\epsilon_2}{\pi K}-\frac{\Delta}{8K}$.
 We then reduce the squeezing amplitude, and thus the associated anomalous heating, to $\epsilon_2/K = \msepsOK$ and set $\Delta / 2\pi = \msDelta$ to increase the number of captured states.  In this condition, we measure a lifetime of $ T_X=\msTauX$---an increase of  $T_X/T^*_2\sim 440$ times---while maintaining coherent control over the cat states for a time $T_{YZ} =\msTcat$.
 We show these results in Figure \ref{fig:1ms} \textbf{E} and \textbf{F}.
The finite cat lifetime and the large bias ($T_X / T_{YZ} \sim 1450$) makes this system an excellent ancilla for fault-tolerant error syndrome detection \cite{puri2019}. Note, nonetheless, that interference effects between quantum phase-space trajectories take place for $\Delta<0$ \cite{dykman2007,roberts2020} and contribute also to the lifetime enhancement. This provides the basis for a new type a cat-qubit protection. A detailed experimental study of lifetime control with the detuning parameter $\Delta$ will be communicated elsewhere \cite{inprep2022}.

In conclusion, we have experimentally realized a squeezed Kerr oscillator by applying a strong microwave drive to a SNAIL-transmon. The corresponding static effective Hamiltonian correctly accounts for the observed spectrum. We measured the spectroscopic fingerprint of pairwise kissing in agreement with theory without adjustable parameters. We also measured the related staircase-shaped lifetime increase of the coherent ground-state and its deviation from the RWA model for dissipation. We showed that Bohr's action quantization can be applied to an out-of-equilibrium driven system. Our experiment realizes a new cat-qubit with a coherent state lifetime beyond a millisecond while maintaining full quantum control and a QND readout fidelities surpassing 99\%. The physics uncovered in our experiment may have a direct impact in the engineering of parametric gates and readout as well as in the engineering of alternative qubits.

Short term applications of our setup include the realization of a pair of strongly driven squeezed Kerr oscillators operated simultaneously and in interaction (see Figure \ref{fig:setup_metapot}).
This Kerr-cat molecule will be of great interest for the implementation of single and multi-qubit gates in a quantum processor with reduced hardware overhead \cite{Guillaud2019,puri2020,bonilla_ataides_xzzx_2021,darmawan_practical_2021,chamberland_building_2022}.
Another immediate use of our system is as a fault-tolerant ancilla \cite{puri2019} for other bosonic error correction codes \cite{Campagne-Ibarcq2020, grimsmo_quantum_2021}.
The large error bias and the possibility to perform fast Raman gates via the excited state spectrum are enabling assets of our implementation \cite{xu_engineering_2022, kanao_quantum_2022, chono_two-qubit_2022}, leaving the squeezed Kerr oscillator poised for use in quantum information processing.

In the longer term, the setup presented in this work can be used to perform new experiments of fundamental interest. Among them we mention the demonstration of new dynamical Casimir effects and quantum heating \cite{Wilson2011,dykmanBook2012}, dynamical tunneling and interference in the classically forbidden region \cite{dykman2007}, quantum simulation of excited-state phase transitions in nuclear and molecular systems \cite{caprio2008,ribeiro2008}, and the exploration of quantum chaos in systems with a direct classical correspondence \cite{milburn1991, Goto2021,Monteoliva2000, zurek1998}. 

%.The cat-qubit encoded in the ground state allows for few operating points of interest. In particular, we explored a new degree of freedom (the detuning) and demonstrated a new operating point with more than a millisecond coherence along one axis of the Bloch sphere  %Remotely similar conditions in nonlinear driven systems have received theoretical attention in the past \cite{dykman2007,roberts2020} but their consequences in the context of bosonic cat-codes has been completely overlooked until our work. We have realized the first in depth experimental demonstration of this rich parameter regime and it will soon be published elsewhere.

%Upon the completion of our experiments we learned that two other groups have developed related theoretical interests \cite{putterman_stabilizing_2022,gautier_combined_2022}.

NEF designed and fabricated the superconducting device and package with input from CL, BJC, SP and MHD. NEF and RGC designed experiments, built the setup, and took and analyzed the data.
BJC and VRJ fabricated the quantum-limited amplifier used for readout.
JV and XX proposed and insisted the on importance of off-resonant pumping and proposed the dynamical Bohr quantization to our system. QS and RGC developed numerical tools to analyze decoherence with support from NEF, SP, SMG, and MHD.
RGC, NEF, JV, and MHD wrote the paper with input from all authors.

We acknowledge the contributions of L.~Frunzio, A.~Grimm, and V.~V.~Sivak. RGC acknowledges illuminating discussions with Lea Santos and Francisco Perez-Bernal.
Facilities use was supported by YINQE and the Yale SEAS cleanroom.
We also acknowledge the support of the Yale Quantum Institute.
This research was sponsored by the Army Research Office (ARO) and was accomplished under the grant numbers W911NF-18-1-0212 and W911NF-16-1-0349, by the Air Force Office of Scientific Research (AFOSR)  under award number FA9550-19-1-0399, and by the National Science Foundation (NSF) under grant number 1941583 and Centers for Chemical Innovation (CCI) grant 2124511. The views and conclusions contained in this document belong to the authors and should not be interpreted as representing the official policies, either expressed or implied, of the grant agencies, or the U.S. Government. The U.S. Government is authorized to reproduce and distribute reprints for Government purposes notwithstanding any copyright notation herein.

\bibliography{bib}

\begin{thebibliography}{49}
\expandafter\ifx\csname natexlab\endcsname\relax\def\natexlab#1{#1}\fi
\expandafter\ifx\csname bibnamefont\endcsname\relax
  \def\bibnamefont#1{#1}\fi
\expandafter\ifx\csname bibfnamefont\endcsname\relax
  \def\bibfnamefont#1{#1}\fi
\expandafter\ifx\csname citenamefont\endcsname\relax
  \def\citenamefont#1{#1}\fi
\expandafter\ifx\csname url\endcsname\relax
  \def\url#1{\texttt{#1}}\fi
\expandafter\ifx\csname urlprefix\endcsname\relax\def\urlprefix{URL }\fi
\providecommand{\bibinfo}[2]{#2}
\providecommand{\eprint}[2][]{\url{#2}}

\bibitem[{\citenamefont{Yurke and Stoler}(1986)}]{yurke1986}
\bibinfo{author}{\bibfnamefont{B.}~\bibnamefont{Yurke}} \bibnamefont{and}
  \bibinfo{author}{\bibfnamefont{D.}~\bibnamefont{Stoler}},
  \bibinfo{journal}{Phys. Rev. Lett.} \textbf{\bibinfo{volume}{57}},
  \bibinfo{pages}{13} (\bibinfo{year}{1986}),
  \urlprefix\url{https://link.aps.org/doi/10.1103/PhysRevLett.57.13}.

\bibitem[{\citenamefont{Haroche and Raimond}(2006)}]{haroche2006}
\bibinfo{author}{\bibfnamefont{S.}~\bibnamefont{Haroche}} \bibnamefont{and}
  \bibinfo{author}{\bibfnamefont{J.-M.} \bibnamefont{Raimond}},
  \emph{\bibinfo{title}{Exploring the quantum: atoms, cavities, and photons}}
  (\bibinfo{publisher}{Oxford University Press}, \bibinfo{year}{2006}).

\bibitem[{\citenamefont{Kirchmair et~al.}(2013)\citenamefont{Kirchmair,
  Vlastakis, Leghtas, Nigg, Paik, Ginossar, Mirrahimi, Frunzio, Girvin, and
  Schoelkopf}}]{kirchmair2013}
\bibinfo{author}{\bibfnamefont{G.}~\bibnamefont{Kirchmair}},
  \bibinfo{author}{\bibfnamefont{B.}~\bibnamefont{Vlastakis}},
  \bibinfo{author}{\bibfnamefont{Z.}~\bibnamefont{Leghtas}},
  \bibinfo{author}{\bibfnamefont{S.~E.} \bibnamefont{Nigg}},
  \bibinfo{author}{\bibfnamefont{H.}~\bibnamefont{Paik}},
  \bibinfo{author}{\bibfnamefont{E.}~\bibnamefont{Ginossar}},
  \bibinfo{author}{\bibfnamefont{M.}~\bibnamefont{Mirrahimi}},
  \bibinfo{author}{\bibfnamefont{L.}~\bibnamefont{Frunzio}},
  \bibinfo{author}{\bibfnamefont{S.~M.} \bibnamefont{Girvin}},
  \bibnamefont{and} \bibinfo{author}{\bibfnamefont{R.~J.}
  \bibnamefont{Schoelkopf}}, \bibinfo{journal}{Nature}
  \textbf{\bibinfo{volume}{495}}, \bibinfo{pages}{205} (\bibinfo{year}{2013}),
  ISSN \bibinfo{issn}{1476-4687},
  \urlprefix\url{https://doi.org/10.1038/nature11902}.

\bibitem[{\citenamefont{Landau and Lifshitz}(1976)}]{landau1976}
\bibinfo{author}{\bibfnamefont{L.~D.} \bibnamefont{Landau}} \bibnamefont{and}
  \bibinfo{author}{\bibfnamefont{E.~M.} \bibnamefont{Lifshitz}},
  \emph{\bibinfo{title}{Mechanics: Volume 1}}, vol.~\bibinfo{volume}{1}
  (\bibinfo{publisher}{Butterworth-Heinemann}, \bibinfo{year}{1976}).

\bibitem[{\citenamefont{Venkatraman et~al.}(2021)\citenamefont{Venkatraman,
  Xiao, Cortiñas, Eickbusch, and Devoret}}]{venkatraman2021}
\bibinfo{author}{\bibfnamefont{J.}~\bibnamefont{Venkatraman}},
  \bibinfo{author}{\bibfnamefont{X.}~\bibnamefont{Xiao}},
  \bibinfo{author}{\bibfnamefont{R.~G.} \bibnamefont{Cortiñas}},
  \bibinfo{author}{\bibfnamefont{A.}~\bibnamefont{Eickbusch}},
  \bibnamefont{and} \bibinfo{author}{\bibfnamefont{M.~H.}
  \bibnamefont{Devoret}} (\bibinfo{year}{2021}),
  \urlprefix\url{https://doi.org/10.48550/arXiv.2108.02861}.

\bibitem[{\citenamefont{Blais et~al.}(2004)\citenamefont{Blais, Huang,
  Wallraff, Girvin, and Schoelkopf}}]{blais2004}
\bibinfo{author}{\bibfnamefont{A.}~\bibnamefont{Blais}},
  \bibinfo{author}{\bibfnamefont{R.-S.} \bibnamefont{Huang}},
  \bibinfo{author}{\bibfnamefont{A.}~\bibnamefont{Wallraff}},
  \bibinfo{author}{\bibfnamefont{S.~M.} \bibnamefont{Girvin}},
  \bibnamefont{and} \bibinfo{author}{\bibfnamefont{R.~J.}
  \bibnamefont{Schoelkopf}}, \bibinfo{journal}{Phys. Rev. A}
  \textbf{\bibinfo{volume}{69}}, \bibinfo{pages}{062320}
  (\bibinfo{year}{2004}),
  \urlprefix\url{https://link.aps.org/doi/10.1103/PhysRevA.69.062320}.

\bibitem[{\citenamefont{Blais et~al.}(2021)\citenamefont{Blais, Grimsmo,
  Girvin, and Wallraff}}]{Blais2021}
\bibinfo{author}{\bibfnamefont{A.}~\bibnamefont{Blais}},
  \bibinfo{author}{\bibfnamefont{A.~L.} \bibnamefont{Grimsmo}},
  \bibinfo{author}{\bibfnamefont{S.~M.} \bibnamefont{Girvin}},
  \bibnamefont{and} \bibinfo{author}{\bibfnamefont{A.}~\bibnamefont{Wallraff}},
  \bibinfo{journal}{Rev. Mod. Phys.} \textbf{\bibinfo{volume}{93}},
  \bibinfo{pages}{025005} (\bibinfo{year}{2021}),
  \urlprefix\url{https://link.aps.org/doi/10.1103/RevModPhys.93.025005}.

\bibitem[{\citenamefont{Frattini et~al.}(2017)\citenamefont{Frattini, Vool,
  Shankar, Narla, Sliwa, and Devoret}}]{frattini2017}
\bibinfo{author}{\bibfnamefont{N.}~\bibnamefont{Frattini}},
  \bibinfo{author}{\bibfnamefont{U.}~\bibnamefont{Vool}},
  \bibinfo{author}{\bibfnamefont{S.}~\bibnamefont{Shankar}},
  \bibinfo{author}{\bibfnamefont{A.}~\bibnamefont{Narla}},
  \bibinfo{author}{\bibfnamefont{K.}~\bibnamefont{Sliwa}}, \bibnamefont{and}
  \bibinfo{author}{\bibfnamefont{M.}~\bibnamefont{Devoret}},
  \bibinfo{journal}{Applied Physics Letters} \textbf{\bibinfo{volume}{110}},
  \bibinfo{pages}{222603} (\bibinfo{year}{2017}).

\bibitem[{\citenamefont{Grimm et~al.}(2020)\citenamefont{Grimm, Frattini, Puri,
  Mundhada, Touzard, Mirrahimi, Girvin, Shankar, and Devoret}}]{grimm2020}
\bibinfo{author}{\bibfnamefont{A.}~\bibnamefont{Grimm}},
  \bibinfo{author}{\bibfnamefont{N.~E.} \bibnamefont{Frattini}},
  \bibinfo{author}{\bibfnamefont{S.}~\bibnamefont{Puri}},
  \bibinfo{author}{\bibfnamefont{S.~O.} \bibnamefont{Mundhada}},
  \bibinfo{author}{\bibfnamefont{S.}~\bibnamefont{Touzard}},
  \bibinfo{author}{\bibfnamefont{M.}~\bibnamefont{Mirrahimi}},
  \bibinfo{author}{\bibfnamefont{S.~M.} \bibnamefont{Girvin}},
  \bibinfo{author}{\bibfnamefont{S.}~\bibnamefont{Shankar}}, \bibnamefont{and}
  \bibinfo{author}{\bibfnamefont{M.~H.} \bibnamefont{Devoret}},
  \bibinfo{journal}{Nature} \textbf{\bibinfo{volume}{584}},
  \bibinfo{pages}{205} (\bibinfo{year}{2020}).

\bibitem[{\citenamefont{Mirrahimi et~al.}(2014)\citenamefont{Mirrahimi,
  Leghtas, Albert, Touzard, Schoelkopf, Jiang, and Devoret}}]{mirrahimi2014}
\bibinfo{author}{\bibfnamefont{M.}~\bibnamefont{Mirrahimi}},
  \bibinfo{author}{\bibfnamefont{Z.}~\bibnamefont{Leghtas}},
  \bibinfo{author}{\bibfnamefont{V.~V.} \bibnamefont{Albert}},
  \bibinfo{author}{\bibfnamefont{S.}~\bibnamefont{Touzard}},
  \bibinfo{author}{\bibfnamefont{R.~J.} \bibnamefont{Schoelkopf}},
  \bibinfo{author}{\bibfnamefont{L.}~\bibnamefont{Jiang}}, \bibnamefont{and}
  \bibinfo{author}{\bibfnamefont{M.~H.} \bibnamefont{Devoret}},
  \bibinfo{journal}{New Journal of Physics} \textbf{\bibinfo{volume}{16}},
  \bibinfo{pages}{045014} (\bibinfo{year}{2014}).

\bibitem[{\citenamefont{Leghtas et~al.}(2015)\citenamefont{Leghtas, Touzard,
  Pop, Kou, Vlastakis, Petrenko, Sliwa, Narla, Shankar, Hatridge
  et~al.}}]{leghtas2015}
\bibinfo{author}{\bibfnamefont{Z.}~\bibnamefont{Leghtas}},
  \bibinfo{author}{\bibfnamefont{S.}~\bibnamefont{Touzard}},
  \bibinfo{author}{\bibfnamefont{I.~M.} \bibnamefont{Pop}},
  \bibinfo{author}{\bibfnamefont{A.}~\bibnamefont{Kou}},
  \bibinfo{author}{\bibfnamefont{B.}~\bibnamefont{Vlastakis}},
  \bibinfo{author}{\bibfnamefont{A.}~\bibnamefont{Petrenko}},
  \bibinfo{author}{\bibfnamefont{K.~M.} \bibnamefont{Sliwa}},
  \bibinfo{author}{\bibfnamefont{A.}~\bibnamefont{Narla}},
  \bibinfo{author}{\bibfnamefont{S.}~\bibnamefont{Shankar}},
  \bibinfo{author}{\bibfnamefont{M.~J.} \bibnamefont{Hatridge}},
  \bibnamefont{et~al.}, \bibinfo{journal}{Science}
  \textbf{\bibinfo{volume}{347}}, \bibinfo{pages}{853} (\bibinfo{year}{2015}).

\bibitem[{\citenamefont{Touzard et~al.}(2018)\citenamefont{Touzard, Grimm,
  Leghtas, Mundhada, Reinhold, Axline, Reagor, Chou, Blumoff, Sliwa
  et~al.}}]{touzard2018}
\bibinfo{author}{\bibfnamefont{S.}~\bibnamefont{Touzard}},
  \bibinfo{author}{\bibfnamefont{A.}~\bibnamefont{Grimm}},
  \bibinfo{author}{\bibfnamefont{Z.}~\bibnamefont{Leghtas}},
  \bibinfo{author}{\bibfnamefont{S.~O.} \bibnamefont{Mundhada}},
  \bibinfo{author}{\bibfnamefont{P.}~\bibnamefont{Reinhold}},
  \bibinfo{author}{\bibfnamefont{C.}~\bibnamefont{Axline}},
  \bibinfo{author}{\bibfnamefont{M.}~\bibnamefont{Reagor}},
  \bibinfo{author}{\bibfnamefont{K.}~\bibnamefont{Chou}},
  \bibinfo{author}{\bibfnamefont{J.}~\bibnamefont{Blumoff}},
  \bibinfo{author}{\bibfnamefont{K.~M.} \bibnamefont{Sliwa}},
  \bibnamefont{et~al.}, \bibinfo{journal}{Phys. Rev. X}
  \textbf{\bibinfo{volume}{8}}, \bibinfo{pages}{021005} (\bibinfo{year}{2018}),
  \urlprefix\url{https://link.aps.org/doi/10.1103/PhysRevX.8.021005}.

\bibitem[{\citenamefont{Lescanne et~al.}(2020)\citenamefont{Lescanne, Villiers,
  Peronnin, Sarlette, Delbecq, Huard, Kontos, Mirrahimi, and
  Leghtas}}]{lescanne2020}
\bibinfo{author}{\bibfnamefont{R.}~\bibnamefont{Lescanne}},
  \bibinfo{author}{\bibfnamefont{M.}~\bibnamefont{Villiers}},
  \bibinfo{author}{\bibfnamefont{T.}~\bibnamefont{Peronnin}},
  \bibinfo{author}{\bibfnamefont{A.}~\bibnamefont{Sarlette}},
  \bibinfo{author}{\bibfnamefont{M.}~\bibnamefont{Delbecq}},
  \bibinfo{author}{\bibfnamefont{B.}~\bibnamefont{Huard}},
  \bibinfo{author}{\bibfnamefont{T.}~\bibnamefont{Kontos}},
  \bibinfo{author}{\bibfnamefont{M.}~\bibnamefont{Mirrahimi}},
  \bibnamefont{and} \bibinfo{author}{\bibfnamefont{Z.}~\bibnamefont{Leghtas}},
  \bibinfo{journal}{Nature Physics} \textbf{\bibinfo{volume}{16}},
  \bibinfo{pages}{509} (\bibinfo{year}{2020}).

\bibitem[{\citenamefont{Berdou et~al.}(2022)\citenamefont{Berdou, Murani,
  Reglade, Smith, Villiers, Palomo, Rosticher, Denis, Morfin, Delbecq
  et~al.}}]{berdou_one_2022}
\bibinfo{author}{\bibfnamefont{C.}~\bibnamefont{Berdou}},
  \bibinfo{author}{\bibfnamefont{A.}~\bibnamefont{Murani}},
  \bibinfo{author}{\bibfnamefont{U.}~\bibnamefont{Reglade}},
  \bibinfo{author}{\bibfnamefont{W.~C.} \bibnamefont{Smith}},
  \bibinfo{author}{\bibfnamefont{M.}~\bibnamefont{Villiers}},
  \bibinfo{author}{\bibfnamefont{J.}~\bibnamefont{Palomo}},
  \bibinfo{author}{\bibfnamefont{M.}~\bibnamefont{Rosticher}},
  \bibinfo{author}{\bibfnamefont{A.}~\bibnamefont{Denis}},
  \bibinfo{author}{\bibfnamefont{P.}~\bibnamefont{Morfin}},
  \bibinfo{author}{\bibfnamefont{M.}~\bibnamefont{Delbecq}},
  \bibnamefont{et~al.}, \bibinfo{journal}{arXiv:2204.09128 [quant-ph]}
  (\bibinfo{year}{2022}), \bibinfo{note}{arXiv: 2204.09128},
  \urlprefix\url{http://arxiv.org/abs/2204.09128}.

\bibitem[{\citenamefont{Harris}(1969)}]{Harris1969}
\bibinfo{author}{\bibfnamefont{S.}~\bibnamefont{Harris}},
  \bibinfo{journal}{Proceedings of the IEEE} \textbf{\bibinfo{volume}{57}},
  \bibinfo{pages}{2096} (\bibinfo{year}{1969}).

\bibitem[{\citenamefont{Milburn and Holmes}(1991)}]{milburn1991}
\bibinfo{author}{\bibfnamefont{G.~J.} \bibnamefont{Milburn}} \bibnamefont{and}
  \bibinfo{author}{\bibfnamefont{C.~A.} \bibnamefont{Holmes}},
  \bibinfo{journal}{Phys. Rev. A} \textbf{\bibinfo{volume}{44}},
  \bibinfo{pages}{4704} (\bibinfo{year}{1991}),
  \urlprefix\url{https://link.aps.org/doi/10.1103/PhysRevA.44.4704}.

\bibitem[{\citenamefont{Cochrane et~al.}(1999)\citenamefont{Cochrane, Milburn,
  and Munro}}]{cochrane1999}
\bibinfo{author}{\bibfnamefont{P.~T.} \bibnamefont{Cochrane}},
  \bibinfo{author}{\bibfnamefont{G.~J.} \bibnamefont{Milburn}},
  \bibnamefont{and} \bibinfo{author}{\bibfnamefont{W.~J.} \bibnamefont{Munro}},
  \bibinfo{journal}{Phys. Rev. A} \textbf{\bibinfo{volume}{59}},
  \bibinfo{pages}{2631} (\bibinfo{year}{1999}),
  \urlprefix\url{https://link.aps.org/doi/10.1103/PhysRevA.59.2631}.

\bibitem[{\citenamefont{Wielinga and Milburn}(1993)}]{wielinga1993}
\bibinfo{author}{\bibfnamefont{B.}~\bibnamefont{Wielinga}} \bibnamefont{and}
  \bibinfo{author}{\bibfnamefont{G.~J.} \bibnamefont{Milburn}},
  \bibinfo{journal}{Phys. Rev. A} \textbf{\bibinfo{volume}{48}},
  \bibinfo{pages}{2494} (\bibinfo{year}{1993}),
  \urlprefix\url{https://link.aps.org/doi/10.1103/PhysRevA.48.2494}.

\bibitem[{\citenamefont{Goto}(2016)}]{goto2016}
\bibinfo{author}{\bibfnamefont{H.}~\bibnamefont{Goto}},
  \bibinfo{journal}{Scientific Reports} \textbf{\bibinfo{volume}{6}},
  \bibinfo{pages}{21686} (\bibinfo{year}{2016}), ISSN
  \bibinfo{issn}{2045-2322}, \urlprefix\url{https://doi.org/10.1038/srep21686}.

\bibitem[{\citenamefont{Zhang and Dykman}(2017)}]{zhang2017}
\bibinfo{author}{\bibfnamefont{Y.}~\bibnamefont{Zhang}} \bibnamefont{and}
  \bibinfo{author}{\bibfnamefont{M.~I.} \bibnamefont{Dykman}},
  \bibinfo{journal}{Phys. Rev. A} \textbf{\bibinfo{volume}{95}},
  \bibinfo{pages}{053841} (\bibinfo{year}{2017}),
  \urlprefix\url{https://link.aps.org/doi/10.1103/PhysRevA.95.053841}.

\bibitem[{\citenamefont{Puri et~al.}(2017)\citenamefont{Puri, Boutin, and
  Blais}}]{puri2017}
\bibinfo{author}{\bibfnamefont{S.}~\bibnamefont{Puri}},
  \bibinfo{author}{\bibfnamefont{S.}~\bibnamefont{Boutin}}, \bibnamefont{and}
  \bibinfo{author}{\bibfnamefont{A.}~\bibnamefont{Blais}},
  \bibinfo{journal}{npj Quantum Information} \textbf{\bibinfo{volume}{3}},
  \bibinfo{pages}{1} (\bibinfo{year}{2017}).

\bibitem[{\citenamefont{Kinsler and Drummond}(1991)}]{Kinsler1991}
\bibinfo{author}{\bibfnamefont{P.}~\bibnamefont{Kinsler}} \bibnamefont{and}
  \bibinfo{author}{\bibfnamefont{P.~D.} \bibnamefont{Drummond}},
  \bibinfo{journal}{Phys. Rev. A} \textbf{\bibinfo{volume}{43}},
  \bibinfo{pages}{6194} (\bibinfo{year}{1991}),
  \urlprefix\url{https://link.aps.org/doi/10.1103/PhysRevA.43.6194}.

\bibitem[{\citenamefont{Guillaud and Mirrahimi}(2019)}]{Guillaud2019}
\bibinfo{author}{\bibfnamefont{J.}~\bibnamefont{Guillaud}} \bibnamefont{and}
  \bibinfo{author}{\bibfnamefont{M.}~\bibnamefont{Mirrahimi}},
  \bibinfo{journal}{Phys. Rev. X} \textbf{\bibinfo{volume}{9}},
  \bibinfo{pages}{041053} (\bibinfo{year}{2019}),
  \urlprefix\url{https://link.aps.org/doi/10.1103/PhysRevX.9.041053}.

\bibitem[{\citenamefont{Puri et~al.}(2020)\citenamefont{Puri, St-Jean, Gross,
  Grimm, Frattini, Iyer, Krishna, Touzard, Jiang, Blais et~al.}}]{puri2020}
\bibinfo{author}{\bibfnamefont{S.}~\bibnamefont{Puri}},
  \bibinfo{author}{\bibfnamefont{L.}~\bibnamefont{St-Jean}},
  \bibinfo{author}{\bibfnamefont{J.~A.} \bibnamefont{Gross}},
  \bibinfo{author}{\bibfnamefont{A.}~\bibnamefont{Grimm}},
  \bibinfo{author}{\bibfnamefont{N.~E.} \bibnamefont{Frattini}},
  \bibinfo{author}{\bibfnamefont{P.~S.} \bibnamefont{Iyer}},
  \bibinfo{author}{\bibfnamefont{A.}~\bibnamefont{Krishna}},
  \bibinfo{author}{\bibfnamefont{S.}~\bibnamefont{Touzard}},
  \bibinfo{author}{\bibfnamefont{L.}~\bibnamefont{Jiang}},
  \bibinfo{author}{\bibfnamefont{A.}~\bibnamefont{Blais}},
  \bibnamefont{et~al.}, \bibinfo{journal}{Science Advances}
  \textbf{\bibinfo{volume}{6}}, \bibinfo{pages}{eaay5901}
  (\bibinfo{year}{2020}),
  \eprint{https://www.science.org/doi/pdf/10.1126/sciadv.aay5901},
  \urlprefix\url{https://www.science.org/doi/abs/10.1126/sciadv.aay5901}.

\bibitem[{\citenamefont{Bonilla~Ataides
  et~al.}(2021)\citenamefont{Bonilla~Ataides, Tuckett, Bartlett, Flammia, and
  Brown}}]{bonilla_ataides_xzzx_2021}
\bibinfo{author}{\bibfnamefont{J.~P.} \bibnamefont{Bonilla~Ataides}},
  \bibinfo{author}{\bibfnamefont{D.~K.} \bibnamefont{Tuckett}},
  \bibinfo{author}{\bibfnamefont{S.~D.} \bibnamefont{Bartlett}},
  \bibinfo{author}{\bibfnamefont{S.~T.} \bibnamefont{Flammia}},
  \bibnamefont{and} \bibinfo{author}{\bibfnamefont{B.~J.} \bibnamefont{Brown}},
  \bibinfo{journal}{Nature Communications} \textbf{\bibinfo{volume}{12}},
  \bibinfo{pages}{2172} (\bibinfo{year}{2021}), ISSN \bibinfo{issn}{2041-1723},
  \urlprefix\url{https://www.nature.com/articles/s41467-021-22274-1}.

\bibitem[{\citenamefont{Darmawan et~al.}(2021)\citenamefont{Darmawan, Brown,
  Grimsmo, Tuckett, and Puri}}]{darmawan_practical_2021}
\bibinfo{author}{\bibfnamefont{A.~S.} \bibnamefont{Darmawan}},
  \bibinfo{author}{\bibfnamefont{B.~J.} \bibnamefont{Brown}},
  \bibinfo{author}{\bibfnamefont{A.~L.} \bibnamefont{Grimsmo}},
  \bibinfo{author}{\bibfnamefont{D.~K.} \bibnamefont{Tuckett}},
  \bibnamefont{and} \bibinfo{author}{\bibfnamefont{S.}~\bibnamefont{Puri}},
  \bibinfo{journal}{PRX Quantum} \textbf{\bibinfo{volume}{2}},
  \bibinfo{pages}{030345} (\bibinfo{year}{2021}), \bibinfo{note}{publisher:
  American Physical Society},
  \urlprefix\url{https://link.aps.org/doi/10.1103/PRXQuantum.2.030345}.

\bibitem[{\citenamefont{Chamberland et~al.}(2022)\citenamefont{Chamberland,
  Noh, Arrangoiz-Arriola, Campbell, Hann, Iverson, Putterman, Bohdanowicz,
  Flammia, Keller et~al.}}]{chamberland_building_2022}
\bibinfo{author}{\bibfnamefont{C.}~\bibnamefont{Chamberland}},
  \bibinfo{author}{\bibfnamefont{K.}~\bibnamefont{Noh}},
  \bibinfo{author}{\bibfnamefont{P.}~\bibnamefont{Arrangoiz-Arriola}},
  \bibinfo{author}{\bibfnamefont{E.~T.} \bibnamefont{Campbell}},
  \bibinfo{author}{\bibfnamefont{C.~T.} \bibnamefont{Hann}},
  \bibinfo{author}{\bibfnamefont{J.}~\bibnamefont{Iverson}},
  \bibinfo{author}{\bibfnamefont{H.}~\bibnamefont{Putterman}},
  \bibinfo{author}{\bibfnamefont{T.~C.} \bibnamefont{Bohdanowicz}},
  \bibinfo{author}{\bibfnamefont{S.~T.} \bibnamefont{Flammia}},
  \bibinfo{author}{\bibfnamefont{A.}~\bibnamefont{Keller}},
  \bibnamefont{et~al.}, \bibinfo{journal}{PRX Quantum}
  \textbf{\bibinfo{volume}{3}}, \bibinfo{pages}{010329} (\bibinfo{year}{2022}),
  \bibinfo{note}{publisher: American Physical Society},
  \urlprefix\url{https://link.aps.org/doi/10.1103/PRXQuantum.3.010329}.

\bibitem[{\citenamefont{Puri et~al.}(2019)\citenamefont{Puri, Grimm,
  Campagne-Ibarcq, Eickbusch, Noh, Roberts, Jiang, Mirrahimi, Devoret, and
  Girvin}}]{puri2019}
\bibinfo{author}{\bibfnamefont{S.}~\bibnamefont{Puri}},
  \bibinfo{author}{\bibfnamefont{A.}~\bibnamefont{Grimm}},
  \bibinfo{author}{\bibfnamefont{P.}~\bibnamefont{Campagne-Ibarcq}},
  \bibinfo{author}{\bibfnamefont{A.}~\bibnamefont{Eickbusch}},
  \bibinfo{author}{\bibfnamefont{K.}~\bibnamefont{Noh}},
  \bibinfo{author}{\bibfnamefont{G.}~\bibnamefont{Roberts}},
  \bibinfo{author}{\bibfnamefont{L.}~\bibnamefont{Jiang}},
  \bibinfo{author}{\bibfnamefont{M.}~\bibnamefont{Mirrahimi}},
  \bibinfo{author}{\bibfnamefont{M.~H.} \bibnamefont{Devoret}},
  \bibnamefont{and} \bibinfo{author}{\bibfnamefont{S.~M.}
  \bibnamefont{Girvin}}, \bibinfo{journal}{Phys. Rev. X}
  \textbf{\bibinfo{volume}{9}}, \bibinfo{pages}{041009} (\bibinfo{year}{2019}),
  \urlprefix\url{https://link.aps.org/doi/10.1103/PhysRevX.9.041009}.

\bibitem[{\citenamefont{Putterman et~al.}(2022)\citenamefont{Putterman,
  Iverson, Xu, Jiang, Painter, Brandão, and Noh}}]{putterman_stabilizing_2022}
\bibinfo{author}{\bibfnamefont{H.}~\bibnamefont{Putterman}},
  \bibinfo{author}{\bibfnamefont{J.}~\bibnamefont{Iverson}},
  \bibinfo{author}{\bibfnamefont{Q.}~\bibnamefont{Xu}},
  \bibinfo{author}{\bibfnamefont{L.}~\bibnamefont{Jiang}},
  \bibinfo{author}{\bibfnamefont{O.}~\bibnamefont{Painter}},
  \bibinfo{author}{\bibfnamefont{F.~G.} \bibnamefont{Brandão}},
  \bibnamefont{and} \bibinfo{author}{\bibfnamefont{K.}~\bibnamefont{Noh}},
  \bibinfo{journal}{Physical Review Letters} \textbf{\bibinfo{volume}{128}},
  \bibinfo{pages}{110502} (\bibinfo{year}{2022}), \bibinfo{note}{publisher:
  American Physical Society},
  \urlprefix\url{https://link.aps.org/doi/10.1103/PhysRevLett.128.110502}.

\bibitem[{\citenamefont{Gautier et~al.}(2022)\citenamefont{Gautier, Sarlette,
  and Mirrahimi}}]{gautier_combined_2022}
\bibinfo{author}{\bibfnamefont{R.}~\bibnamefont{Gautier}},
  \bibinfo{author}{\bibfnamefont{A.}~\bibnamefont{Sarlette}}, \bibnamefont{and}
  \bibinfo{author}{\bibfnamefont{M.}~\bibnamefont{Mirrahimi}},
  \bibinfo{journal}{arXiv:2112.05545 [quant-ph]}  (\bibinfo{year}{2022}),
  \bibinfo{note}{arXiv: 2112.05545},
  \urlprefix\url{http://arxiv.org/abs/2112.05545}.

\bibitem[{\citenamefont{Curtright et~al.}(2013)\citenamefont{Curtright,
  Fairlie, and Zachos}}]{curtright2013}
\bibinfo{author}{\bibfnamefont{T.~L.} \bibnamefont{Curtright}},
  \bibinfo{author}{\bibfnamefont{D.~B.} \bibnamefont{Fairlie}},
  \bibnamefont{and} \bibinfo{author}{\bibfnamefont{C.~K.}
  \bibnamefont{Zachos}}, \emph{\bibinfo{title}{A concise treatise on quantum
  mechanics in phase space}} (\bibinfo{publisher}{World Scientific Publishing
  Company}, \bibinfo{year}{2013}).

\bibitem[{\citenamefont{Breuer and Holthaus}(1991)}]{BREUER1991249}
\bibinfo{author}{\bibfnamefont{H.}~\bibnamefont{Breuer}} \bibnamefont{and}
  \bibinfo{author}{\bibfnamefont{M.}~\bibnamefont{Holthaus}},
  \bibinfo{journal}{Annals of Physics} \textbf{\bibinfo{volume}{211}},
  \bibinfo{pages}{249} (\bibinfo{year}{1991}), ISSN \bibinfo{issn}{0003-4916},
  \urlprefix\url{https://www.sciencedirect.com/science/article/pii/000349169190206N}.

\bibitem[{\citenamefont{Frattini}(2021)}]{Frattini2021}
\bibinfo{author}{\bibfnamefont{N.}~\bibnamefont{Frattini}},
  \emph{\bibinfo{title}{Three-wave Mixing in Superconducting Circuits:
  Stabilizing Cats with SNAILs}} (\bibinfo{publisher}{Yale University, thesis},
  \bibinfo{year}{2021}).

\bibitem[{\citenamefont{Petrescu et~al.}(2020)\citenamefont{Petrescu,
  Malekakhlagh, and T\"ureci}}]{petrescu2020}
\bibinfo{author}{\bibfnamefont{A.}~\bibnamefont{Petrescu}},
  \bibinfo{author}{\bibfnamefont{M.}~\bibnamefont{Malekakhlagh}},
  \bibnamefont{and} \bibinfo{author}{\bibfnamefont{H.~E.}
  \bibnamefont{T\"ureci}}, \bibinfo{journal}{Phys. Rev. B}
  \textbf{\bibinfo{volume}{101}}, \bibinfo{pages}{134510}
  (\bibinfo{year}{2020}).

\bibitem[{\citenamefont{Marthaler and Dykman}(2007)}]{dykman2007}
\bibinfo{author}{\bibfnamefont{M.}~\bibnamefont{Marthaler}} \bibnamefont{and}
  \bibinfo{author}{\bibfnamefont{M.~I.} \bibnamefont{Dykman}},
  \bibinfo{journal}{Phys. Rev. A} \textbf{\bibinfo{volume}{76}},
  \bibinfo{pages}{010102} (\bibinfo{year}{2007}),
  \urlprefix\url{https://link.aps.org/doi/10.1103/PhysRevA.76.010102}.

\bibitem[{\citenamefont{Roberts and Clerk}(2020)}]{roberts2020}
\bibinfo{author}{\bibfnamefont{D.}~\bibnamefont{Roberts}} \bibnamefont{and}
  \bibinfo{author}{\bibfnamefont{A.~A.} \bibnamefont{Clerk}},
  \bibinfo{journal}{Phys. Rev. X} \textbf{\bibinfo{volume}{10}},
  \bibinfo{pages}{021022} (\bibinfo{year}{2020}),
  \urlprefix\url{https://link.aps.org/doi/10.1103/PhysRevX.10.021022}.

\bibitem[{\citenamefont{Venkatraman et~al.}(2022)\citenamefont{Venkatraman,
  Cortiñas, Frattini, Xiao, and Devoret}}]{inprep2022}
\bibinfo{author}{\bibfnamefont{J.}~\bibnamefont{Venkatraman}},
  \bibinfo{author}{\bibfnamefont{R.}~\bibnamefont{Cortiñas}},
  \bibinfo{author}{\bibfnamefont{N.~E.} \bibnamefont{Frattini}},
  \bibinfo{author}{\bibfnamefont{X.}~\bibnamefont{Xiao}}, \bibnamefont{and}
  \bibinfo{author}{\bibfnamefont{M.~H.} \bibnamefont{Devoret}},
  \bibinfo{journal}{In preparation}  (\bibinfo{year}{2022}).

\bibitem[{\citenamefont{Campagne-Ibarcq
  et~al.}(2020)\citenamefont{Campagne-Ibarcq, Eickbusch, Touzard, Zalys-Geller,
  Frattini, Sivak, Reinhold, Puri, Shankar, Schoelkopf
  et~al.}}]{Campagne-Ibarcq2020}
\bibinfo{author}{\bibfnamefont{P.}~\bibnamefont{Campagne-Ibarcq}},
  \bibinfo{author}{\bibfnamefont{A.}~\bibnamefont{Eickbusch}},
  \bibinfo{author}{\bibfnamefont{S.}~\bibnamefont{Touzard}},
  \bibinfo{author}{\bibfnamefont{E.}~\bibnamefont{Zalys-Geller}},
  \bibinfo{author}{\bibfnamefont{N.~E.} \bibnamefont{Frattini}},
  \bibinfo{author}{\bibfnamefont{V.~V.} \bibnamefont{Sivak}},
  \bibinfo{author}{\bibfnamefont{P.}~\bibnamefont{Reinhold}},
  \bibinfo{author}{\bibfnamefont{S.}~\bibnamefont{Puri}},
  \bibinfo{author}{\bibfnamefont{S.}~\bibnamefont{Shankar}},
  \bibinfo{author}{\bibfnamefont{R.~J.} \bibnamefont{Schoelkopf}},
  \bibnamefont{et~al.}, \bibinfo{journal}{Nature}
  \textbf{\bibinfo{volume}{584}}, \bibinfo{pages}{368} (\bibinfo{year}{2020}),
  ISSN \bibinfo{issn}{1476-4687},
  \urlprefix\url{https://doi.org/10.1038/s41586-020-2603-3}.

\bibitem[{\citenamefont{Grimsmo and Puri}(2021)}]{grimsmo_quantum_2021}
\bibinfo{author}{\bibfnamefont{A.~L.} \bibnamefont{Grimsmo}} \bibnamefont{and}
  \bibinfo{author}{\bibfnamefont{S.}~\bibnamefont{Puri}}, \bibinfo{journal}{PRX
  Quantum} \textbf{\bibinfo{volume}{2}}, \bibinfo{pages}{020101}
  (\bibinfo{year}{2021}), \bibinfo{note}{publisher: American Physical Society},
  \urlprefix\url{https://link.aps.org/doi/10.1103/PRXQuantum.2.020101}.

\bibitem[{\citenamefont{Xu et~al.}(2022)\citenamefont{Xu, Iverson, Brandão,
  and Jiang}}]{xu_engineering_2022}
\bibinfo{author}{\bibfnamefont{Q.}~\bibnamefont{Xu}},
  \bibinfo{author}{\bibfnamefont{J.~K.} \bibnamefont{Iverson}},
  \bibinfo{author}{\bibfnamefont{F.~G. S.~L.} \bibnamefont{Brandão}},
  \bibnamefont{and} \bibinfo{author}{\bibfnamefont{L.}~\bibnamefont{Jiang}},
  \bibinfo{journal}{Physical Review Research} \textbf{\bibinfo{volume}{4}},
  \bibinfo{pages}{013082} (\bibinfo{year}{2022}), \bibinfo{note}{publisher:
  American Physical Society},
  \urlprefix\url{https://link.aps.org/doi/10.1103/PhysRevResearch.4.013082}.

\bibitem[{\citenamefont{Kanao et~al.}(2022)\citenamefont{Kanao, Masuda,
  Kawabata, and Goto}}]{kanao_quantum_2022}
\bibinfo{author}{\bibfnamefont{T.}~\bibnamefont{Kanao}},
  \bibinfo{author}{\bibfnamefont{S.}~\bibnamefont{Masuda}},
  \bibinfo{author}{\bibfnamefont{S.}~\bibnamefont{Kawabata}}, \bibnamefont{and}
  \bibinfo{author}{\bibfnamefont{H.}~\bibnamefont{Goto}},
  \bibinfo{journal}{Physical Review Applied} \textbf{\bibinfo{volume}{18}},
  \bibinfo{pages}{014019} (\bibinfo{year}{2022}), \bibinfo{note}{publisher:
  American Physical Society},
  \urlprefix\url{https://link.aps.org/doi/10.1103/PhysRevApplied.18.014019}.

\bibitem[{\citenamefont{Chono et~al.}(2022)\citenamefont{Chono, Kanao, and
  Goto}}]{chono_two-qubit_2022}
\bibinfo{author}{\bibfnamefont{H.}~\bibnamefont{Chono}},
  \bibinfo{author}{\bibfnamefont{T.}~\bibnamefont{Kanao}}, \bibnamefont{and}
  \bibinfo{author}{\bibfnamefont{H.}~\bibnamefont{Goto}},
  \bibinfo{journal}{arXiv:2204.03347 [cond-mat, physics:physics,
  physics:quant-ph]}  (\bibinfo{year}{2022}), \bibinfo{note}{arXiv:
  2204.03347}, \urlprefix\url{http://arxiv.org/abs/2204.03347}.

\bibitem[{\citenamefont{Wilson et~al.}(2011)\citenamefont{Wilson, Johansson,
  Pourkabirian, Simoen, Johansson, Duty, Nori, and Delsing}}]{Wilson2011}
\bibinfo{author}{\bibfnamefont{C.~M.} \bibnamefont{Wilson}},
  \bibinfo{author}{\bibfnamefont{G.}~\bibnamefont{Johansson}},
  \bibinfo{author}{\bibfnamefont{A.}~\bibnamefont{Pourkabirian}},
  \bibinfo{author}{\bibfnamefont{M.}~\bibnamefont{Simoen}},
  \bibinfo{author}{\bibfnamefont{J.~R.} \bibnamefont{Johansson}},
  \bibinfo{author}{\bibfnamefont{T.}~\bibnamefont{Duty}},
  \bibinfo{author}{\bibfnamefont{F.}~\bibnamefont{Nori}}, \bibnamefont{and}
  \bibinfo{author}{\bibfnamefont{P.}~\bibnamefont{Delsing}},
  \bibinfo{journal}{Nature} \textbf{\bibinfo{volume}{479}},
  \bibinfo{pages}{376} (\bibinfo{year}{2011}), ISSN \bibinfo{issn}{1476-4687},
  \urlprefix\url{https://doi.org/10.1038/nature10561}.

\bibitem[{\citenamefont{Dykman}(2012)}]{dykmanBook2012}
\bibinfo{author}{\bibfnamefont{M.}~\bibnamefont{Dykman}},
  \emph{\bibinfo{title}{Fluctuating Nonlinear Oscillators: From Nanomechanics
  to Quantum Superconducting Circuits}} (\bibinfo{publisher}{Oxford},
  \bibinfo{year}{2012}),
  \urlprefix\url{DOI:10.1093/acprof:oso/9780199691388.001.0001}.

\bibitem[{\citenamefont{Caprio et~al.}(2008)\citenamefont{Caprio, Cejnar, and
  Iachello}}]{caprio2008}
\bibinfo{author}{\bibfnamefont{M.}~\bibnamefont{Caprio}},
  \bibinfo{author}{\bibfnamefont{P.}~\bibnamefont{Cejnar}}, \bibnamefont{and}
  \bibinfo{author}{\bibfnamefont{F.}~\bibnamefont{Iachello}},
  \bibinfo{journal}{Annals of Physics} \textbf{\bibinfo{volume}{323}},
  \bibinfo{pages}{1106} (\bibinfo{year}{2008}), ISSN \bibinfo{issn}{0003-4916},
  \urlprefix\url{https://www.sciencedirect.com/science/article/pii/S0003491607001042}.

\bibitem[{\citenamefont{Ribeiro et~al.}(2008)\citenamefont{Ribeiro, Vidal, and
  Mosseri}}]{ribeiro2008}
\bibinfo{author}{\bibfnamefont{P.}~\bibnamefont{Ribeiro}},
  \bibinfo{author}{\bibfnamefont{J.}~\bibnamefont{Vidal}}, \bibnamefont{and}
  \bibinfo{author}{\bibfnamefont{R.}~\bibnamefont{Mosseri}},
  \bibinfo{journal}{Phys. Rev. E} \textbf{\bibinfo{volume}{78}},
  \bibinfo{pages}{021106} (\bibinfo{year}{2008}),
  \urlprefix\url{https://link.aps.org/doi/10.1103/PhysRevE.78.021106}.

\bibitem[{\citenamefont{Goto and Kanao}(2021)}]{Goto2021}
\bibinfo{author}{\bibfnamefont{H.}~\bibnamefont{Goto}} \bibnamefont{and}
  \bibinfo{author}{\bibfnamefont{T.}~\bibnamefont{Kanao}},
  \bibinfo{journal}{Phys. Rev. Research} \textbf{\bibinfo{volume}{3}},
  \bibinfo{pages}{043196} (\bibinfo{year}{2021}),
  \urlprefix\url{https://link.aps.org/doi/10.1103/PhysRevResearch.3.043196}.

\bibitem[{\citenamefont{Monteoliva and Paz}(2000)}]{Monteoliva2000}
\bibinfo{author}{\bibfnamefont{D.}~\bibnamefont{Monteoliva}} \bibnamefont{and}
  \bibinfo{author}{\bibfnamefont{J.~P.} \bibnamefont{Paz}},
  \bibinfo{journal}{Phys. Rev. Lett.} \textbf{\bibinfo{volume}{85}},
  \bibinfo{pages}{3373} (\bibinfo{year}{2000}),
  \urlprefix\url{https://link.aps.org/doi/10.1103/PhysRevLett.85.3373}.

\bibitem[{\citenamefont{Habib et~al.}(1998)\citenamefont{Habib, Shizume, and
  Zurek}}]{zurek1998}
\bibinfo{author}{\bibfnamefont{S.}~\bibnamefont{Habib}},
  \bibinfo{author}{\bibfnamefont{K.}~\bibnamefont{Shizume}}, \bibnamefont{and}
  \bibinfo{author}{\bibfnamefont{W.~H.} \bibnamefont{Zurek}},
  \bibinfo{journal}{Phys. Rev. Lett.} \textbf{\bibinfo{volume}{80}},
  \bibinfo{pages}{4361} (\bibinfo{year}{1998}),
  \urlprefix\url{https://link.aps.org/doi/10.1103/PhysRevLett.80.4361}.

\end{thebibliography}


\begin{thebibliography}{43}
\expandafter\ifx\csname natexlab\endcsname\relax\def\natexlab#1{#1}\fi
\expandafter\ifx\csname bibnamefont\endcsname\relax
  \def\bibnamefont#1{#1}\fi
\expandafter\ifx\csname bibfnamefont\endcsname\relax
  \def\bibfnamefont#1{#1}\fi
\expandafter\ifx\csname citenamefont\endcsname\relax
  \def\citenamefont#1{#1}\fi
\expandafter\ifx\csname url\endcsname\relax
  \def\url#1{\texttt{#1}}\fi
\expandafter\ifx\csname urlprefix\endcsname\relax\def\urlprefix{URL }\fi
\providecommand{\bibinfo}[2]{#2}
\providecommand{\eprint}[2][]{\url{#2}}

\bibitem[{\citenamefont{Frattini et~al.}(2017)\citenamefont{Frattini, Vool,
  Shankar, Narla, Sliwa, and Devoret}}]{frattini2017}
\bibinfo{author}{\bibfnamefont{N.}~\bibnamefont{Frattini}},
  \bibinfo{author}{\bibfnamefont{U.}~\bibnamefont{Vool}},
  \bibinfo{author}{\bibfnamefont{S.}~\bibnamefont{Shankar}},
  \bibinfo{author}{\bibfnamefont{A.}~\bibnamefont{Narla}},
  \bibinfo{author}{\bibfnamefont{K.}~\bibnamefont{Sliwa}}, \bibnamefont{and}
  \bibinfo{author}{\bibfnamefont{M.}~\bibnamefont{Devoret}},
  \bibinfo{journal}{Applied Physics Letters} \textbf{\bibinfo{volume}{110}},
  \bibinfo{pages}{222603} (\bibinfo{year}{2017}).

\bibitem[{\citenamefont{Venkatraman et~al.}(2021)\citenamefont{Venkatraman,
  Xiao, Cortiñas, Eickbusch, and Devoret}}]{venkatraman2021}
\bibinfo{author}{\bibfnamefont{J.}~\bibnamefont{Venkatraman}},
  \bibinfo{author}{\bibfnamefont{X.}~\bibnamefont{Xiao}},
  \bibinfo{author}{\bibfnamefont{R.~G.} \bibnamefont{Cortiñas}},
  \bibinfo{author}{\bibfnamefont{A.}~\bibnamefont{Eickbusch}},
  \bibnamefont{and} \bibinfo{author}{\bibfnamefont{M.~H.}
  \bibnamefont{Devoret}} (\bibinfo{year}{2021}),
  \urlprefix\url{https://doi.org/10.48550/arXiv.2108.02861}.

\bibitem[{\citenamefont{Frattini et~al.}(2018)\citenamefont{Frattini, Sivak,
  Lingenfelter, Shankar, and Devoret}}]{frattini_optimizing_2018}
\bibinfo{author}{\bibfnamefont{N.~E.} \bibnamefont{Frattini}},
  \bibinfo{author}{\bibfnamefont{V.~V.} \bibnamefont{Sivak}},
  \bibinfo{author}{\bibfnamefont{A.}~\bibnamefont{Lingenfelter}},
  \bibinfo{author}{\bibfnamefont{S.}~\bibnamefont{Shankar}}, \bibnamefont{and}
  \bibinfo{author}{\bibfnamefont{M.~H.} \bibnamefont{Devoret}},
  \bibinfo{journal}{Physical Review Applied} \textbf{\bibinfo{volume}{10}},
  \bibinfo{pages}{054020} (\bibinfo{year}{2018}),
  \urlprefix\url{https://link.aps.org/doi/10.1103/PhysRevApplied.10.054020}.

\bibitem[{\citenamefont{Sivak et~al.}(2019)\citenamefont{Sivak, Frattini,
  Joshi, Lingenfelter, Shankar, and Devoret}}]{sivak_kerr-free_2019}
\bibinfo{author}{\bibfnamefont{V.}~\bibnamefont{Sivak}},
  \bibinfo{author}{\bibfnamefont{N.}~\bibnamefont{Frattini}},
  \bibinfo{author}{\bibfnamefont{V.}~\bibnamefont{Joshi}},
  \bibinfo{author}{\bibfnamefont{A.}~\bibnamefont{Lingenfelter}},
  \bibinfo{author}{\bibfnamefont{S.}~\bibnamefont{Shankar}}, \bibnamefont{and}
  \bibinfo{author}{\bibfnamefont{M.}~\bibnamefont{Devoret}},
  \bibinfo{journal}{Physical Review Applied} \textbf{\bibinfo{volume}{11}},
  \bibinfo{pages}{054060} (\bibinfo{year}{2019}),
  \urlprefix\url{https://link.aps.org/doi/10.1103/PhysRevApplied.11.054060}.

\bibitem[{\citenamefont{Grimm et~al.}(2020)\citenamefont{Grimm, Frattini, Puri,
  Mundhada, Touzard, Mirrahimi, Girvin, Shankar, and Devoret}}]{grimm2020}
\bibinfo{author}{\bibfnamefont{A.}~\bibnamefont{Grimm}},
  \bibinfo{author}{\bibfnamefont{N.~E.} \bibnamefont{Frattini}},
  \bibinfo{author}{\bibfnamefont{S.}~\bibnamefont{Puri}},
  \bibinfo{author}{\bibfnamefont{S.~O.} \bibnamefont{Mundhada}},
  \bibinfo{author}{\bibfnamefont{S.}~\bibnamefont{Touzard}},
  \bibinfo{author}{\bibfnamefont{M.}~\bibnamefont{Mirrahimi}},
  \bibinfo{author}{\bibfnamefont{S.~M.} \bibnamefont{Girvin}},
  \bibinfo{author}{\bibfnamefont{S.}~\bibnamefont{Shankar}}, \bibnamefont{and}
  \bibinfo{author}{\bibfnamefont{M.~H.} \bibnamefont{Devoret}},
  \bibinfo{journal}{Nature} \textbf{\bibinfo{volume}{584}},
  \bibinfo{pages}{205} (\bibinfo{year}{2020}).

\bibitem[{\citenamefont{Frattini}(2021)}]{Frattini2021}
\bibinfo{author}{\bibfnamefont{N.}~\bibnamefont{Frattini}},
  \emph{\bibinfo{title}{Three-wave Mixing in Superconducting Circuits:
  Stabilizing Cats with SNAILs}} (\bibinfo{publisher}{Yale University, thesis},
  \bibinfo{year}{2021}).

\bibitem[{\citenamefont{Puri et~al.}(2020)\citenamefont{Puri, St-Jean, Gross,
  Grimm, Frattini, Iyer, Krishna, Touzard, Jiang, Blais et~al.}}]{puri2020}
\bibinfo{author}{\bibfnamefont{S.}~\bibnamefont{Puri}},
  \bibinfo{author}{\bibfnamefont{L.}~\bibnamefont{St-Jean}},
  \bibinfo{author}{\bibfnamefont{J.~A.} \bibnamefont{Gross}},
  \bibinfo{author}{\bibfnamefont{A.}~\bibnamefont{Grimm}},
  \bibinfo{author}{\bibfnamefont{N.~E.} \bibnamefont{Frattini}},
  \bibinfo{author}{\bibfnamefont{P.~S.} \bibnamefont{Iyer}},
  \bibinfo{author}{\bibfnamefont{A.}~\bibnamefont{Krishna}},
  \bibinfo{author}{\bibfnamefont{S.}~\bibnamefont{Touzard}},
  \bibinfo{author}{\bibfnamefont{L.}~\bibnamefont{Jiang}},
  \bibinfo{author}{\bibfnamefont{A.}~\bibnamefont{Blais}},
  \bibnamefont{et~al.}, \bibinfo{journal}{Science Advances}
  \textbf{\bibinfo{volume}{6}}, \bibinfo{pages}{eaay5901}
  (\bibinfo{year}{2020}),
  \eprint{https://www.science.org/doi/pdf/10.1126/sciadv.aay5901},
  \urlprefix\url{https://www.science.org/doi/abs/10.1126/sciadv.aay5901}.

\bibitem[{\citenamefont{Nigg et~al.}(2012)\citenamefont{Nigg, Paik, Vlastakis,
  Kirchmair, Shankar, Frunzio, Devoret, Schoelkopf, and
  Girvin}}]{nigg_black-box_2012}
\bibinfo{author}{\bibfnamefont{S.~E.} \bibnamefont{Nigg}},
  \bibinfo{author}{\bibfnamefont{H.}~\bibnamefont{Paik}},
  \bibinfo{author}{\bibfnamefont{B.}~\bibnamefont{Vlastakis}},
  \bibinfo{author}{\bibfnamefont{G.}~\bibnamefont{Kirchmair}},
  \bibinfo{author}{\bibfnamefont{S.}~\bibnamefont{Shankar}},
  \bibinfo{author}{\bibfnamefont{L.}~\bibnamefont{Frunzio}},
  \bibinfo{author}{\bibfnamefont{M.~H.} \bibnamefont{Devoret}},
  \bibinfo{author}{\bibfnamefont{R.~J.} \bibnamefont{Schoelkopf}},
  \bibnamefont{and} \bibinfo{author}{\bibfnamefont{S.~M.}
  \bibnamefont{Girvin}}, \bibinfo{journal}{Physical Review Letters}
  \textbf{\bibinfo{volume}{108}}, \bibinfo{pages}{240502}
  (\bibinfo{year}{2012}),
  \urlprefix\url{https://link.aps.org/doi/10.1103/PhysRevLett.108.240502}.

\bibitem[{\citenamefont{Minev et~al.}(2021)\citenamefont{Minev, Leghtas,
  Mundhada, Christakis, Pop, and Devoret}}]{minev_energy-participation_2021}
\bibinfo{author}{\bibfnamefont{Z.~K.} \bibnamefont{Minev}},
  \bibinfo{author}{\bibfnamefont{Z.}~\bibnamefont{Leghtas}},
  \bibinfo{author}{\bibfnamefont{S.~O.} \bibnamefont{Mundhada}},
  \bibinfo{author}{\bibfnamefont{L.}~\bibnamefont{Christakis}},
  \bibinfo{author}{\bibfnamefont{I.~M.} \bibnamefont{Pop}}, \bibnamefont{and}
  \bibinfo{author}{\bibfnamefont{M.~H.} \bibnamefont{Devoret}},
  \bibinfo{journal}{arXiv:2010.00620 [cond-mat, physics:quant-ph]}
  (\bibinfo{year}{2021}), \bibinfo{note}{arXiv: 2010.00620},
  \urlprefix\url{http://arxiv.org/abs/2010.00620}.

\bibitem[{\citenamefont{Axline et~al.}(2016)\citenamefont{Axline, Reagor,
  Heeres, Reinhold, Wang, Shain, Pfaff, Chu, Frunzio, and
  Schoelkopf}}]{axline_architecture_2016}
\bibinfo{author}{\bibfnamefont{C.}~\bibnamefont{Axline}},
  \bibinfo{author}{\bibfnamefont{M.}~\bibnamefont{Reagor}},
  \bibinfo{author}{\bibfnamefont{R.}~\bibnamefont{Heeres}},
  \bibinfo{author}{\bibfnamefont{P.}~\bibnamefont{Reinhold}},
  \bibinfo{author}{\bibfnamefont{C.}~\bibnamefont{Wang}},
  \bibinfo{author}{\bibfnamefont{K.}~\bibnamefont{Shain}},
  \bibinfo{author}{\bibfnamefont{W.}~\bibnamefont{Pfaff}},
  \bibinfo{author}{\bibfnamefont{Y.}~\bibnamefont{Chu}},
  \bibinfo{author}{\bibfnamefont{L.}~\bibnamefont{Frunzio}}, \bibnamefont{and}
  \bibinfo{author}{\bibfnamefont{R.~J.} \bibnamefont{Schoelkopf}},
  \bibinfo{journal}{Applied Physics Letters} \textbf{\bibinfo{volume}{109}},
  \bibinfo{pages}{042601} (\bibinfo{year}{2016}), ISSN
  \bibinfo{issn}{0003-6951},
  \urlprefix\url{https://aip.scitation.org/doi/10.1063/1.4959241}.

\bibitem[{\citenamefont{Xu et~al.}(2022)\citenamefont{Xu, Iverson, Brandão,
  and Jiang}}]{xu_engineering_2022}
\bibinfo{author}{\bibfnamefont{Q.}~\bibnamefont{Xu}},
  \bibinfo{author}{\bibfnamefont{J.~K.} \bibnamefont{Iverson}},
  \bibinfo{author}{\bibfnamefont{F.~G. S.~L.} \bibnamefont{Brandão}},
  \bibnamefont{and} \bibinfo{author}{\bibfnamefont{L.}~\bibnamefont{Jiang}},
  \bibinfo{journal}{Physical Review Research} \textbf{\bibinfo{volume}{4}},
  \bibinfo{pages}{013082} (\bibinfo{year}{2022}), \bibinfo{note}{publisher:
  American Physical Society},
  \urlprefix\url{https://link.aps.org/doi/10.1103/PhysRevResearch.4.013082}.

\bibitem[{\citenamefont{Didier et~al.}(2015)\citenamefont{Didier, Bourassa, and
  Blais}}]{didier_fast_2015}
\bibinfo{author}{\bibfnamefont{N.}~\bibnamefont{Didier}},
  \bibinfo{author}{\bibfnamefont{J.}~\bibnamefont{Bourassa}}, \bibnamefont{and}
  \bibinfo{author}{\bibfnamefont{A.}~\bibnamefont{Blais}},
  \bibinfo{journal}{Physical Review Letters} \textbf{\bibinfo{volume}{115}},
  \bibinfo{pages}{203601} (\bibinfo{year}{2015}), \bibinfo{note}{publisher:
  American Physical Society},
  \urlprefix\url{https://link.aps.org/doi/10.1103/PhysRevLett.115.203601}.

\bibitem[{\citenamefont{Touzard et~al.}(2019)\citenamefont{Touzard, Kou,
  Frattini, Sivak, Puri, Grimm, Frunzio, Shankar, and Devoret}}]{Touzard2019}
\bibinfo{author}{\bibfnamefont{S.}~\bibnamefont{Touzard}},
  \bibinfo{author}{\bibfnamefont{A.}~\bibnamefont{Kou}},
  \bibinfo{author}{\bibfnamefont{N.~E.} \bibnamefont{Frattini}},
  \bibinfo{author}{\bibfnamefont{V.~V.} \bibnamefont{Sivak}},
  \bibinfo{author}{\bibfnamefont{S.}~\bibnamefont{Puri}},
  \bibinfo{author}{\bibfnamefont{A.}~\bibnamefont{Grimm}},
  \bibinfo{author}{\bibfnamefont{L.}~\bibnamefont{Frunzio}},
  \bibinfo{author}{\bibfnamefont{S.}~\bibnamefont{Shankar}}, \bibnamefont{and}
  \bibinfo{author}{\bibfnamefont{M.~H.} \bibnamefont{Devoret}},
  \bibinfo{journal}{Phys. Rev. Lett.} \textbf{\bibinfo{volume}{122}},
  \bibinfo{pages}{080502} (\bibinfo{year}{2019}),
  \urlprefix\url{https://link.aps.org/doi/10.1103/PhysRevLett.122.080502}.

\bibitem[{\citenamefont{Campagne-Ibarcq
  et~al.}(2016)\citenamefont{Campagne-Ibarcq, Six, Bretheau, Sarlette,
  Mirrahimi, Rouchon, and Huard}}]{campagne-ibarcq_observing_2016}
\bibinfo{author}{\bibfnamefont{P.}~\bibnamefont{Campagne-Ibarcq}},
  \bibinfo{author}{\bibfnamefont{P.}~\bibnamefont{Six}},
  \bibinfo{author}{\bibfnamefont{L.}~\bibnamefont{Bretheau}},
  \bibinfo{author}{\bibfnamefont{A.}~\bibnamefont{Sarlette}},
  \bibinfo{author}{\bibfnamefont{M.}~\bibnamefont{Mirrahimi}},
  \bibinfo{author}{\bibfnamefont{P.}~\bibnamefont{Rouchon}}, \bibnamefont{and}
  \bibinfo{author}{\bibfnamefont{B.}~\bibnamefont{Huard}},
  \bibinfo{journal}{Physical Review X} \textbf{\bibinfo{volume}{6}},
  \bibinfo{pages}{011002} (\bibinfo{year}{2016}),
  \urlprefix\url{https://link.aps.org/doi/10.1103/PhysRevX.6.011002}.

\bibitem[{\citenamefont{Pfaff et~al.}(2017)\citenamefont{Pfaff, Axline,
  Burkhart, Vool, Reinhold, Frunzio, Jiang, Devoret, and
  Schoelkopf}}]{pfaff_controlled_2017}
\bibinfo{author}{\bibfnamefont{W.}~\bibnamefont{Pfaff}},
  \bibinfo{author}{\bibfnamefont{C.~J.} \bibnamefont{Axline}},
  \bibinfo{author}{\bibfnamefont{L.~D.} \bibnamefont{Burkhart}},
  \bibinfo{author}{\bibfnamefont{U.}~\bibnamefont{Vool}},
  \bibinfo{author}{\bibfnamefont{P.}~\bibnamefont{Reinhold}},
  \bibinfo{author}{\bibfnamefont{L.}~\bibnamefont{Frunzio}},
  \bibinfo{author}{\bibfnamefont{L.}~\bibnamefont{Jiang}},
  \bibinfo{author}{\bibfnamefont{M.~H.} \bibnamefont{Devoret}},
  \bibnamefont{and} \bibinfo{author}{\bibfnamefont{R.~J.}
  \bibnamefont{Schoelkopf}}, \bibinfo{journal}{Nature Physics}
  \textbf{\bibinfo{volume}{13}}, \bibinfo{pages}{882} (\bibinfo{year}{2017}),
  ISSN \bibinfo{issn}{1745-2481},
  \urlprefix\url{https://www.nature.com/articles/nphys4143}.

\bibitem[{\citenamefont{Sank et~al.}(2016)\citenamefont{Sank, Chen, Khezri,
  Kelly, Barends, Campbell, Chen, Chiaro, Dunsworth, Fowler et~al.}}]{sank2016}
\bibinfo{author}{\bibfnamefont{D.}~\bibnamefont{Sank}},
  \bibinfo{author}{\bibfnamefont{Z.}~\bibnamefont{Chen}},
  \bibinfo{author}{\bibfnamefont{M.}~\bibnamefont{Khezri}},
  \bibinfo{author}{\bibfnamefont{J.}~\bibnamefont{Kelly}},
  \bibinfo{author}{\bibfnamefont{R.}~\bibnamefont{Barends}},
  \bibinfo{author}{\bibfnamefont{B.}~\bibnamefont{Campbell}},
  \bibinfo{author}{\bibfnamefont{Y.}~\bibnamefont{Chen}},
  \bibinfo{author}{\bibfnamefont{B.}~\bibnamefont{Chiaro}},
  \bibinfo{author}{\bibfnamefont{A.}~\bibnamefont{Dunsworth}},
  \bibinfo{author}{\bibfnamefont{A.}~\bibnamefont{Fowler}},
  \bibnamefont{et~al.}, \bibinfo{journal}{Phys. Rev. Lett.}
  \textbf{\bibinfo{volume}{117}}, \bibinfo{pages}{190503}
  (\bibinfo{year}{2016}),
  \urlprefix\url{https://link.aps.org/doi/10.1103/PhysRevLett.117.190503}.

\bibitem[{\citenamefont{Minev et~al.}(2019)\citenamefont{Minev, Mundhada,
  Shankar, Reinhold, Guti{\'e}rrez-J{\'a}uregui, Schoelkopf, Mirrahimi,
  Carmichael, and Devoret}}]{minev2019}
\bibinfo{author}{\bibfnamefont{Z.~K.} \bibnamefont{Minev}},
  \bibinfo{author}{\bibfnamefont{S.~O.} \bibnamefont{Mundhada}},
  \bibinfo{author}{\bibfnamefont{S.}~\bibnamefont{Shankar}},
  \bibinfo{author}{\bibfnamefont{P.}~\bibnamefont{Reinhold}},
  \bibinfo{author}{\bibfnamefont{R.}~\bibnamefont{Guti{\'e}rrez-J{\'a}uregui}},
  \bibinfo{author}{\bibfnamefont{R.~J.} \bibnamefont{Schoelkopf}},
  \bibinfo{author}{\bibfnamefont{M.}~\bibnamefont{Mirrahimi}},
  \bibinfo{author}{\bibfnamefont{H.~J.} \bibnamefont{Carmichael}},
  \bibnamefont{and} \bibinfo{author}{\bibfnamefont{M.~H.}
  \bibnamefont{Devoret}}, \bibinfo{journal}{Nature}
  \textbf{\bibinfo{volume}{570}}, \bibinfo{pages}{200} (\bibinfo{year}{2019}),
  ISSN \bibinfo{issn}{1476-4687},
  \urlprefix\url{https://doi.org/10.1038/s41586-019-1287-z}.

\bibitem[{\citenamefont{Lescanne et~al.}(2019)\citenamefont{Lescanne, Verney,
  Ficheux, Devoret, Huard, Mirrahimi, and Leghtas}}]{lescanne_escape_2019}
\bibinfo{author}{\bibfnamefont{R.}~\bibnamefont{Lescanne}},
  \bibinfo{author}{\bibfnamefont{L.}~\bibnamefont{Verney}},
  \bibinfo{author}{\bibfnamefont{Q.}~\bibnamefont{Ficheux}},
  \bibinfo{author}{\bibfnamefont{M.~H.} \bibnamefont{Devoret}},
  \bibinfo{author}{\bibfnamefont{B.}~\bibnamefont{Huard}},
  \bibinfo{author}{\bibfnamefont{M.}~\bibnamefont{Mirrahimi}},
  \bibnamefont{and} \bibinfo{author}{\bibfnamefont{Z.}~\bibnamefont{Leghtas}},
  \bibinfo{journal}{Physical Review Applied} \textbf{\bibinfo{volume}{11}},
  \bibinfo{pages}{014030} (\bibinfo{year}{2019}), \bibinfo{note}{publisher:
  American Physical Society},
  \urlprefix\url{https://link.aps.org/doi/10.1103/PhysRevApplied.11.014030}.

\bibitem[{\citenamefont{Touzard et~al.}(2018)\citenamefont{Touzard, Grimm,
  Leghtas, Mundhada, Reinhold, Axline, Reagor, Chou, Blumoff, Sliwa
  et~al.}}]{touzard2018}
\bibinfo{author}{\bibfnamefont{S.}~\bibnamefont{Touzard}},
  \bibinfo{author}{\bibfnamefont{A.}~\bibnamefont{Grimm}},
  \bibinfo{author}{\bibfnamefont{Z.}~\bibnamefont{Leghtas}},
  \bibinfo{author}{\bibfnamefont{S.~O.} \bibnamefont{Mundhada}},
  \bibinfo{author}{\bibfnamefont{P.}~\bibnamefont{Reinhold}},
  \bibinfo{author}{\bibfnamefont{C.}~\bibnamefont{Axline}},
  \bibinfo{author}{\bibfnamefont{M.}~\bibnamefont{Reagor}},
  \bibinfo{author}{\bibfnamefont{K.}~\bibnamefont{Chou}},
  \bibinfo{author}{\bibfnamefont{J.}~\bibnamefont{Blumoff}},
  \bibinfo{author}{\bibfnamefont{K.~M.} \bibnamefont{Sliwa}},
  \bibnamefont{et~al.}, \bibinfo{journal}{Phys. Rev. X}
  \textbf{\bibinfo{volume}{8}}, \bibinfo{pages}{021005} (\bibinfo{year}{2018}),
  \urlprefix\url{https://link.aps.org/doi/10.1103/PhysRevX.8.021005}.

\bibitem[{\citenamefont{McCoy}(1932)}]{McCoy1932}
\bibinfo{author}{\bibfnamefont{N.~H.} \bibnamefont{McCoy}},
  \bibinfo{journal}{Proceedings of the National Academy of Sciences}
  \textbf{\bibinfo{volume}{18}}, \bibinfo{pages}{674} (\bibinfo{year}{1932}).

\bibitem[{\citenamefont{Hillery et~al.}(1984)\citenamefont{Hillery, O'Connell,
  Scully, and Wigner}}]{hillery1984}
\bibinfo{author}{\bibfnamefont{M.}~\bibnamefont{Hillery}},
  \bibinfo{author}{\bibfnamefont{R.}~\bibnamefont{O'Connell}},
  \bibinfo{author}{\bibfnamefont{M.}~\bibnamefont{Scully}}, \bibnamefont{and}
  \bibinfo{author}{\bibfnamefont{E.}~\bibnamefont{Wigner}},
  \bibinfo{journal}{Physics Reports} \textbf{\bibinfo{volume}{106}},
  \bibinfo{pages}{121} (\bibinfo{year}{1984}), ISSN \bibinfo{issn}{0370-1573}.

\bibitem[{\citenamefont{Groenewold}(1946)}]{groenewold1946}
\bibinfo{author}{\bibfnamefont{H.~J.} \bibnamefont{Groenewold}}, in
  \emph{\bibinfo{booktitle}{On the Principles of Elementary Quantum Mechanics}}
  (\bibinfo{publisher}{Springer}, \bibinfo{year}{1946}), pp.
  \bibinfo{pages}{1--56}.

\bibitem[{\citenamefont{Curtright et~al.}(2013)\citenamefont{Curtright,
  Fairlie, and Zachos}}]{curtright2013}
\bibinfo{author}{\bibfnamefont{T.~L.} \bibnamefont{Curtright}},
  \bibinfo{author}{\bibfnamefont{D.~B.} \bibnamefont{Fairlie}},
  \bibnamefont{and} \bibinfo{author}{\bibfnamefont{C.~K.}
  \bibnamefont{Zachos}}, \emph{\bibinfo{title}{A concise treatise on quantum
  mechanics in phase space}} (\bibinfo{publisher}{World Scientific Publishing
  Company}, \bibinfo{year}{2013}).

\bibitem[{\citenamefont{Moyal}(1949)}]{moyal1949}
\bibinfo{author}{\bibfnamefont{J.~E.} \bibnamefont{Moyal}},
  \bibinfo{journal}{Mathematical Proceedings of the Cambridge Philosophical
  Society} \textbf{\bibinfo{volume}{45}}, \bibinfo{pages}{99–124}
  (\bibinfo{year}{1949}).

\bibitem[{\citenamefont{Dirac}(1945)}]{dirac1945}
\bibinfo{author}{\bibfnamefont{P.~A.~M.} \bibnamefont{Dirac}},
  \bibinfo{journal}{Reviews of Modern Physics} \textbf{\bibinfo{volume}{17}},
  \bibinfo{pages}{195} (\bibinfo{year}{1945}).

\bibitem[{\citenamefont{Wielinga and Milburn}(1993)}]{wielinga1993}
\bibinfo{author}{\bibfnamefont{B.}~\bibnamefont{Wielinga}} \bibnamefont{and}
  \bibinfo{author}{\bibfnamefont{G.~J.} \bibnamefont{Milburn}},
  \bibinfo{journal}{Phys. Rev. A} \textbf{\bibinfo{volume}{48}},
  \bibinfo{pages}{2494} (\bibinfo{year}{1993}),
  \urlprefix\url{https://link.aps.org/doi/10.1103/PhysRevA.48.2494}.

\bibitem[{\citenamefont{Habib et~al.}(1998)\citenamefont{Habib, Shizume, and
  Zurek}}]{zurek1998}
\bibinfo{author}{\bibfnamefont{S.}~\bibnamefont{Habib}},
  \bibinfo{author}{\bibfnamefont{K.}~\bibnamefont{Shizume}}, \bibnamefont{and}
  \bibinfo{author}{\bibfnamefont{W.~H.} \bibnamefont{Zurek}},
  \bibinfo{journal}{Phys. Rev. Lett.} \textbf{\bibinfo{volume}{80}},
  \bibinfo{pages}{4361} (\bibinfo{year}{1998}),
  \urlprefix\url{https://link.aps.org/doi/10.1103/PhysRevLett.80.4361}.

\bibitem[{\citenamefont{Zurek and Paz}(1999)}]{zurek1999}
\bibinfo{author}{\bibfnamefont{W.~H.} \bibnamefont{Zurek}} \bibnamefont{and}
  \bibinfo{author}{\bibfnamefont{J.~P.} \bibnamefont{Paz}}, in
  \emph{\bibinfo{booktitle}{Epistemological and Experimental Perspectives on
  Quantum Physics}} (\bibinfo{publisher}{Springer}, \bibinfo{year}{1999}), pp.
  \bibinfo{pages}{167--177}.

\bibitem[{\citenamefont{Brune et~al.}(1996)\citenamefont{Brune, Hagley, Dreyer,
  Ma\^{\i}tre, Maali, Wunderlich, Raimond, and Haroche}}]{brune1996}
\bibinfo{author}{\bibfnamefont{M.}~\bibnamefont{Brune}},
  \bibinfo{author}{\bibfnamefont{E.}~\bibnamefont{Hagley}},
  \bibinfo{author}{\bibfnamefont{J.}~\bibnamefont{Dreyer}},
  \bibinfo{author}{\bibfnamefont{X.}~\bibnamefont{Ma\^{\i}tre}},
  \bibinfo{author}{\bibfnamefont{A.}~\bibnamefont{Maali}},
  \bibinfo{author}{\bibfnamefont{C.}~\bibnamefont{Wunderlich}},
  \bibinfo{author}{\bibfnamefont{J.~M.} \bibnamefont{Raimond}},
  \bibnamefont{and} \bibinfo{author}{\bibfnamefont{S.}~\bibnamefont{Haroche}},
  \bibinfo{journal}{Phys. Rev. Lett.} \textbf{\bibinfo{volume}{77}},
  \bibinfo{pages}{4887} (\bibinfo{year}{1996}),
  \urlprefix\url{https://link.aps.org/doi/10.1103/PhysRevLett.77.4887}.

\bibitem[{\citenamefont{Deleglise et~al.}(2008)\citenamefont{Deleglise,
  Dotsenko, Sayrin, Bernu, Brune, Raimond, and Haroche}}]{deleglise2008}
\bibinfo{author}{\bibfnamefont{S.}~\bibnamefont{Deleglise}},
  \bibinfo{author}{\bibfnamefont{I.}~\bibnamefont{Dotsenko}},
  \bibinfo{author}{\bibfnamefont{C.}~\bibnamefont{Sayrin}},
  \bibinfo{author}{\bibfnamefont{J.}~\bibnamefont{Bernu}},
  \bibinfo{author}{\bibfnamefont{M.}~\bibnamefont{Brune}},
  \bibinfo{author}{\bibfnamefont{J.-M.} \bibnamefont{Raimond}},
  \bibnamefont{and} \bibinfo{author}{\bibfnamefont{S.}~\bibnamefont{Haroche}},
  \bibinfo{journal}{Nature} \textbf{\bibinfo{volume}{455}},
  \bibinfo{pages}{510} (\bibinfo{year}{2008}).

\bibitem[{\citenamefont{Haroche and Raimond}(2006)}]{haroche2006}
\bibinfo{author}{\bibfnamefont{S.}~\bibnamefont{Haroche}} \bibnamefont{and}
  \bibinfo{author}{\bibfnamefont{J.-M.} \bibnamefont{Raimond}},
  \emph{\bibinfo{title}{Exploring the quantum: atoms, cavities, and photons}}
  (\bibinfo{publisher}{Oxford University Press}, \bibinfo{year}{2006}).

\bibitem[{\citenamefont{Leghtas et~al.}(2015)\citenamefont{Leghtas, Touzard,
  Pop, Kou, Vlastakis, Petrenko, Sliwa, Narla, Shankar, Hatridge
  et~al.}}]{leghtas2015}
\bibinfo{author}{\bibfnamefont{Z.}~\bibnamefont{Leghtas}},
  \bibinfo{author}{\bibfnamefont{S.}~\bibnamefont{Touzard}},
  \bibinfo{author}{\bibfnamefont{I.~M.} \bibnamefont{Pop}},
  \bibinfo{author}{\bibfnamefont{A.}~\bibnamefont{Kou}},
  \bibinfo{author}{\bibfnamefont{B.}~\bibnamefont{Vlastakis}},
  \bibinfo{author}{\bibfnamefont{A.}~\bibnamefont{Petrenko}},
  \bibinfo{author}{\bibfnamefont{K.~M.} \bibnamefont{Sliwa}},
  \bibinfo{author}{\bibfnamefont{A.}~\bibnamefont{Narla}},
  \bibinfo{author}{\bibfnamefont{S.}~\bibnamefont{Shankar}},
  \bibinfo{author}{\bibfnamefont{M.~J.} \bibnamefont{Hatridge}},
  \bibnamefont{et~al.}, \bibinfo{journal}{Science}
  \textbf{\bibinfo{volume}{347}}, \bibinfo{pages}{853} (\bibinfo{year}{2015}).

\bibitem[{\citenamefont{Signoles et~al.}(2014)\citenamefont{Signoles, Facon,
  Grosso, Dotsenko, Haroche, Raimond, Brune, and Gleyzes}}]{Signoles2014}
\bibinfo{author}{\bibfnamefont{A.}~\bibnamefont{Signoles}},
  \bibinfo{author}{\bibfnamefont{A.}~\bibnamefont{Facon}},
  \bibinfo{author}{\bibfnamefont{D.}~\bibnamefont{Grosso}},
  \bibinfo{author}{\bibfnamefont{I.}~\bibnamefont{Dotsenko}},
  \bibinfo{author}{\bibfnamefont{S.}~\bibnamefont{Haroche}},
  \bibinfo{author}{\bibfnamefont{J.-M.} \bibnamefont{Raimond}},
  \bibinfo{author}{\bibfnamefont{M.}~\bibnamefont{Brune}}, \bibnamefont{and}
  \bibinfo{author}{\bibfnamefont{S.}~\bibnamefont{Gleyzes}},
  \bibinfo{journal}{Nature Physics} \textbf{\bibinfo{volume}{10}},
  \bibinfo{pages}{715} (\bibinfo{year}{2014}), ISSN \bibinfo{issn}{1745-2481},
  \urlprefix\url{https://doi.org/10.1038/nphys3076}.

\bibitem[{\citenamefont{Yurke and Stoler}(1986)}]{Yurke1986}
\bibinfo{author}{\bibfnamefont{B.}~\bibnamefont{Yurke}} \bibnamefont{and}
  \bibinfo{author}{\bibfnamefont{D.}~\bibnamefont{Stoler}},
  \bibinfo{journal}{Phys. Rev. Lett.} \textbf{\bibinfo{volume}{57}},
  \bibinfo{pages}{13} (\bibinfo{year}{1986}),
  \urlprefix\url{https://link.aps.org/doi/10.1103/PhysRevLett.57.13}.

\bibitem[{\citenamefont{Yurke and Stoler}(1988)}]{yurke1988}
\bibinfo{author}{\bibfnamefont{B.}~\bibnamefont{Yurke}} \bibnamefont{and}
  \bibinfo{author}{\bibfnamefont{D.}~\bibnamefont{Stoler}},
  \bibinfo{journal}{Physica B+C} \textbf{\bibinfo{volume}{151}},
  \bibinfo{pages}{298} (\bibinfo{year}{1988}), ISSN \bibinfo{issn}{0378-4363},
  \urlprefix\url{https://www.sciencedirect.com/science/article/pii/0378436388901817}.

\bibitem[{\citenamefont{Kirchmair et~al.}(2013)\citenamefont{Kirchmair,
  Vlastakis, Leghtas, Nigg, Paik, Ginossar, Mirrahimi, Frunzio, Girvin, and
  Schoelkopf}}]{kirchmair2013}
\bibinfo{author}{\bibfnamefont{G.}~\bibnamefont{Kirchmair}},
  \bibinfo{author}{\bibfnamefont{B.}~\bibnamefont{Vlastakis}},
  \bibinfo{author}{\bibfnamefont{Z.}~\bibnamefont{Leghtas}},
  \bibinfo{author}{\bibfnamefont{S.~E.} \bibnamefont{Nigg}},
  \bibinfo{author}{\bibfnamefont{H.}~\bibnamefont{Paik}},
  \bibinfo{author}{\bibfnamefont{E.}~\bibnamefont{Ginossar}},
  \bibinfo{author}{\bibfnamefont{M.}~\bibnamefont{Mirrahimi}},
  \bibinfo{author}{\bibfnamefont{L.}~\bibnamefont{Frunzio}},
  \bibinfo{author}{\bibfnamefont{S.~M.} \bibnamefont{Girvin}},
  \bibnamefont{and} \bibinfo{author}{\bibfnamefont{R.~J.}
  \bibnamefont{Schoelkopf}}, \bibinfo{journal}{Nature}
  \textbf{\bibinfo{volume}{495}}, \bibinfo{pages}{205} (\bibinfo{year}{2013}),
  ISSN \bibinfo{issn}{1476-4687},
  \urlprefix\url{https://doi.org/10.1038/nature11902}.

\bibitem[{\citenamefont{Assemat et~al.}(2019)\citenamefont{Assemat, Grosso,
  Signoles, Facon, Dotsenko, Haroche, Raimond, Brune, and
  Gleyzes}}]{Assemat2019}
\bibinfo{author}{\bibfnamefont{F.}~\bibnamefont{Assemat}},
  \bibinfo{author}{\bibfnamefont{D.}~\bibnamefont{Grosso}},
  \bibinfo{author}{\bibfnamefont{A.}~\bibnamefont{Signoles}},
  \bibinfo{author}{\bibfnamefont{A.}~\bibnamefont{Facon}},
  \bibinfo{author}{\bibfnamefont{I.}~\bibnamefont{Dotsenko}},
  \bibinfo{author}{\bibfnamefont{S.}~\bibnamefont{Haroche}},
  \bibinfo{author}{\bibfnamefont{J.~M.} \bibnamefont{Raimond}},
  \bibinfo{author}{\bibfnamefont{M.}~\bibnamefont{Brune}}, \bibnamefont{and}
  \bibinfo{author}{\bibfnamefont{S.}~\bibnamefont{Gleyzes}},
  \bibinfo{journal}{Phys. Rev. Lett.} \textbf{\bibinfo{volume}{123}},
  \bibinfo{pages}{143605} (\bibinfo{year}{2019}),
  \urlprefix\url{https://link.aps.org/doi/10.1103/PhysRevLett.123.143605}.

\bibitem[{\citenamefont{Zurek}(2003)}]{Zurek2003}
\bibinfo{author}{\bibfnamefont{W.~H.} \bibnamefont{Zurek}},
  \bibinfo{journal}{Rev. Mod. Phys.} \textbf{\bibinfo{volume}{75}},
  \bibinfo{pages}{715} (\bibinfo{year}{2003}),
  \urlprefix\url{https://link.aps.org/doi/10.1103/RevModPhys.75.715}.

\bibitem[{\citenamefont{Marthaler and Dykman}(2007)}]{Marthaler2007}
\bibinfo{author}{\bibfnamefont{M.}~\bibnamefont{Marthaler}} \bibnamefont{and}
  \bibinfo{author}{\bibfnamefont{M.~I.} \bibnamefont{Dykman}},
  \bibinfo{journal}{Phys. Rev. A} \textbf{\bibinfo{volume}{76}},
  \bibinfo{pages}{010102} (\bibinfo{year}{2007}),
  \urlprefix\url{https://link.aps.org/doi/10.1103/PhysRevA.76.010102}.

\bibitem[{\citenamefont{Puri et~al.}(2017)\citenamefont{Puri, Boutin, and
  Blais}}]{puri2017}
\bibinfo{author}{\bibfnamefont{S.}~\bibnamefont{Puri}},
  \bibinfo{author}{\bibfnamefont{S.}~\bibnamefont{Boutin}}, \bibnamefont{and}
  \bibinfo{author}{\bibfnamefont{A.}~\bibnamefont{Blais}},
  \bibinfo{journal}{npj Quantum Information} \textbf{\bibinfo{volume}{3}},
  \bibinfo{pages}{1} (\bibinfo{year}{2017}).

\bibitem[{\citenamefont{Lescanne et~al.}(2020)\citenamefont{Lescanne, Villiers,
  Peronnin, Sarlette, Delbecq, Huard, Kontos, Mirrahimi, and
  Leghtas}}]{lescanne2020}
\bibinfo{author}{\bibfnamefont{R.}~\bibnamefont{Lescanne}},
  \bibinfo{author}{\bibfnamefont{M.}~\bibnamefont{Villiers}},
  \bibinfo{author}{\bibfnamefont{T.}~\bibnamefont{Peronnin}},
  \bibinfo{author}{\bibfnamefont{A.}~\bibnamefont{Sarlette}},
  \bibinfo{author}{\bibfnamefont{M.}~\bibnamefont{Delbecq}},
  \bibinfo{author}{\bibfnamefont{B.}~\bibnamefont{Huard}},
  \bibinfo{author}{\bibfnamefont{T.}~\bibnamefont{Kontos}},
  \bibinfo{author}{\bibfnamefont{M.}~\bibnamefont{Mirrahimi}},
  \bibnamefont{and} \bibinfo{author}{\bibfnamefont{Z.}~\bibnamefont{Leghtas}},
  \bibinfo{journal}{Nature Physics} \textbf{\bibinfo{volume}{16}},
  \bibinfo{pages}{509} (\bibinfo{year}{2020}).

\bibitem[{\citenamefont{Gautier et~al.}(2022)\citenamefont{Gautier, Sarlette,
  and Mirrahimi}}]{gautier_combined_2022}
\bibinfo{author}{\bibfnamefont{R.}~\bibnamefont{Gautier}},
  \bibinfo{author}{\bibfnamefont{A.}~\bibnamefont{Sarlette}}, \bibnamefont{and}
  \bibinfo{author}{\bibfnamefont{M.}~\bibnamefont{Mirrahimi}},
  \bibinfo{journal}{arXiv:2112.05545 [quant-ph]}  (\bibinfo{year}{2022}),
  \bibinfo{note}{arXiv: 2112.05545},
  \urlprefix\url{http://arxiv.org/abs/2112.05545}.

\bibitem[{\citenamefont{Putterman et~al.}(2022)\citenamefont{Putterman,
  Iverson, Xu, Jiang, Painter, Brandão, and Noh}}]{putterman_stabilizing_2022}
\bibinfo{author}{\bibfnamefont{H.}~\bibnamefont{Putterman}},
  \bibinfo{author}{\bibfnamefont{J.}~\bibnamefont{Iverson}},
  \bibinfo{author}{\bibfnamefont{Q.}~\bibnamefont{Xu}},
  \bibinfo{author}{\bibfnamefont{L.}~\bibnamefont{Jiang}},
  \bibinfo{author}{\bibfnamefont{O.}~\bibnamefont{Painter}},
  \bibinfo{author}{\bibfnamefont{F.~G.} \bibnamefont{Brandão}},
  \bibnamefont{and} \bibinfo{author}{\bibfnamefont{K.}~\bibnamefont{Noh}},
  \bibinfo{journal}{Physical Review Letters} \textbf{\bibinfo{volume}{128}},
  \bibinfo{pages}{110502} (\bibinfo{year}{2022}), \bibinfo{note}{publisher:
  American Physical Society},
  \urlprefix\url{https://link.aps.org/doi/10.1103/PhysRevLett.128.110502}.

\end{thebibliography}

\end{document}

% \title{Spectroscopy of the \textcolor{red}{Kerr-cat} \textcolor{red}{metapotential}}
% \title{Period doubling of the \textcolor{red}{Kerr-cat} eigenspectrum}
\title{Supplementary Material for \\ The squeezed Kerr oscillator: spectral kissing and phase-flip robustness}
\author{Nicholas E. Frattini}
 \thanks{These authors contributed equally}
\email{nicholas.frattini@colorado.edu, rodrigo.cortinas@yale.edu.}
\affiliation{Department of Applied Physics and Physics, Yale University, New Haven, CT 06520, USA}
\author{Rodrigo G. Corti\~nas}
\thanks{These authors contributed equally}
\email{nicholas.frattini@colorado.edu, rodrigo.cortinas@yale.edu.}
\affiliation{Department of Applied Physics and Physics, Yale University, New Haven, CT 06520, USA}
\author{Jayameenakshi Venkatraman}
\affiliation{Department of Applied Physics and Physics, Yale University, New Haven, CT 06520, USA}
\author{Xiao Xu}
\affiliation{Department of Applied Physics and Physics, Yale University, New Haven, CT 06520, USA}
\author{Qile Su}
\affiliation{Department of Applied Physics and Physics, Yale University, New Haven, CT 06520, USA}
\author{Chan U Lei}
\affiliation{Department of Applied Physics and Physics, Yale University, New Haven, CT 06520, USA}
\author{Benjamin J. Chapman}
\affiliation{Department of Applied Physics and Physics, Yale University, New Haven, CT 06520, USA}
\author{Vidul R. Joshi}
\affiliation{Department of Applied Physics and Physics, Yale University, New Haven, CT 06520, USA}
\author{S. M. Girvin}
\affiliation{Department of Applied Physics and Physics, Yale University, New Haven, CT 06520, USA}
\author{Robert J. Schoelkopf}
\affiliation{Department of Applied Physics and Physics, Yale University, New Haven, CT 06520, USA}
\author{Shruti Puri}
\affiliation{Department of Applied Physics and Physics, Yale University, New Haven, CT 06520, USA}
\author{Michel H. Devoret}
\email{michel.devoret@yale.edu}
\affiliation{Department of Applied Physics and Physics, Yale University, New Haven, CT 06520, USA}
\date{\today}

\maketitle

\section{Derivation of the static-effective squeezed Kerr Hamiltonian}
\label{supp:1}
The SNAIL-transmon Josephson circuit \cite{frattini2017} is well-modelled as a nonlinear oscillator. To compute the static-effective squeezed Kerr oscillator Hamiltonian, we consider the Hamiltonian of a driven nonlinear oscillator in the bosonic basis,
\begin{align}
\label{eq:H-t}
\frac{\hat{H}(t)}{\hbar} = \omega_o \hat{a}^{\dagger} \hat{a} + \sum_{m \ge 3} \frac{g_m}{m} (\hat{a} + \hat{a}^\dagger + \Pi e^{-i \omega_d t} + \Pi^* e^{i \omega_d t})^m,
\end{align}
where the oscillator is characterized by $\omega_o$, its bare frequency, $g_m/m$, its $m$th order nonlinearity, and $\hat{a}^{\dagger}$ the bosonic creation operator satisfying the standard commutation relation between bosonic operators: $[\hat{a}, \hat{a}^{\dagger}] = 1$. The Hamiltonian in  Eq.~(1) is related to Eq.~(\ref{eq:H-t}) by a displacement transformation into the linear response of the oscillator, where the effective amplitude of the displacement is $\Pi = \frac{4\Omega_d}{3\omega_d}$.

Next, we seek a transformed frame in which the nonlinear Hamiltonian rates are small compared to the drive subharmonic frequency $\omega_d/2 \approx \omega_o$.
We achieve this by going into a rotating frame induced by $\frac{\omega_d}{2} \hat{a}^{\dagger} \hat{a}$, transforming Eq.~(\ref{eq:H-t}) to
\begin{align}
\label{eq:H-tr}
\frac{\hat{H}(t)}{\hbar} = -\delta \hat{a}^{\dagger} \hat{a} +  \sum_{n \ge 3} \frac{g_m}{m} (\hat{a} e^{-i \omega_d t/2}+ \hat{a}^\dagger e^{i \omega_d t/2}+ \Pi e^{-i \omega_d t} + \Pi^* e^{i \omega_d t})^m,
\end{align}
where $\delta = \frac{\omega_d}{2} - \omega_o \ll \omega_d $. 

From Eq.~(\ref{eq:H-tr}), we compute, perturbatively, an effective Hamiltonian that captures the relevant dynamics in this displaced rotating frame. Here, we introduce $\varphi_{\mathrm{zps}} \ll 1$, the zero point spread of the phase across the Josephson junction which is the perturbative parameter of our expansion. The source of the nonlinearity being the Josephson potential guarantees a hierarchy in the Hamiltonian nonlinearities. The $m$-th nonlinearity is order $m-2$ in the perturbation, i.e., $g_m/m = \mathcal{O}(\omega_o \varphi_{\mathrm{zps}}^{m-2})$ for $m \ge 3$.  With the placement of our drive, we ensure that $\delta \lesssim g_3$.

Following \cite{venkatraman2021}, we find an effective Hamiltonian with following structure:
\begin{align}
\label{eq:Heff-KC}
\hat{H}_{\mathrm{eff}} %&= -\Delta \hat{a}^{\dagger} \hat{a} - K \hat{a}^{\dagger 2} \hat{a}^2 + \epsilon_2 \hat{a}^{\dagger 2} + \epsilon_2^* \hat{a}^{2},\\
=  \sum_{n \ge 1} \hat{H}_{\mathrm{eff}}^{(n)}; \qquad
\hat{H}_{\mathrm{eff}}^{(n)} = \sum_{k \ge 0} \hat{H}_{\mathrm{eff}\, [k]}^{(n)} |\Pi|^{2k},
\end{align}
where the superscript $n$ denotes the order in the perturbation parameter and the subscript $k$ denotes the number of participating drive excitation pairs.

%where, here and in the following text, $\square^{(n)}$ denotes the $n$th order contribution to a coefficient $\square$ and $\square_{[k]}$ denotes an effective Hamiltonian coefficient containing $k$ pairs of drive excitations, i.e., $\square = \sum_k\square_{[k]}  |\Pi|^{2k}$.

At orders 1 and 2, we find the effective Hamiltonian takes the form
\begin{align}
\frac{\hat{H}_{\mathrm{eff}}}{\hbar} = -\Delta \hat{a}^{\dagger} \hat{a} - K \hat{a}^{\dagger 2} \hat{a}^2 + \epsilon_2 \hat{a}^{\dagger 2} + \epsilon_2^* \hat{a}^{2} + \mathcal{O}(\omegaa {\varphi_{\mathrm{zps}}}^3),
\end{align}
with
\begin{align}
\begin{split}
\Delta =   \sum_{n = 1,2} \Delta^{(n)}  \qquad K = \sum_{n = 1,2} K^{(n)},  \qquad \epsilon_2 = \sum_{n = 1,2} \epsilon_2^{(n)},
\end{split}
\end{align}
where, %as series in $\hbar$, 
we find:
\begin{align}
\begin{split}
-\Delta^{(1)}  &= -\delta, \qquad -\Delta^{(2)} = -\sum_{k = 0,1} \Delta^{(2)}_{[k]}  |\Pi|^{2k}, \qquad -\Delta^{(2)}_{[0]} = 3g_4 + \frac{10}{3} \frac{g_3^2}{\omega_a} , \qquad
-\Delta^{(2)}_{[1]} = 6g_4  + \frac{9}{2} \frac{g_3^2}{\omega_a} \\
-K^{(1)} &= 0 \qquad -K^{(2)} = \frac{3 g_4}{2} + \frac{5}{3} \frac{g_3^2}{\omega_a} , \\
\epsilon_2^{(1)} &= g_3 \Pi \qquad \epsilon_2^{(2)} = 0.
\end{split}
\end{align}
We remark that $\Delta^{(2)}_{[0]}$ and $\Delta^{(2)}_{[1]}$ are the well-known Lamb shift and AC Stark shift respectively.
Similarly, the dependence of $K^{(2)}$ and $\epsilon_2^{(1)}$ on external flux have been previously verified in SNAIL parametric amplifiers \cite{frattini_optimizing_2018, sivak_kerr-free_2019}.

At order 3, following the notation introduced above, we find the effective Hamiltonian as
\begin{align}
\frac{\hat{H}_{\mathrm{eff}}^{(3)}}{\hbar} = - \Delta^{(3)} \hat{a}^{\dagger} \hat{a} - K^{(3)} \hat{a}^{\dagger 2} \hat{a}^2 + \epsilon_2^{(3)} \hat{a}^{\dagger 2} + \epsilon_2^{(3)*} \hat{a}^{2}  + \epsilon_2^{\prime (3)} \hat{a}^{\dagger 3} \hat{a} + \epsilon_2^{\prime * (3)} \hat{a}^{\dagger} \hat{a}^{3},
\end{align}
where
\begin{align}
\Delta^{(3)} &= 0 \qquad
K^{(3)} = 0 \\
\epsilon_2^{(3)} &= \left(6 g_5  + \frac{141}{20} \frac{g_3 g_4}{\omega_a}  \right)  |\Pi|^2 \Pi^* +\left( 6 g_5 + \frac{63}{8} \frac{g_3 g_4}{\omega_a}  \right) \Pi \\ %+ \left(4 g_5 \Pi + \frac{21}{4} \frac{g_3 g_4}{\omega_a} \Pi \hbar \right) ( \hat{a}^{\dagger} \hat{a} -2)
\epsilon_2^{\prime (3)} &= \left( 4 g_5  + \frac{21}{4} \frac{g_3 g_4}{\omega_a} \right) \Pi.
\end{align}

Note that the Hamiltonian terms $\propto \hat{a}^{\dagger 3} \hat{a} + \hat{a}^{\dagger } \hat{a}^3 =  \hat{a}^{\dagger 2} (\hat{a}^{\dagger} \hat{a}) + (\hat{a}^{\dagger } \hat{a}) \hat{a}^2$ represent a photon-number dependent squeezing interaction. 

With this understanding, we remark that even at the third order the effective Hamiltonian contains no parasitic terms, a quantum manifestation of the robustness of the period-doubling bifurcation.

% With this understanding, we remark that even at the third order the effective Hamiltonian contains no parasitic parametric terms other than the squeezing, a quantum manifestation of the robustness of the period-doubling bifurcation. \sout{\textcolor{blue}{which in this expansion manifest as the absence of unwanted nonlinear resonances up to this order}}.

At order 4, we find a four-photon drive, and the first non-squeezing drive term, in the effective Hamiltonian:
\begin{align}
\label{eq:H-4}
\frac{\hat{H}_{\mathrm{eff}}^{(4)}}{\hbar} = -\Delta^{(4)} \hat{a}^{\dagger} \hat{a} - K^{(4)} \hat{a}^{\dagger 2} \hat{a}^2  - \lambda^{(4)} \hat{a}^{\dagger 3} \hat{a}^3 + \epsilon_4^{(4)} \hat{a}^{\dagger 4} + \epsilon_4^* \hat{a}^{(4)} ,
\end{align}
where $\Delta^{(4)} =  \sum_{k = 0, 1, 2} \Delta^{(4)}_{[k]}  |\Pi|^{2k}$, $K^{(4) = }\sum_{k = 0, 1} K^{(4)}_{[k]}  |\Pi|^{2k}$, and displayed below, are analytical expressions for all the coefficients of Eq.~(\ref{eq:H-4}):
\begin{align}
\begin{split}
-\Delta^{(4)}_{[0]} &=  15 g_6 +  \frac{9 g_4^2}{\omega_a} + \frac{110}{3} \frac{g_3 g_5}{\omega_a}   + 47 \frac{g_3^2 g_4}{\omega_a^2}    - \frac{6269}{324} \frac{g_3^4}{\omega_a^3}          \\
-\Delta^{(4)}_{[1]} &=  60 g_6  + \frac{54}{5}  \frac{g_4^2}{\omega_a}  + 116 \frac{g_3 g_5}{\omega_a}     + \frac{671}{10}  \frac{g_3^2 g_4}{\omega_a^2 }      + \frac{113}{360}  \frac{g_3^4}{\omega_a^3}        \\
-\Delta^{(4)}_{[2]} &=  30 g_6 -    \frac{9}{2} \frac{g_4^2}{\omega_a}  + \frac{322}{5} \frac{g_3 g_5}{\omega_a}      + \frac{15113}{600} \frac{g_3^2 g_4}{\omega_a^2}      -\frac{297947}{32400} \frac{g_3^4}{\omega_a^3}     \\
-K^{(4)}_{[0]} &= 15 g_6 + \frac{153}{16} \frac{g_4^2}{\omega_a} + 42 \frac{g_3 g_5}{\omega_a}     + \frac{225}{4} \frac{g_3^2 g_4}{\omega_a^2}     +  \frac{805}{36} \frac{g_3^4}{\omega_a^3}    \\
-K^{(4)}_{[1]} &=  30 g_6 +  \frac{27}{5} \frac{g_4^2}{\omega_a} + 58 \frac{g_3 g_5}{\omega_a}     + \frac{671}{20} \frac{g_3^2 g_4}{\omega_a^2}      +  \frac{113}{720}\frac{g_3^4}{ \omega_a^3}     \\
-\lambda^{(4)} &=  \frac{10}{6} g_6 + \frac{17}{8} \frac{g_4^2}{\omega_a} + \frac{28}{3} \frac{g_3 g_5}{\omega_a}     +  \frac{25}{2} \frac{g_3^2 g_4}{ \omega_a^2}    + \frac{805}{162} \frac{g_3^4}{\omega_a^3}     \\
\epsilon_4^{(4)} &=\left(\frac{5}{2} g_6 + \frac{33}{8} \frac{g_4^2}{\omega_a} -\frac{1}{15}  \frac{g_3 g_5}{\omega_a}    -\frac{101}{96} \frac{g_3^2 g_4}{\omega_a}     -\frac{2009}{1296} \frac{g_3^4}{\omega_a^3}     \right)\Pi^2.
\end{split}
\end{align}

All in all, the squeezed Kerr Hamiltonian is not modified structurally by contributions beyond the rotating wave approximation (RWA), up to the fourth order.
This theoretical understanding explains our clean experimental realization of the squeezed Kerr Hamiltonian with a driven SNAIL.

 \section{The Bloch-sphere for the Kerr-cat qubit}
\begin{figure}[h!]
    \centering
    \includegraphics[width=0.8\textwidth]{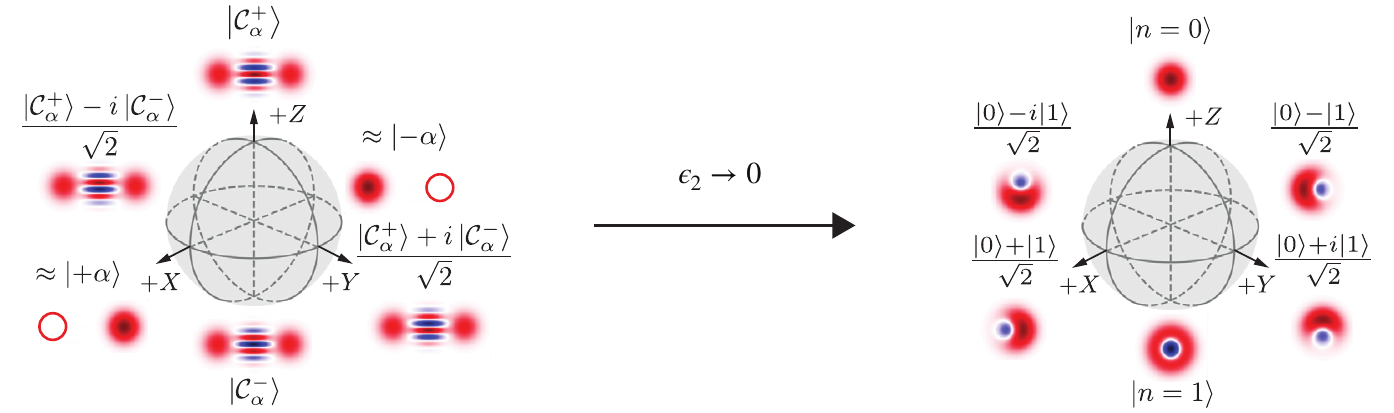}
    \caption{\textbf{The adiabatic mapping of the Kerr-cat qubit Bloch sphere to the Fock encoding}. The logical states in the Kerr-cat qubit and their parity conserving adiabatic mapping to the two-level transmon Bloch sphere.}
    \label{fig:Blochsphere}
    %fig generated with : '/Users/rodrigo/Dropbox/KerrCatMetapot/SupMat/Python/gBS_cal.py')
\end{figure}

The ground states of Hamiltonian  Eq.(2) in the main text are exactly degenerate Schr\"odinger cat states $|\mathcal C_{\alpha}^{\pm}\rangle$.
Restricting our system to this two ground states we define a computational Kerr-cat qubit subspace.
Its Bloch sphere is shown in Figure \ref{fig:Blochsphere}.
The even and odd Schr\"odinger cat states are taken to be in the north and south poles of the sphere since they adiabatically map to the zero and one Fock states of the transmon Kerr oscillator  as $\epsilon_2\rightarrow0$.
This provides a physically realizable mapping \cite{grimm2020} to standard Bloch sphere used in the  superconducting circuit platform. 

Under this convention we define the basis for the computational space by %$|\pm Z\rangle = |\mathcal C_{\alpha}^{\pm}\rangle \propto |+\alpha\rangle \pm|-\alpha\rangle$, $|\pm Y\rangle = |\mathcal C_{\alpha}^{\pm i}\rangle \propto |+\alpha\rangle \pm i|-\alpha\rangle$, and $|\pm X\rangle =\frac{1}{\sqrt{2}}\left( |\mathcal C_{\alpha}^{+}\rangle \pm |\mathcal C_{\alpha}^{-}\rangle\right)= |\pm\alpha\rangle +\mathcal O(e^{-2|\alpha|^2}).$
\begin{equation}
\begin{aligned}
|\pm Z\rangle &= \catpm \\
&=\mathcal{N}_{\alpha}^{\pm}(|+\alpha\rangle \pm|-\alpha\rangle) \\
&=\mathcal{N}_{\alpha}^{\pm} e^{-|\alpha|^{2} / 2} \sum_{n=0}^{\infty}\left(1 \pm(-1)^{n}\right) \frac{\alpha^{n}}{\sqrt{n !}}|n\rangle \\
\mathcal{N}_{\alpha}^{\pm} &=1 / \sqrt{2\left(1 \pm e^{-2|\alpha|^{2}}\right)}.
\end{aligned}
\end{equation}

The mean photon number in these states is given by $\bar{n}_{\pm}=\left\langle\mathcal{C}_{\alpha}^{\pm}\left|\hat a^{\dagger} \hat a\right| \mathcal{C}_{\alpha}^{\pm}\right\rangle=r^{\pm 2}|\alpha|^{2}$, and 
\begin{equation}
r=\frac{\mathcal{N}_{\alpha}^{+}}{\mathcal{N}_{\alpha}^{-}}=\frac{\sqrt{1-e^{-2|\alpha|^{2}}}}{\sqrt{1+e^{-2|\alpha|^{2}}}} \rightarrow 1-e^{-2|\alpha|^{2}}, \label{eq:ratioNorm}
\end{equation}
where the arrow corresponds to the expressions in the limit $\epsilon_2\gg K$ $(|\alpha|^2\gg1)$. We thus define the other two complementary and mutually non-biased Pauli basis as 
\begin{equation}
\begin{aligned}
|\pm X\rangle &=\frac{1}{\sqrt{2}}\left( \catp \pm \catm \right) \rightarrow|\pm \alpha\rangle \\
|\pm Y\rangle &=\frac{1}{\sqrt{2}}\left( \catp \pm i \catm \right) \rightarrow \frac{1}{\sqrt{2}}(|+\alpha\rangle \mp i|-\alpha\rangle) .
\end{aligned}
\end{equation}

The Pauli operators are then defined as 
\begin{equation} \label{eq:catPauliOps}
\begin{aligned}
& \op{X}=\left|\mathcal{C}_{\alpha}^{+}\right\rangle\left\langle\mathcal{C}_{\alpha}^{-}|+| \mathcal{C}_{\alpha}^{-}\right\rangle\left\langle\mathcal{C}_{\alpha}^{+}\right| \\
& \op{Y} =-i\left|\mathcal{C}_{\alpha}^{+}\right\rangle\left\langle\mathcal{C}_{\alpha}^{-}|+i| \mathcal{C}_{\alpha}^{-}\right\rangle\left\langle\mathcal{C}_{\alpha}^{+}\right| \\
& \op{Z} =\left|\mathcal{C}_{\alpha}^{+}\right\rangle\left\langle\mathcal{C}_{\alpha}^{+}|-| \mathcal{C}_{\alpha}^{-}\right\rangle\left\langle\mathcal{C}_{\alpha}^{-}\right|
\end{aligned}
\end{equation}
and their scaling with $\alpha$ is shown in Table \ref{tab:1}.

\begin{table}[h!]
\label{tab:1}
\centering
\begin{tabular}{c|ccc}
\hline \hline  & $\alpha \rightarrow 0$ & $\alpha$ & Large $|\alpha|$ \\
\hline $\hat a$ & $\frac{1}{2}(\hat{X}+i \hat{Y})$\;\;\;\; & $\alpha\left(\frac{r+r^{-1}}{2}\right) \hat{X}-i \alpha\left(\frac{r-r^{-1}}{2}\right) \hat{Y}$ \;\;\;\;& $\alpha \hat{X}+i \alpha e^{-2|\alpha|^{2}} \hat{Y}$ \\
$\hat a^{\dagger}$ & $\frac{1}{2}(\hat{X}-i \hat{Y})$ \;\;\;\;& $\alpha^{*}\left(\frac{r+r^{-1}}{2}\right) \hat{X}+i \alpha^{*}\left(\frac{r-r^{-1}}{2}\right) \hat{Y}$ \;\;\;\;& $\alpha^{*} \hat{X}-i \alpha^{*} e^{-2|\alpha|^{2}} \hat{Y}$ \\
$\hat a^{\dagger} \hat a$ & $\frac{1}{2}(\hat{I}-\hat{Z})$ \;\;\;\;& $|\alpha|^{2}\left(\frac{r^{2}+r^{-2}}{2}\right) \hat{I}+|\alpha|^{2}\left(\frac{r^{2}-r^{-2}}{2}\right) \hat{Z}$ \;\;\;\;& $|\alpha|^{2} \hat{I}-2|\alpha|^{2} e^{-2|\alpha|^{2}} \hat{Z}$ \\
\hline \hline
\end{tabular}
\caption{\textbf{Projection of some ``local'' (low order in $\hat a$ and $\hat a^{\dagger}$) phase space operators over the oscillator to the Kerr-cat logical manifold}. They represent the main source of errors in bosonic codes. Besides the exact expression for all values of $\alpha$ we provide instructive limits for large and small photon-number. Note that in the large cat limit the operators contain corrections of at most $\mathcal{O}(\alpha^2e^{-4|\alpha|^2})$. }
\end{table}

\subsection{Definition of autonomous error correction gain}
Given that we have defined the Bloch sphere, we may also define the average coherence of the Bloch sphere as
\begin{equation}
    \gamma = \frac{1}{6} \sum_{i={\pm X, \pm Y, \pm Z}} \frac{1}{T_i}
\end{equation}
where $T_i$ is the decay time constant for preparing the system in one of the six cardinal states ($\ket{i}$) of the Bloch sphere.

For the Fock qubit, we use the measured Ramsey decay to calculate $(1/T_{+X} + 1/T_{-X} + 1/T_{+Y} + 1/T_{-Y})/4 = 1/T_{2R}$, and the measured single excitation decay time to calculate $1/T_{+Z} + 1/T_{-Z} = 1/ T_1$; this defines the average coherence of the Fock qubit.

For the Kerr-cat qubit, we either independently measure each cardinal state's coherence or use $T_{YZ}$ as in Figure 3 \textbf{E} to summarize the coherence on the $YZ$ plane.
The combination of these measurements allows us to define the autonomous quantum error correction gain factor $\gamma_{\mathrm{Fock}} / \gamma_{\mathrm{cat}}$, which efficiently summarizes the Kerr-cat qubit's performance as a quantum memory relative to the bare Fock qubit encoding.
At the bias point shown in Figure 4 \textbf{C-D}, this leads to an autonomous quantum error correction gain factor of $\befactor$ beyond break-even.

\subsection{Photon loss}
It is interesting to notice that in a realistic scenario, the presence of dissipation will slightly deform the computational states. We analyze this by considering only single and two photon loss. We generalize the Hamiltonian in Eq.(2) by casting the no-jump Hamiltonian of the open quantum system as 
\begin{equation}
\label{eq:HKCnojump}
    \hat H_{\textrm{SK}}^{\textrm{no-j}}/\hbar = - \tilde \Delta \hat a^{\dagger} \hat a - \tilde K\hat a^{\dagger2} \hat a^2+ \epsilon_2(\hat a^{\dagger2} + \hat a^{2}).
\end{equation}
Here the Hamiltonian constants are complex numbers: generalized detuning $\tilde \Delta  = \Delta +i\kappa_1/2$, and generalized Kerr $\tilde K  = K +i\kappa_2/2$, where $\kappa_1$ and $\kappa_2$ are the single and two-photon loss rates in the system.
The phase space center of mass of the new eigenstates is obtained by displacing this no-jump Hamiltonian by a complex amplitude $\tilde\alpha = |\tilde \alpha| e^{-i (\arg{\epsilon_2 / \tilde{K}})/2} e^{i\tilde\phi}$ and solving for null-linear terms in the displaced Hamiltonian.
The dissipative eigenstates are found to be centered at 
\begin{equation}
 |\tilde \alpha|^2 =\frac{1}{2|\tilde{K}|^{2}}\left(-\Delta K+\frac{\kappa_{1} \kappa_{2}}{4}\right) \pm \sqrt{\left|\frac{\epsilon_{2}}{\tilde{K}}\right|^{2}-\left(\frac{K \kappa_{1}-\Delta \kappa_{2}}{4|\tilde{K}|^{2}}\right)^{2}},
\end{equation}
\begin{equation}
\sin \left(2 \tilde{\phi}\right)=\frac{K \kappa_{1}-\Delta \kappa_{2}}{4\left|\epsilon_{2}\right||\tilde{K}|}.
\end{equation}
where the $+$ solution is valid when $|\epsilon_2|^2 \geq (\Delta^2 + \kappa_1^2/4)/4$, which is beyond the parametric instability exploited for parametric amplification and inside the double-well regime of bifurcation relevant to this work (see \cite{Frattini2021} for in depth discussion). Note that the cat size also increases with $-\Delta$, and thus increases the protection of the coherent states to perturbations.

Examining the equation for $\tilde{\phi}$, the remarkable $\pi$-periodicity of the solutions is evident.
This is the quantum manifestation of a classically robust period-doubling bifurcation and it contains the essence of the noise-resilient properties of the our cat qubit.
The solutions are always maximally spaced in phase space forming an angle of $180^{\circ}$.
For $\Delta = 0$ and in the bifurcation regime, we have that the ground-state manifold is spanned by the coherent states $|\pm \tilde \alpha\rangle$ even in presence of dissipation.

\section{The superconducting package and design choices}

In this section, we outline the key design principles for the superconducting package and cQED system.
The overarching design goal was to incorporate two capacitively coupled SNAIL-transmon circuits that could each be individually stabilized with a squeezing drive and independently read out.
With such a system, we plan to implement two qubit gates between two Kerr-cat qubits: specifically, a noise-bias preserving CNOT gate \cite{puri2020}.
For the work presented here, we address only one of the two chips with microwave drives and the modes from the other chip play the role of spectators.
Table \ref{tab:deviceParams} lists a few key design parameters.

\begin{table}[h!]
\label{tab:deviceParams}
\centering
\begin{tabular}{l| c | l}
\hline \hline  
\textbf{Parameter} & \textbf{Value} & \textbf{Method of estimate or measurement} \\
\hline
Oscillator dipole capacitance $E_C/h$ & 60 MHz & Design simulation \\ %for 6.15 GHz at 3 nH
Oscillator number of SNAILs & 2 & Design \\
Oscillator SNAIL asymmetry $\alpha$ & 0.1 & Room temperature resistance msmt.\\
Oscillator inductance of 1 large junction & 0.6 nH & Room temperature resistance msmt.\\
Oscillator oscillator inductance of 1 small junction & 6 nH & Room temperature resistance msmt.\\
Oscillator oscillator frequency at $\Phi /\Phi_0 = 0$ & 6.668 GHz & Two-tone spectroscopy\\ % need 2.55 nH inductance according to HFSS
Oscillator oscillator frequency at $\Phi /\Phi_0 = 0.5$ & 5.815 GHz & Two-tone spectroscopy\\ % need 3.55 nH, HFSS predicts 200 MHz lower than actual so need a bit more inductance IRL to compensate
\hline
External flux bias point $\Phi / \Phi_0$ & 0.33 & Two-tone spectroscopy \\
Oscillator frequency $\omegaa/ 2\pi$ & 6.079 GHz & Two-tone spectroscopy \\
Oscillator cubic nonlinearity $g_3/ 3 / 2\pi$ & $\approx10$ MHz  & Design simulation \\
Oscillator self-Kerr nonlinearity $K/2\pi$ & $\kerr$ & Coherent state refocusing experiment (Fig.~\ref{fig:supmat_FreeKerr}) \\
Oscillator single-photon decay time $T_1$ & $\fockTone$ & Std.~coherence msmt.~(w/ fluorescence readout) \\
Oscillator Ramsey decay time $T_2^*$ & $\FockTtwostar$ & Std.~Ramsey coherence msmt.~(w/ fluorescence readout) \\
Oscillator Hahn echo decay time $T_{2E}$ & 13~\textmu s & Std.~echo coherence msmt.~(w/ fluorescence readout) \\
Readout resonator frequency $\omegar / 2\pi$ & 8.506 GHz & Direct RF reflection measurement \\
Readout resonator linewidth $\kappar / 2\pi$ & $\kapparValue$ & Direct RF reflection measurement \\
Readout resonator internal linewidth  & $< 0.04$ MHz  & Direct RF reflection measurement \\
Readout to oscillator cross-Kerr $\chi_{ab}/2\pi$ & $\sim 10$ kHz & Design simulation \\ 
Purcell filter frequency & 8.703 GHz & Direct RF reflection measurement \\
Purcell filter linewidth & 25 MHz & Direct RF reflection measurement \\
Lowest box mode frequency & 12.46 GHz & Design simulation\\ %box 1 (long) 4mm x 28 mm x 11 mm
\hline
Spectator SNAIL-transmon frequency & 6.302 GHz & Two-tone spectroscopy \\ %7.035 at 0-flux
Capacitive coupling btwn.~oscillator \& spectator & 9.4 MHz & Avoided crossing in two-tone spec.~vs.~flux \\
Spectator SNAIL-transmon self-Kerr & 2.8 MHz & Two-tone spectroscopy \\ % 211221
Spectator SNAIL-transmon single-photon decay time & 20 \textmu s & Std.~coherence msmt.~(w/ fluorescence readout)\\
Spectator SNAIL-transmon Ramsey decay time & 2 \textmu s & Std.~Ramsey coherence msmt.~(w/ fluorescence readout) \\
Spectator to oscillator cross-Kerr & $\approx 0.03$ MHz & Design simulation \\ % chi = 2 (4g/Delta)^2 [anharm_a + anharm_b]
Spectator readout resonator frequency & 8.775 GHz & Direct RF reflection measurement \\
Spectator readout resonator linewidth & 0.58 MHz & Direct RF reflection measurement \\
Spectator Purcell filter frequency & 8.890 GHz & Direct RF reflection measurement \\
Spectator Purcell filter linewidth & 14 MHz & Direct RF reflection measurement \\
Lowest spectator box mode frequency & 12.78 GHz & Design simulation\\ % box 2 (short) 4mm x 23mm x 11mm
Coupling between lowest box modes & $\approx 20$ MHz & Design simulation\\ % Box sep: 0.05in, box_coup: 5mm wide and 3/4 of 11mm height
\hline \hline
\end{tabular}
\caption{\textbf{Summary of device parameters}. All design simulations were performed with Ansys HFSS and black box quantization \cite{nigg_black-box_2012, minev_energy-participation_2021} including corrections to Kerr from cubic nonlinearities \cite{Frattini2021}, which follow from similar corrections in lumped-element calculations \cite{frattini_optimizing_2018}. All parameters in the lower two sections correspond to the particular flux bias point $\Phi/\Phi_0 = 0.33$ common to all data in this work.}
\end{table}
The design closely follows the coaxline architecture that includes a seam between the two package pieces for easy assembly and multiplexing \cite{axline_architecture_2016}.
The package structure consists of two rectangular waveguide cavities designed with lowest box mode at 12.46 GHz for the addressed system and 12.78 GHz for the spectator system.
An aperture between the two cavities couples the two spatially separate box modes weakly compared to their detuning, which helps keep the box eigenmodes---and thus strong microwave drives applied to them--- spatially separated while still allowing for direct capacitive coupling between the two SNAIL-transmons.
The aspect ratio of the boxes (tall and narrow) restricts the on-chip eigenmodes from having significant participation in either the seam or the copper bottom piece.
The copper bottom piece is necessary (as opposed to aluminum) to allow penetration of magnetic flux to bias the superconducting SNAIL loops via two small coils mounted within counterbored holes in the copper piece.
Despite superconducting aluminum's propensity to reject magnetic fields, the necessary applied field to bias the SNAIL loops at one $\Phi_0$ is similar to the same geometry made entirely of copper; as long as the coil area is smaller than magnetic field's entry hole in the aluminum, the magnetic field lines have a convenient return path back through the same hole implying the presence of the aluminum actually has a \emph{focusing} effect of the field toward the SNAIL loops.
This results in independent flux-biasing capabilities for two qubits in a 3D superconducting architecture with field crosstalk of order $10^{-2}$.

Each sapphire chip is clamped to two copper posts with beryllium copper clips, and hosts three electromagnetic modes of interest: SNAIL-transmon, readout resonator, and Purcell filter.
A beryllium copper pin inserted into the cavity defines the readout port and sets the linewidth of the readout resonator and Purcell filter; a second weakly coupled pin serves for the application of all microwave drives.

Focusing on the design of the addressed SNAIL-transmon, the design change compared to previous work \cite{grimm2020}, where the coherent state lifetime saturated at $|\alpha|^2 = 2.6$, is a twenty-fold reduction in self-Kerr nonlinearity.
We actuated this reduction by moving to a series array of $M=2$ SNAILs, reducing the SNAIL junction asymmetry parameter \cite{frattini2017}, and working at a larger magnetic flux bias $\Phi/\Phi_0 = 0.33$ closer to (but not at) the Kerr-free flux point \cite{frattini_optimizing_2018}.
These changes resulted in coherent state lifetime saturation around $|\alpha|^2 = 10$ in the current device (for $\Delta = 0$).
This trend is consistent across multiple (including unpublished) devices; namely, lower self-Kerr $K$ devices attain larger $|\alpha|^2$ before the coherent state lifetime saturates despite lower third-order nonlinearity $g_3$.
The intuition for this result mirrors a similar intuition as to why SNAIL parametric amplifiers with lower self-Kerr handle more signal power before unwanted saturation effects occur \cite{frattini_optimizing_2018, sivak_kerr-free_2019}.
Specifically, although increasing $M$ reduces both $g_3$ and $K$, it also increases the maximum  number of allowed photons in the nonlinear oscillator, which may be parameterized by $n_{\mathrm{crit}} \approx 15 M^2/p^2 \varphi_{\mathrm{zps}}^2$, where $p$ is the inductive participation ratio of the entire SNAIL array in the electromagnetic mode (see \cite{Frattini2021} for derivation).
With the intuition backed by experiments that drive-induced heating becomes problematic at some fraction of $n_\mathrm{crit}$, we may derive an expected increase in Kerr-cat size (at $\Delta = 0$) at which drive-induced heating becomes problematic:
\begin{align}
    |\alpha|^2 &= |\epsilon_2/K| \nonumber \\
    &= |g_3 \Pi / K| \nonumber \\
    &\propto M \Pi \nonumber \\
    &\propto M^2
\end{align}
In the third line we assumed that $|g_3/K| \propto M$ \cite{frattini_optimizing_2018, Frattini2021} and in the last line that $n_\mathrm{crit}$ scales the limit on $\Pi \propto M$.
From this simple argument, we conclude that increasing the number of SNAILs to decrease Kerr nonlinearity effectively increases the experimentally achievable cat size.

Keeping in mind that gate speeds are limited by the gap in the excited state spectrum $4 K |\alpha|^2$ and that $K \propto 1/M^2$, gate speeds are independent of the Kerr nonlinearity if the expected $|\alpha|^2$ increase is simultaneously achieved.
The exception to this is the Kerr-refocusing gate, which takes time $\pi/2K$ irrespective of $|\alpha|^2$.
Generally, the gate fidelity however will decrease under this optimization since the cat-state lifetime $T_{YZ} = 1/2|\alpha|^2 T_1$.
As such, we expect there to be an optimum Kerr nonlinearity for a given application depending on the tradeoff between achievable gate speed and noise bias.
Improvements in gate design and control techniques will help increase gate speed for a given gap in the excited state spectrum \cite{xu_engineering_2022}, and thereby will allow designs with less nonlinearity to further increase coherent state lifetime and noise bias.

\section{Readout of the squeezed Kerr oscillator}
It is critical for our implementation that we do not rely on the ordinary dispersive readout of the superconducting circuit platform.
This means that our Kerr-cats are prepared, evolved, and detected without relying on standard Fock qubit operations or measurements.
This ensures the high performance of a system with a weak \textit{bare} nonlinearity.
Instead of using the dispersive coupling of the SNAIL-transmon to the readout resonator we use a parametrically activated readout scheme with a large on-off ratio.
The readout is enacted by playing a microwave pulse at the frequency difference in between the squeezed Kerr oscillator frame $\omegap/2$ and readout resonator at $\omegar$ while the stabilization drive is on.
The nonlinear term providing the interaction originates from the SNAIL array that, when activated by a microwave tone of displacement amplitude $\xi_{BS}$ and frequency $\omega_{BS}=\omegar-\omegap/2$, transforms as
\begin{equation}
\begin{aligned}
g_{3}\left(\hat a+\frac{g_{b a}}{\Delta_{b a}} \hat b +\text { H.c. }\right)^{3} & \rightarrow g_{3}\left(\hat a e^{-i \omega_d t / 2}+\xi_{BS} e^{-i \omega_{BS} t}+\frac{g_{b a}}{\Delta_{b a}} \hat b e^{-i \omegar t}+\text { H.c. }\right)^{3} \\
& \approx 6 g_{3} \frac{g_{b a}}{\Delta_{b a}}\left(\xi_{BS} \hat a \hat b^{\dagger}+\xi_{BS}^{*} \hat a^{\dagger} \hat b\right).
\end{aligned}
\end{equation}
Here, we have used the RWA to get rid of fast rotating terms in a displaced and rotating frame for $\hat a$, $g_3$ is the third order nonlinearity of the SNAILs, $g_{ba}$ is the bare capacitive coupling in between the Kerr-oscillator and the readout resonator, $\hat b$ is the annihilation operator of the resonator, and $\Delta_{ba} = \omegar-\omegaa$.
The hybridization is given in the dispersive approximation for simplicity.
Note, however, that the RWA approximation is unjustified---the detuning of the readout drive is of order the oscillator frequencies---and higher orders need to be considered (see section \ref{supp:1}).
To leading order, their effect is a renormalization of the beamsplitter coupling rate $g_{BS}$ in the effective interaction
\begin{equation}
   g_{BS} \ah \opdag{b} + g_{\BS}^* \adag \op{b}
\end{equation}
which is the announced frequency-converting beamsplitter interaction.

When activated in presence of the stabilization drive, the photons in the squeezed Kerr oscillator displace the readout resonator with a resonant drive strength $\pm g_{\BS} \alpha$, which subsequently radiates into a quantum-limited amplifier chain that we demodulate and monitor with a precision microwave measurement setup at room temperature.
For sufficiently weak readout drive strength such that $|g_{BS}|^2  \ll 2 K |\alpha| \kappar$, the photons emitted by the oscillator are replenished by the stabilization drive effectively preserving the photon number in the state.
These photons should be thought as displacing the readout resonator by an amount $\beta = \pm i2g_{BS} \alpha/\kappar$, which is conditional on $\hat a \approx \pm\alpha$ and thus forcing the cat to collapse into one of its coherent state components.
The process enacts then a quantum nondemolition (QND) measurement of the quadrature in the Kerr-cat qubit, thus named cat-quadrature readout (CQR) \cite{grimm2020}.
For nearly degenerate excited states within the metapotential wells, the readout resonator is similarly displaced conditioned on which well $\pm \alpha$.
The readout gains no information on population within each well assuming $\kappar \ll 4 K |\alpha|^2$, as is the case in our experiment.
For excited states outside the metapotential wells, the readout drive is sufficiently off-resonance from any transitions such that the readout resonator is not appreciably displaced from its vacuum state.

Writing the Langevin equation for the coupled system and solving for the steady state amplitude in the readout resonator (see \cite{grimm2020, Frattini2021}), we find, as derived \cite{didier_fast_2015} and verified \cite{Touzard2019} for conditional displacement readout, the voltage signal-to-noise ratio of the measurement, which is the square root of the usual power-defined SNR,
\begin{align}
    \sqrt{\mathrm{SNR}(\tau)} &= \sqrt{32 \eta} \frac{|g_{BS} \alpha|}{\kappar} \left[ \kappar \tau - 4 \left(1 - e^{-\kappar \tau/2} \right) + \left(1-e^{-\kappar \tau} \right) \right]^{1/2} \label{eq:KCexp_SNR_CQR}
\end{align}
where $\tau$ is the square measurement pulse time and $\eta$ is the quantum efficiency of the entire measurement chain.
The expression here assumes the optimal demodulation envelope, as used in the experiment.
The experimentally measured SNR may be read from histograms as $\mathrm{SNR} = |I_{+\alpha} - I_{-\alpha}|^2/2\sigma^2$, where $I_{\pm \alpha}$ is the mean in-phase quadrature value for preparation in $\ket{\pm \alpha}$ and $\sigma$ is the standard deviation of the resultant histogram (c.~f.~main text Figure 2).
Note, all information is aligned to the in-phase quadrature either by the demodulation envelope, or by choosing the readout drive phase such that $\arg(g_{BS}) = +\pi/2$.
Invoking the aforementioned drive strength limitation to prevent leakage to higher excited states, we may bound the SNR as
\begin{equation}
    \textrm{SNR} \propto \frac{|g_{BS} \alpha|^2}{\kappar^2} \kappar \tau
    \ll 2 K |\alpha|^3 \tau
\end{equation}
in the long time $\kappar \tau \gg 1$ limit.

The expression favors large values of the mean photon number $|\alpha|^2$ in the oscillator. This is of technological relevance since obtaining high readout SNR is then consistent with the other two main desiderata for the Kerr-cat qubit. Namely, fast gate speed (limited by the gap to the first excited state in the spectrum measured here to scale as $\approx 4K|\alpha|^2$), and large noise-bias (found to increase with $|\alpha|^2$ in the decoherence studies presented here).

\begin{figure}[h!]
    \centering
    \includegraphics[width=\figwidthWide]{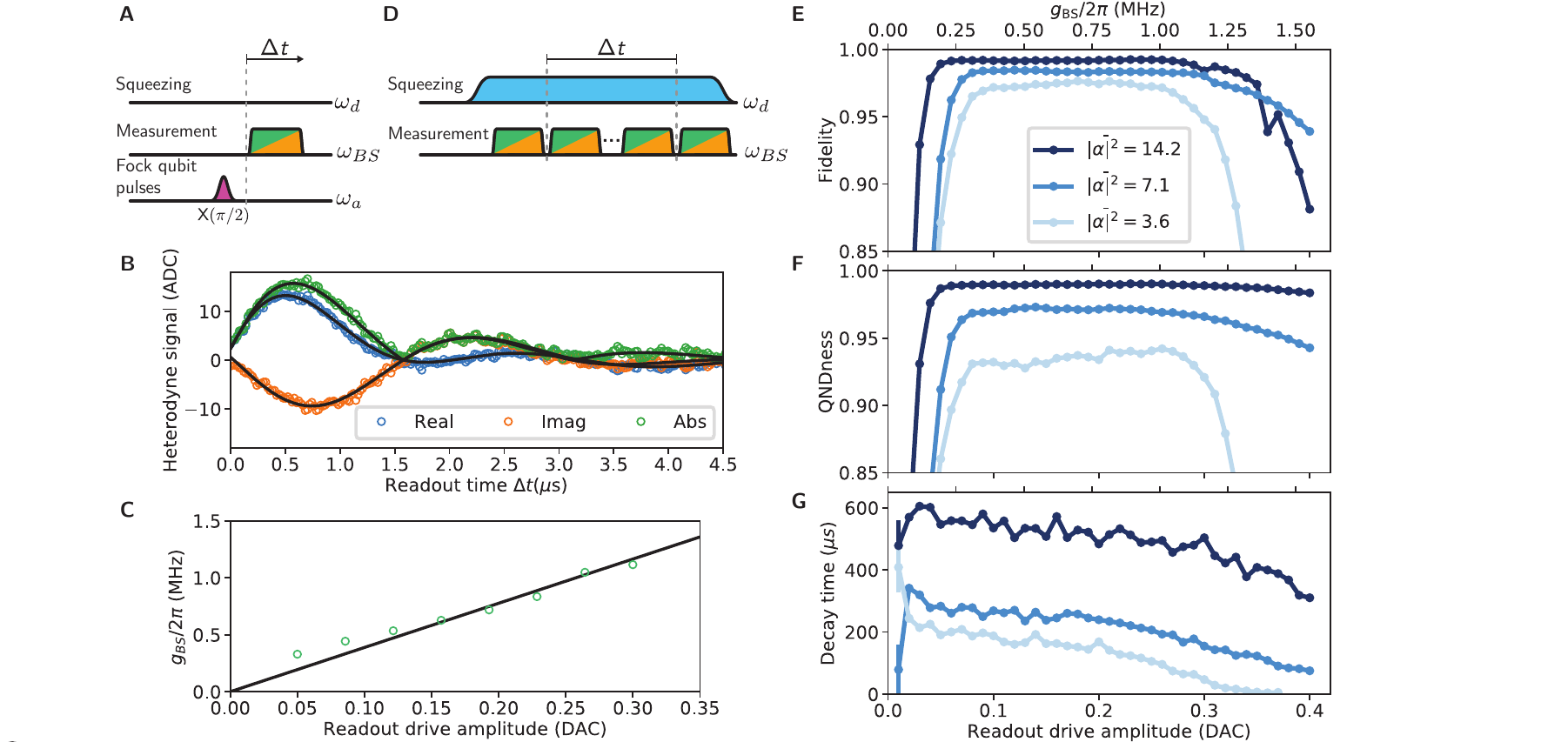}
    \caption{\textbf{ Measurement calibration and robustness to readout power. A}, Pulse sequence for readout drive amplitude calibration experiment (see \textbf{B}, \textbf{C}). Preparing the Fock qubit on the equator of the Bloch sphere (squeezing drive off), a readout pulse at $\omega_{BS} = \omegar - \omegaa$ implements a fluorescence readout.
    \textbf{B}, Example time trace of readout resonator output field recorded as a heterodyne signal after amplification. Black lines are fits to extract beamsplitter strength $g_{BS}$.
    \textbf{C}, Extracted $g_{BS}$ (green circles) as of function of readout pulse amplitude fit to a line (black) with no offset.
    \textbf{D}, Pulse sequence for repeated measurement as in main text, where again $\omega_{BS} = \omegar - \omegap/2$.
    \textbf{E}-\textbf{G}, Fidelity, QNDness, and decay time during repeated measurements respectively as a function of readout drive amplitude for three different squeezing drive strengths on resonance. Lines connecting points are a guide to the eye.
    }
    \label{fig:supmatCQR}
    %fig generated with : '/Users/rodrigo/Dropbox/KerrCatMetapot/SupMat/Python/gBS_cal.py')
\end{figure}
In Figure \ref{fig:supmatCQR}, we examine the measurement quality as function of squeezing drive amplitude and readout drive amplitude.
To this end, we first calibrate the readout drive amplitude in terms of the induced beamsplitter rate $g_{BS}$ with the experimental pulse sequence of Figure \ref{fig:supmatCQR} \textbf{A}.
With the squeezing drive off ($\epsilon_2 = 0$), we prepare the Fock qubit on the equator of its Bloch sphere and turn on an uncalibrated readout drive (frequency $\omega_{\BS} = \omegar - \omegaa$) to pitch the Fock qubit state out the overcoupled port of the readout resonator.
This effectively performs a fluorescence readout of the Fock qubit~\cite{campagne-ibarcq_observing_2016}, which is of higher fidelity than dispersive readout in this system.
An example emitted signal measured via heterodyne detection is shown in Figure \ref{fig:supmatCQR} \textbf{B}.
In the limit $g_{BS} \ll \kappar$, the emitted field's energy would decay exponentially at rate $4g_{BS}^2 / \kappar$, implementing a so-called ``Q-switch" on the Fock qubit often used for cooling.
However, in our experiment $g_{BS} \gtrsim \kappar$ so the photon swaps back and forth at rate $g_{BS}/2$ between the Fock qubit and the readout resonator before being emitted out the detection port.
Fitting these oscillations (black) allows us to extract $g_{BS}$ (see \cite{pfaff_controlled_2017, grimm2020} for model derivation).
Performing this experiment for different readout drive amplitudes, we extract $g_{BS}$ and fit the dependence to a line with no offset (Figure \ref{fig:supmatCQR} \textbf{C}) as expected for this three-wave mixing process.

With readout amplitude calibration in hand, we apply the squeezing drive to assess the quality of readout.
As in the main text, we follow the pulse sequence in Figure \ref{fig:supmatCQR} \textbf{D} to repeatedly perform readout with the squeezing drive on to set the mean number of photons in the squeezed Kerr oscillator to $|\alpha|^2 = \epsilon_2/K$ ($\Delta = 0$).
We examine three readout metrics for three different squeezing strengths as a function of readout drive amplitude: fidelity in \textbf{E}, QNDness in \textbf{F}, and decay time during readout in \textbf{G}.
We calculate the fidelity as $\mathcal{F} = 1 - p(+\alpha | -\alpha) - p(-\alpha | +\alpha)$ where $p(\pm \alpha | \mp\alpha)$ is the probability---as extracted from experimental histograms---of measuring the qubit in $\ket{\pm \alpha}$ after initializing the opposite state $\ket{\mp \alpha}$ with a previous stringently thresholded measurement.
We characterize the QNDness as $\mathcal{Q} = \left( P(+\alpha | +\alpha) + P(-\alpha | -\alpha) \right)/2$ where $P(i|i)$ is the probability of obtaining measurement outcome $i$ in two successive fair measurements.
We extract the decay timescale during measurement by averaging many measurement records together conditioned on the first initialization measurement and fitting to a single exponential decay (see main text Figure 2).

Focusing first on the readout drive amplitude dependence, both the fidelity and QNDness increase sharply at low amplitude as the blobs separate in IQ-space and the bare SNR increases.
For a large region of readout drive amplitude, the fidelity and QNDness are both approximately constant until eventually degrading at higher amplitudes.
We interpret the constant region to correspond the drive amplitudes where the fidelity is no longer limited by the bare SNR of the blob separation, but instead by the decay of one coherent to the other during readout.
This interpretation is consistent with comparing the total time for a single measurement $\tau = 4.44\,\mathrm{\mu s}$ to the measured decay time during readout.
The measured QNDness also saturates since the maximally extractable QNDness for this experimental definition depends on the readout fidelity.
A fidelity-independent way to extract QNDness is to look at the extracted decay time during readout.
The decay time decreases monotonically (ignoring the lowest amplitude points of low QNDness) with increasing readout drive amplitude.
This is reminiscent of the same reduction of lifetime during readout with increased readout power---colloquially often referred to as ``$T_1$ vs. $\bar{n}$"---that plagues most circuit QED systems ~\cite{sank2016, minev2019, lescanne_escape_2019}.

Turning to the cat size $|\alpha|^2$ dependence on readout quality, the data clearly indicate increased fidelity, increased QNDness, and increased lifetime during readout for larger $|\alpha|^2$ in the range shown.
The reason for this trend is three-fold: increasing $|\alpha|^2$ increases bare SNR (c.~f.~Eq.~\ref{eq:KCexp_SNR_CQR}), increasing $|\alpha|^2$ increases coherent state lifetime (c.~f.~main text Figure 3 \textbf{D} and Supplementary Section \ref{sec:SupMatDecoherence}), and increasing $|\alpha|^2$ increases robustness to external drives (c.~f.~Supplementary Section \ref{sec:SupMatTcohUnderRabi}).
Generally, these trends continue until increasing $|\alpha|^2$ no longer increases the coherent state lifetime $T_X$, implying that designing systems with larger $T_X$ will correspondingly increase readout performance even in the regime of ``$T_1$ vs. $\bar{n}$" type effects.
Such considerations might lead future experiments to optimize readout performance by always performing readout at the highest $T_X$ bias point.
For instance, an application that might wish to set $|\alpha|^2$ lower to optimize the breakeven metric may still wish to inflate $|\alpha|^2$ before readout by increasing $\epsilon_2$ adiabatically with respect to $4K|\alpha|^2$.
Without increasing $T_X$, increasing readout speed would further increase readout fidelity (in the flat regime of readout amplitude where fidelity is otherwise limited by lifetime).
The relatively conservative value of $\kappar/2\pi = \kapparValue$ limited our readout time $\tau = 4.44\,\mathrm{\mu s}$ and may be easily increased in future experiments.

\section{Preservation of the bias during continuous cat Rabi gate.} \label{sec:SupMatTcohUnderRabi}
The metapotential describes a double well energy surface with nonlinear components in both the $x$ and $p$ quadratures. Regardless, around its minima $\tilde \alpha$ a harmonic approximation becomes increasingly good as $\epsilon_2\gg K$. Defining the operator $\delta \hat a = \hat a  - \tilde \alpha $ we use the generalized Taylor expansion for Hilbert space operators to write now the non-Hermitian Hamiltonian Eq.(\ref{eq:HKCnojump}) as

\begin{equation}
\hat H^{\textrm{no-j}}_{\textrm{SK}}/\hbar\approx-4 \tilde{K}|\tilde \alpha|^{2} (\delta\hat{a}^{\dagger}) \delta\hat{a}-\tilde{K}(\delta \hat{a}^{\dagger })^2(\delta \hat{a})^{2} - 2 \tilde{K} \tilde \alpha (\delta\hat{a}^{\dagger })^2 \delta\hat{a}+\text { H.c. }
\end{equation}
which stands for an oscillator that becomes stiffer as $\epsilon_2$ grows while its Kerr non-linearity remains constant.
This increasingly harmonic oscillator is described by an energy $\hbar \Delta_{\textrm{gap}} =-\hbar(4 K+i 2 \kappa_{2})|\alpha|^{2}$.
This energy limits the speed of operation of the Kerr-cat qubit and can be thought as the energy gap in between the degenerate logical space of the qubit and the first level out of the code space.
We have reported the experimental verification of this scaling in the case $K \gg \kappa_2$ in the main text. In absence of Kerr ($K\ll \kappa_2$) it has been verified and reported in \cite{touzard2018}.

The operation speed limit is discussed in \cite{grimm2020} but can be understood, alternatively, in the following way.
The cat Rabi drive, in resonance with $\omegap/2$ and of amplitude $\epsilon_x$, can be thought as drive detuned by $\Delta_{\textrm{gap}}$ over the stiff harmonic wells of the metapotential.
In presence of single photon dissipation, such a detuned drive will stabilize, in the harmonic oscillators, a coherent state of amplitude $\epsilon_x / (\Delta_{\textrm{gap}} - i\kappa_1/2) \approx \epsilon_x / \Delta_{\textrm{gap}}$.
The condition of perturbative drive is that the displacement of ground state is much smaller than the vacuum spread: $\epsilon_x / \Delta_{\textrm{gap}} \ll 1$.
This condition ensures that the state will not ``leak'' out of the logical manifold into the excited states and limits the achievable Rabi rate $\Omega_x = \real{4\epsilon_x \alpha^*}$ to $|\Omega_x| \ll 4 |\Delta_{\textrm{gap}} \alpha| = 16 |\alpha|^3 \sqrt{K^2 + \kappa_2^2/4}$.

\begin{figure}[t]
    \centering
    \includegraphics{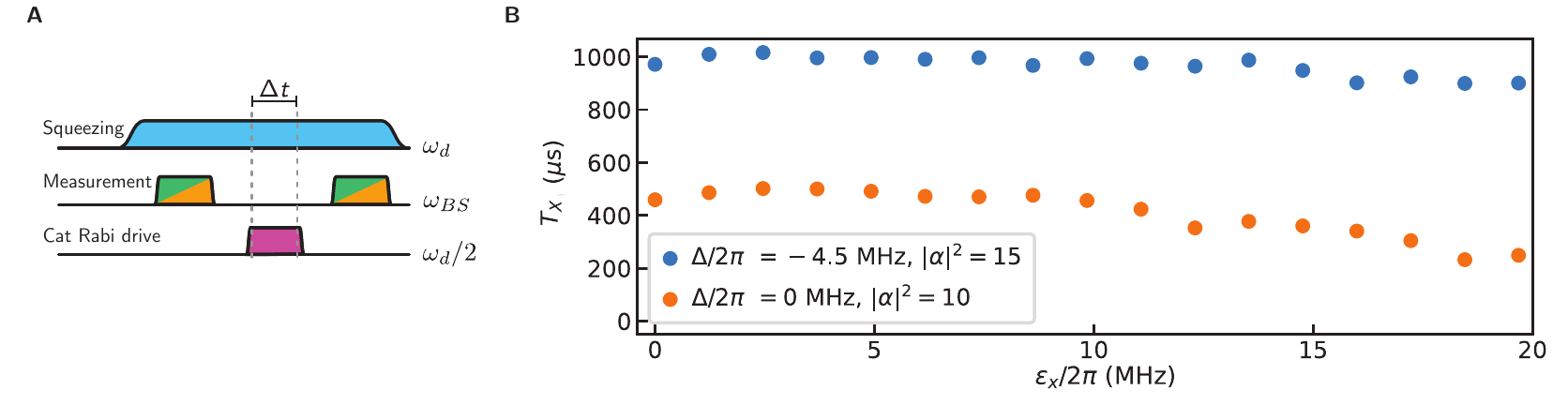}
    \caption{\textbf{Bias robustness during gate operation}. \textbf{A} Pulse sequence to measure $T_X$ in the presence of a strong cat Rabi drive at frequency $\omega_d/2$ with amplitude $\epsilon_x$ and phase set to zero to drive cat Rabi oscillations at rate $\Omega_x = \real{ 4\epsilon_x \alpha^* }$.
    \textbf{B}, The coherent state lifetime $T_X$ is unaffected by the strong Rabi drive. The two sets of points corresponds to a coherent state of $|\alpha|^2= 10$ ($\Delta/2\pi = 0$, orange) and $|\alpha|^2= 15$ ($\Delta/2\pi = -4.5$~MHz, blue). }
    % B) Time evolution of in the Kerr-cat qubit manifold under a single photon drive in resonance with the bare transmon ($|\alpha|^2\sim5$ and $\epsilon_x\sim 2\pi 1.4$~MHz). The phase of the drive (with respect to the squeezing drive) controlled the evolution from inhibited to a fast Rabi oscilation. The details of the image around the values of Rabi phase close to  $\pm 90^{\circ}$ are aliasing caused by the unbounded  spatial frequencies (for long times) of the image around the horizontal axis of the plot. In C) we show a zoom-in of B) the short time signal. In D) we show a vertical cut of panel A) at $ 0^{\circ}$ and $\sim- 90^{\circ}$. In E) we show a zoom-in of panel D). 
    \label{fig:Tcoh_vs_Rabi_amp}
    %Shared.../KerrCatMetapot/SupMat/Tcostate_vs_RabiAmp_2.py
    %Shared.../KerrCatMetapot/SupMat/Python/n_bar_cal_9.py
  %  /Volumes/home/qulab-CC1558-FASAPP/shared/Papers/KerrCat_2021/KerrCat_2022/
     %KerrCatMetapot/SupMat/Python
  %For calibrations see timerabi vs pumpamp 20210713_0
  %freekerr 20210713 (pi/K ~ 1050)
\end{figure}

We further verify that the coherent state lifetime is not degraded by pushing the rate of the continuous cat Rabi gate towards this limit.
In Figure \ref{fig:Tcoh_vs_Rabi_amp}, we observe that even past the largest gap involved in our experiments ($\sim$12~MHz for $|\alpha|^2\sim 10$), the coherent state lifetime $T_X$ is essentially unaffected.
Naturally, for such high Rabi amplitudes the state undergoes coherent oscillations in between the ground state and the excited states of the metapotential (corresponding to leakage) and should be avoided during the operation of the Kerr-cat qubit.
Importantly, at these high amplitudes where the drive is explicitly causing leakage, $T_X$ is unaffected because the leakage is primarily to pairwise degenerate state within the wells.
This relaxes constraints on leakage errors during gates in applications that require the preservation of a large noise bias.
%The one millisecond signal is obtained by $|\alpha|^2\sim 8.3$ and an explicit detuning $\Delta \sim -4.5$~MHz. The half millisecond signal corresponds to $|\alpha|^2\sim 10$ and $\Delta \sim 0$ ($\pm10$~kHz).% This coherent control may allow to use the \textcolor{red}{Kerr-cat} spectrum to do non-bias preserving gates over the Kerr-cat qubit in presence of the stabilization drives and use optimal control theory to beat limits imposed by adiabaticity. This is a unique quality of the \textcolor{red}{Kerr-cat} system and it is considered of interest in applications \cite{GKP}.

\section{Bohr quantization of the squeezed Kerr oscillator}\label{sec:supMatBohrQuant}
In this section we will treat in a semiclassical fashion the squeezed Kerr oscillator and use Bohr's quantization principle to find the number of bound states per well.
We first take the invertible Wigner transform $\mathfrak W$ of the squeezed Kerr Hamiltonian operator in Eq.(2) in main text. The quantum phase space Hamiltonian reads \cite{McCoy1932, hillery1984, groenewold1946,curtright2013, moyal1949}
\begin{equation}
% -\hbar\Delta(x^2+p^2)
\label{eq:h_wig}
H_{\textrm{SK}}(x,p)/\hbar = (K-\Delta/2)(x^2+p^2)-K/4(x^2+p^2)^2+ \epsilon_2(x^2-p^2),
\end{equation}
where we took $\hat a=(\hat x+i\hat p)/\sqrt{2}$ with $[\hat x,\hat p]=i$ and neglected irrelevant additive constants. The frequency shift $K(x^2+p^2)$ is the expected McCoy \cite{McCoy1932,curtright2013} correction and it has its origin in the familiar non-commutativity in between $\hat a $ and $\hat a ^{\dagger}$. This correction is nothing but the Lamb-shift.%Bohr's quantization can be applied over this metapotential surface, but we chose instead to carry out the argument in the classical limit. We do this to avoid an incorrect identification in between the fully quantum metapotential and the semicalssical approach. Also the calculation is easier in the classical case while the main result addressed here is unchanged.\footnote{Applying the quantization argument on the (already fully quantum) metapotential yields $\textrm{Area}/2\pi=\hbar\frac{ \sqrt{2|\alpha|^4-1} + \arccos(-1/2|\alpha|^2)}{2\pi}\sim\hbar(\frac{|\alpha|^2}{\pi}+\frac{1}{4})$ requiring the introduction of a Maslov-like index, where we took $\Delta=0$ for brevity. The generalized EBK \cite{stone2005} quantization condition reads $\frac{1}{2 \pi} \oint Pd  X= \hbar (N+\frac{1}{4}).$ The result is only slightly modified for small quantum numbers and is completely unchanged in the limit of large quantum numbers (achieved here for $|\alpha|^2 \gtrsim1$).} 

Following Dirac's correspondence \cite{dirac1945} we take the classical limit ``$\hbar \rightarrow 0$'' to transform Eq.(\ref{eq:h_wig}) into the fully classical phase space Hamiltonian%\footnote{Understanding Dirac's colloquial classical limit amounts to accepting that ``$\hbar\rightarrow0$'' does not apply to the commutator constant in between the dimensioned coordinates $[\hat X,\hat P]=i\hbar$ (since one must recover the elementary Poisson bracket in the correspondence limit $\frac{1}{i\hbar}[\hat X,\hat P] \rightarrow \{X,P\}=1$) nor to the energy-frequency proportionality constant in $\hbar K$ and $\hbar \epsilon_2$. Only higher order commutators, of order $\hbar^2$ or higher, are taken to zero. This is clarified by the phase space formulation \cite{curtright2013} and discussed in next section. } 
\begin{equation}
% -\hbar\Delta(x^2+p^2)
\label{eq:hkc_classical}
H^{\textrm{cl}}_{\textrm{SK}}(x,p)/\hbar =-\Delta/2(x^2+p^2)-K/4(x^2+p^2)^2+ \epsilon_2(x^2-p^2).
\end{equation}

\begin{figure}[t]
    \centering
    \includegraphics[width=1\textwidth]{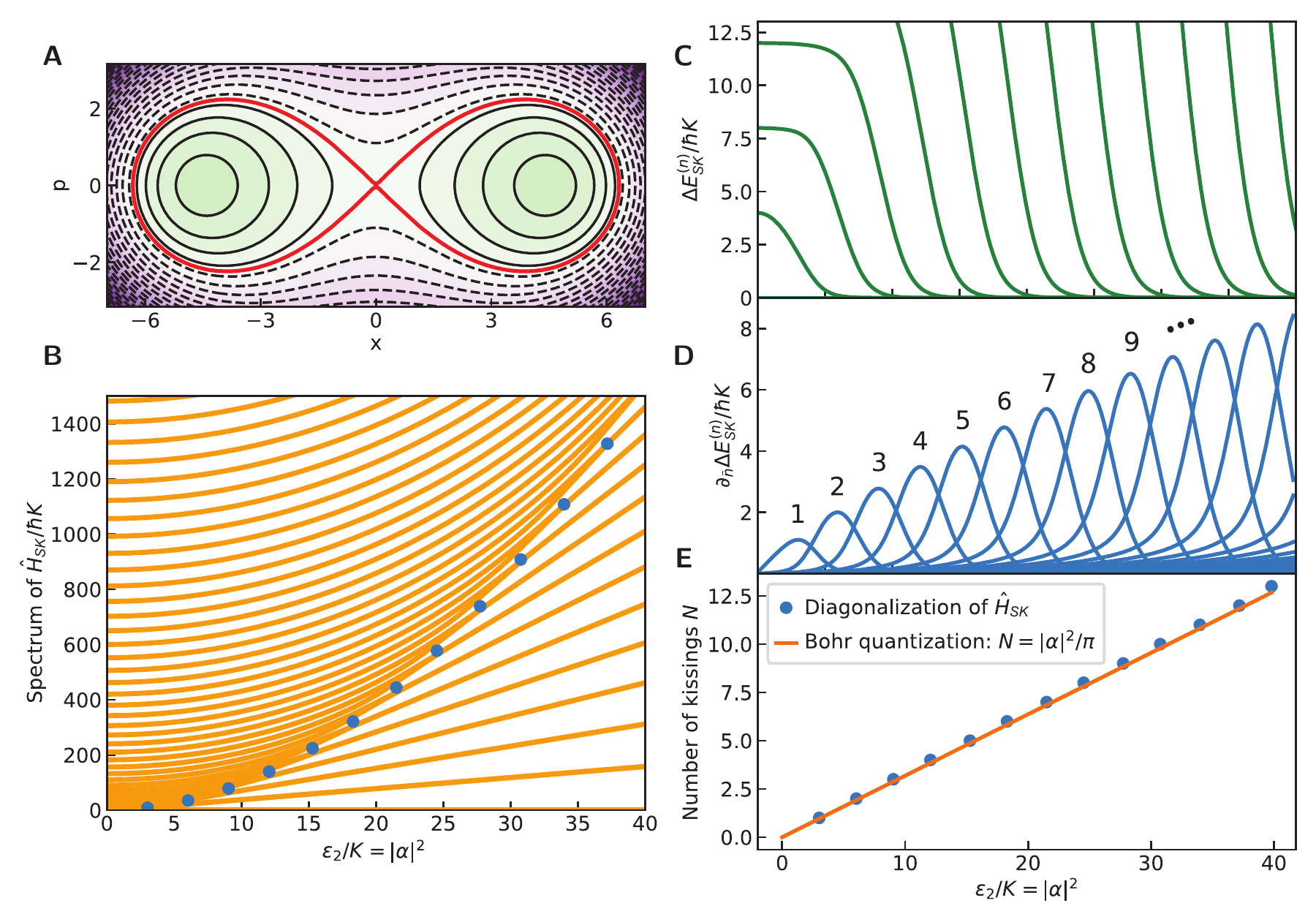}
    \caption{ \textbf{Bohr quantization of the squeezed Kerr oscillator}. \textbf{A}, Equienergy contours of $H^{\textrm{cl}}_{\textrm{SK}}$ describing Cassinian oval orbits, including a lemniscate of Bernoulli (red) taken for $\Delta =0$. \textbf{B}, Quantum spectrum of the squeezed Kerr oscillator and \textbf{C}, the energy gap in between kissing eigenenergies as a function of the squeezing drive amplitude. \textbf{D}, The point of maximal rate of approach is defined as the kissing point and determined by the inflection point in the energy gaps. These are markerd as blue dots in \textbf{B}. \textbf{E}, Comparison in between the quantum Hamiltonian treatment and Bohr's quantization semiclassical argument.}%\textcolor{red}{add EBK}
    \label{fig:supmat1}
    %Figures genereated with '/Users/rodrigo/Dropbox/KerrCatMetapot/SupMat/Python/Energy_cat_spec.py' 
    %and SupMat_Spectroscopy_of_the_kerr_cat_metapot.ipynb
\end{figure}

As one can see in Figure \ref{fig:supmat1} \textbf{A}, the equienergy contours defined by $H_{\textrm{SK}}^{\textrm{cl}}(x,p)=E$ , for $\Delta = 2\epsilon_2$, define trajectories that are Cassinian oval orbits \cite{wielinga1993}. The separatrix dividing the bound states from the out-of-well states is given by $E=0$ and describes a figure ``8'' curve known as Bernoulli's lemniscate. The area enclosed by the tear-drop loop can be analytically computed to be

\begin{equation}
%\frac{1}{2 \pi} \oint p d x=\hbar\frac{\arccos \left(\frac{-1}{\bar{n}}\right)+\bar{n}}{\pi} \sim \hbar\left(\frac{|\alpha|^2}{\pi} +\frac{1}{2}\right),
\frac{1}{2 \pi} \oint\limits_{H_{\textrm{SK}}^{\textrm{cl}}=0} P d X= \hbar \left(\frac{\epsilon_2}{\pi K}+\frac{-\Delta}{8K}\right),
\end{equation}
where $|\alpha|^2=\epsilon_2/K$ and $X$ and $P$ are the dimensionated quadratures. Since Bohr's quantization condition reads $\frac{1}{2 \pi} \oint P d X= \hbar N$, where $N$ is the number of bound states, one finds $N=\left(\frac{\epsilon_2}{\pi K}+\frac{-\Delta}{8K}\right)$. This gives a new rule of thumb to estimate the degree of error protection of the coherent state in the Kerr-cat qubit as a function of its size. 

To verify the semiclassical understanding we compare this prediction with a full quantum treatment that we show in Figure \ref{fig:supmat1} \textbf{B}-\textbf{D}. First we consider the spectrum of $\hat H_{\textrm{SK}}$ (Figure \ref{fig:supmat1} \textbf{B}) and define the energy difference $\Delta E_{\textrm{SK}}^{(n)} = E_{\textrm{SK}}^{(2n+1)}-E_{\textrm{SK}}^{(2n)}$ in between pairs of kissing lines. We see (Figure \ref{fig:supmat1} \textbf{C}) that $\Delta E_{\textrm{SK}}^{(n)} $ describe sigmoidal curves. For relatively small $|\alpha|^2$ the energy gap in between excited states remains approximately constant. At critical $|\alpha|^2$ values an exponential degeneration takes place. The inflection point (Figure \ref{fig:supmat1} \textbf{D}) at which the rate of approach is maximal defines the kissing point after which the marginal splitting left vanishes towards infinity. This defines unambiguously the kissing point. Note, however, that since the approach is so abrupt the argument is largely independent of the exact definition of the kissing point and that any reasonable cutoff imposed over $\Delta E_{\textrm{SK}}^{(n)} $ yields essentially the same result. We compare the fully quantum treatment (for $\Delta =0 $) with the semiclassical argument in Figure \ref{fig:supmat1} \textbf{E} to find remarkable agreement.

% \section{Comment for Doug's consideration}

% \textcolor{red}{Kerr-cat} literature insisted until now in that $N\sim\bar n/4$. The argument consisted in Taylor expanding around the bottom of the well to find a quasi-harmonic oscillator with $\omega  = 4\bar nK$ and using that the energy depth is $E_{\textrm{max}}=\bar n^2 /K$. One finds that the amount of equispaced levels that fit in the wells is $N\sim E_{\textrm{max}}/\omega=\bar n /4.$ 

% \section{Other sections...}
%  \begin{align}
%  \label{eq:h_wig}
% H(x,p) = \hbar^2 K(x^2+p^2)/2-\hbar K(x^2+p^2)^2/4+\hbar \epsilon_2(x^2-p^2).
%  \end{align}

% 

 \section{Decoherence in the squeezed Kerr oscillator}\label{sec:SupMatDecoherence}
 \subsection{The phase space formulation of quantum mechanics}

The phase space Hamiltonian $H_{\textrm{SK}}=\mathfrak{W}(\hat H_{\textrm{SK}})$ in Eq.(\ref{eq:h_wig}) generates evolution of the Wigner function in phase space. The calculation of the expression for the equation of motion is elementary if one uses that the \textit{Hilbert space} product of operators transforms into the \textit{phase space} Groenwold's star-product $\mathfrak{W}(\hat F\hat G) = \mathfrak{W}(\hat F)\star \mathfrak{W}(\hat G)$ \cite{curtright2013}. It is defined to be the exponential of the Poisson bracket over the phase space-functions $f(X,P) =\mathfrak{W}(\hat F) $ and $g(X,P)=\mathfrak{W}(\hat G)$ as

 \begin{align}
     f \star g\;\equiv\; f e^{\frac{i\hbar}{2}\left(\overleftarrow{\partial}_{X} \overrightarrow{\partial}_{P}-\overleftarrow{\partial}_{P} \overrightarrow{\partial}_{X}\right)} g 
     \;=\; fg+\frac{i\hbar}{2}\{f,g\}+\mathcal{O}(\hbar^2).
\label{eq:groenwold}
 \end{align}
 
%  \clearpage

The Schr\"odinger equation transforms as %by transforming via $\mathfrak W$ the equation of motion for the density operator as 

$$\partial_t \hat \rho = \frac{1}{i\hbar}[\hat H_{\textrm{SK}},\hat \rho]\;\;\xrightarrow[]{\mathfrak W} \;\; \partial_t W = \{\!\!\{H_{\textrm{SK}},W\}\!\!\}. $$ 

The double brackets (denoting the Wigner transform of the quantum commutator) are known as the Moyal brackets and shows up naturally as a deformation of the Poisson bracket $\{\!\!\{H_{\textrm{SK}},W\}\!\!\}= \{H_{\textrm{SK}},W\}+\mathcal{O}(\hbar)$. Note that the corrections to the Poisson bracket are directly proportional to the nonlinearities of the Hamiltonian making them essential to exhibit quantum behaviour.

 It is sometimes convenient to express the relationship in terms of $a$ $(=\mathfrak W(\hat a))$ and $a^*$ $(=\mathfrak W(\hat a^{\dagger}))$. Then, the rewriting of the star-product as $\star = \exp\left\{-\frac{1}{2}\left(\overleftarrow{\partial}_{a^*} \overrightarrow{\partial}_{a}-\overleftarrow{\partial}_{a} \overrightarrow{\partial}_{a^*}\right)\right\}$ is in order.
 
% One finds that the \textcolor{red}{Kerr-cat} phase space Hamiltonian $\mathfrak W (\hat H_{\textrm{SK}})$ reads

%  \begin{align}
%  \label{eq:h_wig}
% H_{KC}(x,p) = \hbar^2 K(x^2+p^2)/2-\hbar K(x^2+p^2)^2/4+\hbar \epsilon_2(x^2-p^2),
%  \end{align}
  
% and it generates the evolution of the Wigner function $W(x,p)=\mathfrak W(\hat \rho)$. One can see this by taking the Wigner transform of 

%  We adopt, for notation convenience,  here on-wards the quantum optics convention that $[\hat x, \hat p]=i$, so that $[\hat a, \hat a^{\dagger}]=1$ while keeping $\hat a=(\hat x+i\hat p)/\sqrt{2}$.Groenwold's product transforms as $ \star = \exp\left\{\frac{i}{2}\left(\overleftarrow{\partial}_{x} \overrightarrow{\partial}_{p}-\overleftarrow{\partial}_{p} \overrightarrow{\partial}_{x}\right)\right\}  = \exp\left\{-\frac{1}{2}\left(\overleftarrow{\partial}_{a^*} \overrightarrow{\partial}_{a}-\overleftarrow{\partial}_{a} \overrightarrow{\partial}_{a^*}\right)\right\}$.
  \subsection{Extension to open quantum systems}
 
The phase space formulation of quantum mechanics provides a convenient language for analysing decoherence of Schr\"odinger cats and other bosonic qubits.  We exploit the star-product to transform the Lindblad equation\footnote{We note that phase space Lindbladian can be computed by expanding the star-product directly or by absorbing it as Bopp's shifts \cite{curtright2013}.} for ordinary dissipation (single photon loss) $    \partial_t \hat \rho = \frac{1}{i\hbar}[\hat H,\hat \rho]+\kappa_1\left(\hat a \hat \rho\hat a^{\dagger}-\frac{1}{2}(\hat a^{\dagger}\hat a \hat \rho+\hat \rho \hat a^{ \dagger}\hat a)\right)$ into

\begin{equation}
\label{eq:wigner_evolution}
    \partial_t W = \{\!\!\{H,W\}\!\!\}+\frac{\kappa_1}{2}\left(\partial_xx+\partial_pp+\frac{1}{2}\left(\partial^2_x+\partial^2_p\right)\right) W,
\end{equation}
which we express in dimensionless coordinates for convenience. Note that the dissipation term is identical to the one expected classically \cite{zurek1998,zurek1999}. It is instructive to write the condition to observe the quantum effect of an $n$th order nonlinearity as

\begin{equation}
\label{eq:qcond}
 \hbar^{2n}g_{2 n+1}\left|\partial^{2 n+1} W\right| \gg \kappa_1\left|\partial^{2} W\right|.
\end{equation}
where the vertical bars denote the typical magnitude of the Wigner derivative, $g_m$ is $m$th order nonlinearity of the Hamiltonian, and since the variables $x$ and $p$ are taken to be dimensionless $\hbar$ is understood as the dimensionless quantum of phase space area (sometimes referred to as zero-point spread). Higher order nonlinear effects are then increasingly fragile against dissipation and a strong function of the Wigner function's rugosity.

 In the presence of thermal photons, Eq.(\ref{eq:wigner_evolution}) acquires an additional term that reads
\begin{equation}
\label{eq:d_nth}
\kappa_1 n_{\mathrm{th}}\mathfrak W(\mathcal D[\hat a^{\dagger}]\hat \rho ) = \frac{\kappa_1 n_{\mathrm{th}}}{2}\left(-\partial_xx-\partial_pp+\frac{1}{2}\left(\partial^2_x+\partial^2_p\right)\right) W.
\end{equation}
 where $\mathcal D[\hat{\mathcal{O}}]\hat \rho = \hat{\mathcal{O}} \hat \rho\hat{\mathcal{O}}^{\dagger}-\frac{1}{2}(\hat{\mathcal{O}}^{\dagger}\hat{\mathcal{O}} \hat \rho+\hat \rho \hat{\mathcal{O}}^{ \dagger}\hat{\mathcal{O}})$ is the dissipator associated with jump operator $\hat{\mathcal{O}}$, and $n_{\mathrm{th}}$ the mean number of thermal photons.
 These expressions convey valuable intuition.
 Note, for example, that the diffusion terms in Eq.(\ref{eq:wigner_evolution}) and Eq.(\ref{eq:d_nth}) have the same sign while the drag terms are of opposite signs. These terms appear then with prefactor of $\kappa_1(1\pm n_{\mathrm{th}})$ in the equation of motion. Naturally, in presence of a thermal field the drag towards vacuum is reduced (for $n_{\mathrm{th}}<1$) while the diffusion is always increased. Note that for $n_{\mathrm{th}}>1$ the ``drag term'' now pushes the state away from vacuum.

%To describe the presence of \textit{white} detuning noise and write the contribution $\kappa_{\phi}\mathfrak W(\mathcal D[\hat a^{\dagger}\hat a]\hat \rho )$ one first notes that $\mathfrak W(\hat a^{\dagger}\hat a\rho\hat a^{\dagger}\hat a) = (a^*a-1/2)\star W \star(a^*a-1/2) $ and that $\square \star (a^*a-1/2) = \square (a^*a-1/2) - \frac{1}{2}(\partial_{a*}\square a^* -\partial_a\square a ) -\frac{1}{2}\partial_{a^*}\partial_a \square. $ the master equation further acquires an additional term that reads
% For completeness we give also the expression required in presence of \textit{white} detuning type noise which reads

% \begin{equation} 
% \kappa_{\phi}\mathfrak W(\mathcal D[\hat a^{\dagger}\hat a]\hat \rho ) = \frac{\kappa_{\phi}}{2}\left(\partial_x^2p^2 + \partial_p^2 x^2 -2\partial^2_{xp} xp+\partial_x x+\partial_p p \right) W,
% \label{eq:d_phi}
% %calculation done with SAGE in a Jupiter file called Jaya_commutators
% \end{equation}
% where $\kappa_{\phi}$ is the dephasing rate.

%This expression exposes higher order derivatives responsible for ``non-local'' errors in phase space.
 
% (SAY IT RULES THE TIME AVERAGED QUANTITIES SOMEWHERE IN THE MAIN TEXT?)
 
% (SAY THE \textcolor{red}{Kerr-cat} HAMILTONIAN IS BEYOND RWA SOMEWHERE IN THE MAIN TEXT?)

\subsection{Decoherence for our stabilized Schr\"odinger cat states} \label{sec:catlife}
The evolution driven by mundane dissipation causes the Hamiltonian stabilized Kerr-cat states to decohere at the same rate as for any other cat state \cite{brune1996,deleglise2008,haroche2006, leghtas2015}.
It is easy to see from Eq.(\ref{eq:wigner_evolution}) that the region containing the negative fringes of the cat's Wigner function (producing the highest second order derivatives) is quickly erased by the diffusion terms ($\partial^2_i$).
Using that the inter-fringe distance (their period) is $\pi/2\sqrt{\bar{n}}$ \cite{haroche2006} and that at $t=0$ the fringes saturate the bounds for the Wigner function ($|W|\leq2/\pi$) a back of the envelope calculation to estimate their concavity yields the diffusion rate  $\frac{\kappa_1}{4}|\partial_p^2 W|\sim\frac{64}{\pi^3}\kappa_1\bar{n}\sim2.06\kappa_1\bar{n}$.
A more refined treatment for non-stabilized cats \cite{haroche2006} reveals the exact result for the initial diffusion rate ($t\gtrsim0$) is $2\kappa_1\bar{n}$.
 For large cat states the effect of temperature is only to accelerate this rate as $\kappa_1\rightarrow \kappa_{\textrm{eff}}=\kappa_1(1+2n_{\mathrm{th}})$.
 Equivalently, one can then write the typical coherence time for the cat as $ T_1/2\bar n$.
These results are valid in the case of the Kerr-cat for all times $t>0$ and not only at short timescales since the amplitude of the population of the coherent states does not damp towards the origin of phase space.
This guides the Kerr-cat's coherence to follow an exponential damping while non-stabilized cats follow a more complicated law even in absence of thermal photons ($\sim \mathrm{exp}\{2\bar n (1-e^{-\kappa_1 t})\}$ \cite{haroche2006}).
%Again, a modest amount of thermal photons only gives a rather trivial thermal character to the steady state and does not modify the argument qualitatively. 

\begin{figure}[t]
    \centering
    \includegraphics{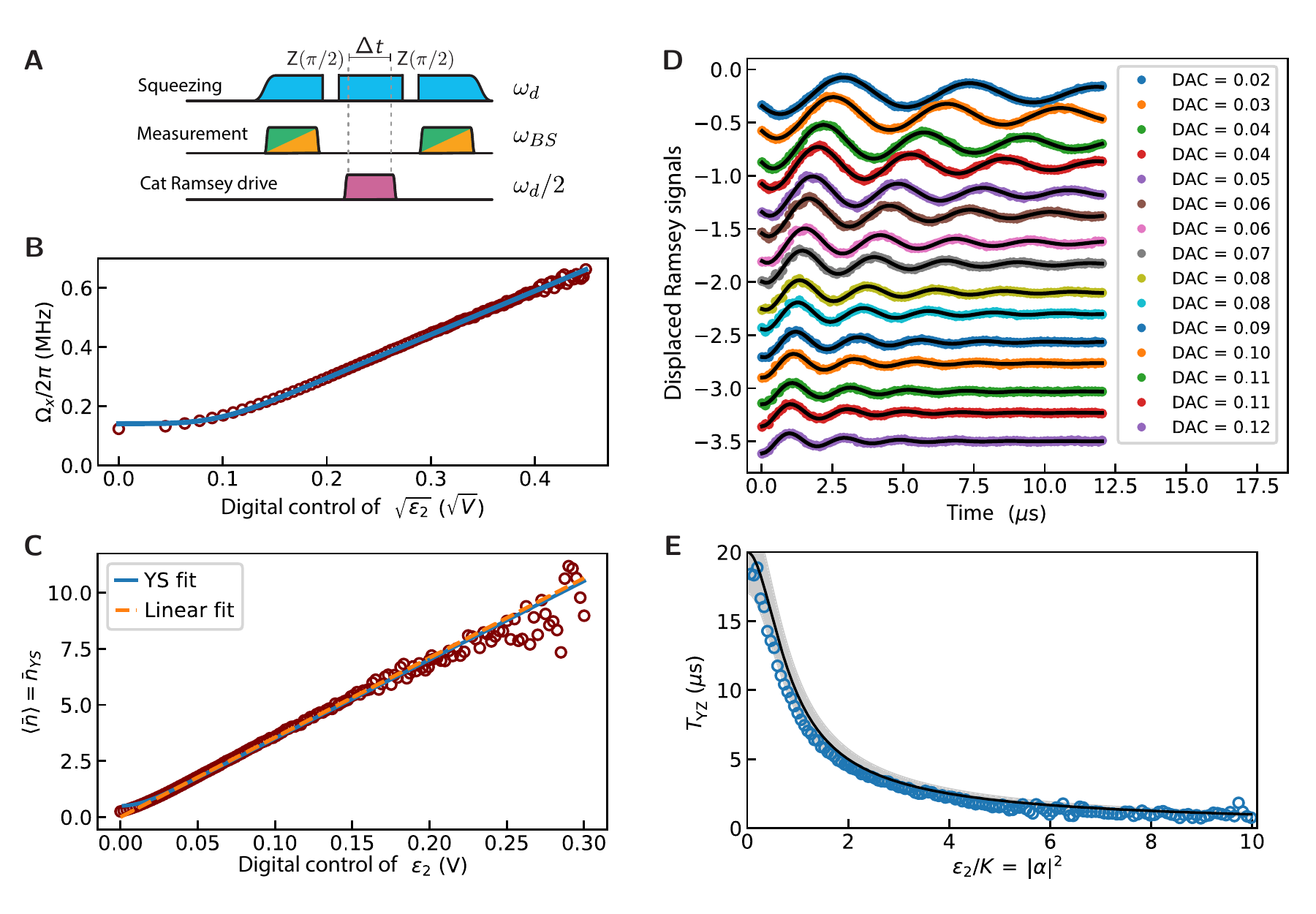}
    \caption{\textbf{cat Rabi as a calibration tool}. \textbf{A},  The experimental sequence used to perform Rabi oscillations in between oscillator states. \textbf{B},  We show 15 of the 101 cat Rabi oscillations, each for a different squeezing amplitude. The frequency of the oscillation is directly proportional to the square root of the cat size and the decoherence constant is inversely proportional to it. \textbf{C},  Fit of the cat Rabi frequency as a function of the voltage control parameter. \textbf{D},  Inverting the formula for the Rabi frequency we access directly by measurement the cat size of the ground state in the metapotential. \textbf{E},  Measurement of the cat-state lifetime. The solid line is a parameter-free theory prediction with shaded errorbars correspondng to the experimental uncertainty on the independently measured single photon lifetime.}
    \label{fig:SupMatTcat}
    %Figure generated with /Volumes/qulab-CC1558-FASAPP/shared/Papers/KerrCat_2021/KerrCat_2022/KerrCatMetapot/SupMat/Python%/n_bar_cal_9.py')
\end{figure}

In Figure \ref{fig:SupMatTcat} we show the coherence damping of Rabi-like fringes while the Kerr-cat qubit oscillates in between the different cat parities as $\mathcal N(t)\left(|\alpha\rangle+e^{i\Omega_xt}|-\alpha\rangle\right)$.
This is achieved by applying a resonant drive on the system while the squeezing drive is maintained on as depicted in Figure \ref{fig:SupMatTcat} \textbf{A}: the stabilization drive is on during a first QND measurement that is used to prepare a coherent state in the metapotential.
The squeezing drive is then turned-off abruptly to perform a Kerr-gate \cite{grimm2020} and prepare a Yurke-Stoler parity-less cat.
After free Kerr evolution time of $\pi/2K$ the stabilization drive is turned on again to ``catch'' the cat state.
Then a resonant Rabi drive is applied on the qubit. The stabilization drive frustrates the displacement of the coherent components but causes the interference fringes of the Kerr-cat qubit to roll changing its parity as it exchanges photons, coherently and one-by-one, with the resonant drive.
This is the mechanism used to perform the continuous $X$ gate in the Kerr-cat qubit \cite{grimm2020}.
This continuous gate in between cat states can be understood as a quantum Zeno evolution \cite{touzard2018} in the cat manifold here enforced not by measurement by the gap in between the ground states and the first excited states \cite{Signoles2014}, not different to the isolation of qubit space in a generic nonlinear oscillator.
The oscillation around the cat meridian of the Kerr-cat qubit Bloch-sphere for different values of the squeezing drive amplitude are show in Figure \ref{fig:SupMatTcat} \textbf{B}.
Since the Rabi dynamics will perform a $\pi$-pulse every time a ``red'' cat-fringe ($W>0$, see Figure 1 in the main text) is displaced to the position previously occupied by a ``blue'' fringe  ($W<0$), one expects the Rabi frequency to be a strong function of $|\alpha|^2$: as the cat states become larger, their fringes become narrower and a smaller displacement is required to enact the coherent parity-flip.\footnote{Note that the cat coherence times drops as $T_{Z}\sim1/\bar n$, the inter-fringe distance decreases as $\sim1/\sqrt{\bar n}$ and the maximal Rabi rate as determined by the gap  scales as $\Delta_{\textrm{gap}}\sim\bar n$. Thus, the amount of coherent operation over the Schr\"odinger cat superposition during a the state's lifetime grows as $\sqrt{\bar n}$.} By fitting the Rabi frequency $\Omega_x(\alpha)$ we can calibrate the size of the cat as a function of digital control amplitude (DAC) controlling the stabilization drive $\epsilon_2$. 

In the large cat limit, a drive term $\hat V_x = \epsilon_x (\hat a ^{\dagger} + \hat a)$ breaks the degeneracy in between the even and odd cat states by \cite{touzard2018, grimm2020}
\begin{equation}
    \Omega_x = 2\langle \alpha|\hat V_x |\alpha\rangle = \mathfrak{Re}(4 \epsilon_x\alpha^* )
\end{equation}
to first order in perturbation theory.

Extending this simple treatment to be valid for smaller values of $|\alpha|^2$ one gets $\Omega_x \approx \mathfrak{Re}(4 \epsilon_x\sqrt{\bar n_{YS}})$, where we have introduced a notation for the mean number of photons in the parity-less cats as $\bar n_{YS} = \frac{1}{2}|\alpha|^2 (r^2+r^{-2})$ with $r$ being the ratio of the normalization constants for the even and odd cats as defined in Eq.~\ref{eq:ratioNorm}.
The extension of this formula relies in the fact that the parity-less cats map naturally to the equatorial superposition of the Fock-state Bloch-sphere, while at large values of $\alpha$ the difference in photon number for different cats is vanishingly small.
The maximal photon-number difference is in between the even and odd cat states and reads $(\bar{n}_{+}-\bar{n}_{-}) \rightarrow-4|\alpha|^{2} e^{-2|\alpha|^{2}}$ in the limit of large $|\alpha|^2$.
In  Figure \ref{fig:SupMatTcat} \textbf{C} we show the fit of the data with the generalized formula. The inverse of the same formula is use to calibrate directly the cat size of the Hamiltonian ground-state as a function of the digital control amplitude for the squeezing drive as shown in Figure \ref{fig:SupMatTcat} \textbf{D}.
As expected the dependency becomes linear already at fairly low values of $|\alpha|^2$.

It is clear that the oscillations shown in Figure \ref{fig:SupMatTcat} \textbf{B} are not fully coherent.
Indeed, their increasing decay rate with $|\alpha|^2$ is a characteristic of Schr\"odinger cat states and a signature of the quantum to classical transition \cite{haroche2006}.
In Figure \ref{fig:SupMatTcat} \textbf{E} we show the measured coherence time of the cat states for different values of $|\alpha|^2 = \epsilon_2/K$.
Since in a Kerr-cat Bloch-sphere with a given $\alpha$ the mean photon number of the cat states is only the same if $|\alpha|^2\gtrsim1$,  measuring the Rabi-like damping of the Kerr-cat Rabi oscillations (instead of monitoring the exponential relaxation of a cat with a given parity) provides us with an effective damping rate representative of the encoding, which we refer to as $T_{YZ}$.
The dots are experimental data and the lines are independently calibrated theory plots.
The solid line represents the parameter-free curve $T_{YZ} = T_1/2\langle \bar n\rangle$, where $\langle \bar n\rangle$ is the Bloch-sphere-average photon-number around the cat-meridian and reads $\langle \bar n\rangle = \bar n _{YS}$.
The dashed lines represent the uncertainty in the independently measured transmon lifetime $T_1 = (20\pm3)~$\textmu s.
%For values $|\alpha|^2>7$ in Figure \ref{fig:SupMatTcat} e) and elsewhere, the calibration in Figure \ref{fig:SupMatTcat} d) is linearly extrapolated

% Note that the average photo-number of cats with a given $\alpha$ and different relative phase (parities) converges exponentially to $|\alpha|^2$ as $\langle \bar n\rangle\rightarrow |\alpha|^2 (1\textcolor{red}{+}e^{-4|\alpha|^2})$.

%the computed fidelity $\frac{2\pi}{\hbar}\int  W(t=0)W(t)\;dxdp$ against time for a free-running cat and stabilized \textcolor{red}{Kerr-cat} both initially with $\bar n =3$. We take for this calculation $\kappa_{\textrm{eff}}=1/T_1$. The single photon lifetime $T_1=20??????$~\textmu s is an independently calibrated experimental parameter of our setup. The points are experimental data. In Figure \ref{fig:app1} b) we show the experimental data (see main text) measuring cat lifetime for stabilized \textcolor{red}{Kerr-cat}s as a function of their mean photon number $\bar n$. The solid line is a parameter-free prediction corresponding to $T_{\textrm{cat}} = \frac{T_1}{2\bar n}$ \cite{haroche2006}.

%The reason becomes clear when one observes that while free cats damp towards vacuum under single photon dissipation, stabilized cats tend to the classical mixture of the undamped coherent state components.$

\subsection{Decoherence of $q$-legged cats during free Kerr evolution (the Kerr-gate)}
During the Kerr-gate \cite{grimm2020} the state evolves under the native (undriven) Kerr Hamiltonian of the system.
Turning off abruptly the squeezing drive, which stabilized the cat manifold, enacts a Yurke-Stoler evolution \cite{Yurke1986} transforming the state from a coherent state into a two-legged parity-less cat state after a time $t=\pi/2K$.
It is interesting to notice that the system visits a series of intermediate $q$-legged cat states every $t=\pi/qK$ \cite{haroche2006}.
These multi-legged cats exhibit a complex pattern of interference fringes in phase space and their sharp phase space features exposes them to accentuated diffusion under dissipation (see Eq.(\ref{eq:wigner_evolution})).
This makes the fidelity of the Kerr-gate a strong function of the mean photon number, and thus a sensitive tool to study the decoherence in our system.
It is important to notice that the high-frequency variation in the Wigner distribution of $q$-legged cat sates is set by the square of their phase space diameter ($D= 2\sqrt{\bar n}$).
Since the Kerr evolution preserves photon number, it follows from Eq.(\ref{eq:wigner_evolution}) that they all decohere at the same typical rate, regardless of how complex the phase space interference patters may be at each time instant.

\begin{figure}[t]
    \centering
    \includegraphics{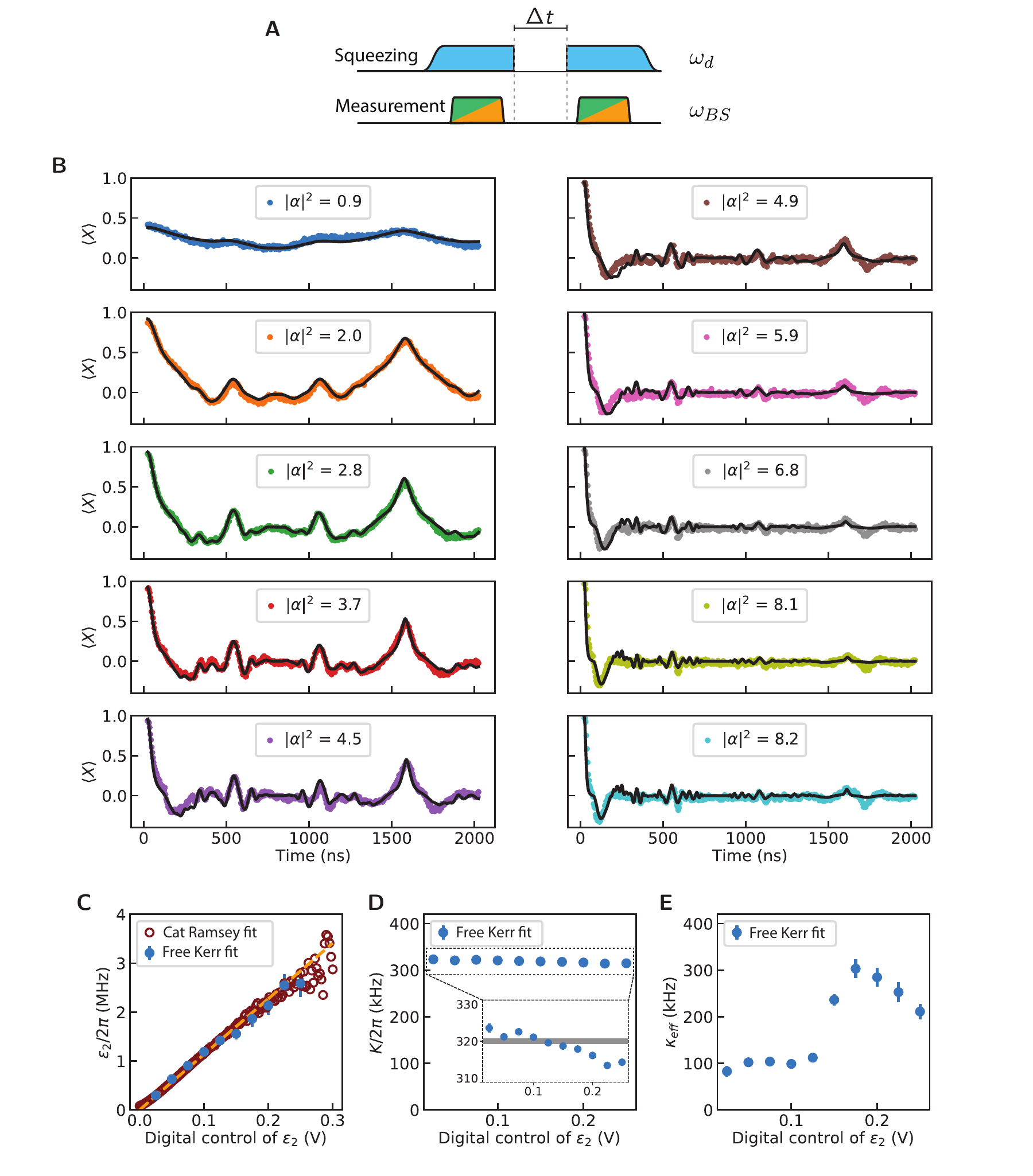}
    \caption{  \textbf{The Yurke and Stoler experiment as a calibration tool}. \textbf{A}, The experimental sequence of pulses allowing for the observation of the phase collapse and the phase revival in free Kerr oscillator. At times equal to $\pi/qK$ the system is in a $q$-legged Schr\"odinger cat state. \textbf{B}, The signal corresponds to the mean vale of the operator $\op{X} \approx |+\alpha\rangle\langle+\alpha| - |-\alpha\rangle\langle-\alpha|$ (c.~f.~Eq.~\ref{eq:catPauliOps}). For large cats the periodicity of the signal is lost due to decoherence. \textbf{C}, Fitting the cat size over the free Kerr evolution signal provides an efficient calibration of the drive parameters which is consistent with that obtained from the cat Rabi experiments discussed in Figure \ref{fig:SupMatTcat}. \textbf{D}, The revival is expected at $\pi/K$ independently of the cat size and thus fitting the signals in \textbf{B,} provides a good calibration for the Kerr parameter $K$. It is found to be $K \sim320~$ kHz in agreement with independently performed spectroscopic experiments (see main text). \textbf{E}, The fitted effective dissipation rate as a function of the cat size. We find a non-trivial dependence on the dissipation rate for larger cats. This suggest an effctive heating of the set-up for the larger squeezing amplitudes suggesting.}
    \label{fig:supmat_FreeKerr}
    %Figure generated with: /Users/rodrigo/Dropbox/KerrCatMetapot/SupMat/Python/n_bar_cal_8.py
    %
\end{figure}

We here study experimentally the decoherence during the free Kerr evolution as a function of the cat size $|\alpha|^2$.
Our measurement configuration allows us to easily access the mean value of the Kerr-cat qubit Pauli operator $ \op{X} \approx |+\alpha\rangle\langle+\alpha| - |-\alpha\rangle\langle-\alpha|$ (see Eq.~\ref{eq:catPauliOps}).
The experimental measurement sequence is shown in Figure \ref{fig:supmat_FreeKerr} \textbf{A}.
A preparation QND measurement performed in presence of the stabilization drive is followed by a period of free Kerr evolution before the final measurement is performed.
The experimental results are show in Figure \ref{fig:supmat_FreeKerr} \textbf{B}.
We see that at the time of the phase refocusing $\pi/K$ the degree of coherence decreases rapidly with the photon number up to the point that the state is largely dephased.
By that time, the phase space distribution is expected to be essentially classical.
The solid black line is a fit via simulation of a master equation including only single photons loss (setting $\kappa_1\rightarrow\kappa_{\textrm{eff}}$ in Eq.(\ref{eq:wigner_evolution})).
Including photon excitation in the simulation with a free thermal photon number as an additional free parameter achieves the same goodness of fit at the cost of a high uncertainty in the fit parameters.
We thus exclude it from the analysis here.
We find that a correct interpretation of the data is achieved by understanding $\kappa_{\textrm{eff}} \sim \kappa_1(1+2n_\mathrm{th})$ as discussed above and verified numerically.
The fitted parameters are shown in Figure \ref{fig:supmat_FreeKerr} \textbf{C}, \textbf{D}, and \textbf{E} . 

By giving the fit function complete freedom over the Hamiltonian parameters $K$ and $\epsilon_2$, we find that only a modulation of $\sim3\%$ is required to reproduce the complex free Kerr evolution for all values in a large range of $|\alpha|^2$.
Moreover, we see from Figure \ref{fig:supmat_FreeKerr} \textbf{C}  that fitted values are consistent with independently performed calibrations discussed in the previous section and in the main text.
We notice that in the free Kerr experiment the Kerr coefficient has a slight decreasing trend as the size of the cat states grows.
As expected for larger values of $|\alpha|^2$, the refocusing signal becomes weaker.
We see nonetheless a significant increase in the effective decoherence rate of $\sim 200\%$ making $2\bar n T_{Y}<T_1 = \kappa_{\textrm{eff}}^{-1}$. The fitted value for $\kappa_{\textrm{eff}}$ is compatible with $\kappa_1=1/20$~\textmu s and $n_{\mathrm{th}}\sim 0.3$ for $|\alpha|^2<4.9.$
%As anticipated, due to the symmetric roles of the dissipators $\mathcal D[\hat a]$ and $\mathcal D[\hat a^{\dagger}]$ over the coherence of the multi-legged cats, a variety of combination in between $n_{\mathrm{th}}$ and $\kappa_1$ fit the data.

 %an effective parametric renormalization due to beyond RWA corrections. We have computed the analytical expression for the \textcolor{red}{Kerr-cat} metapotentialto order five beyond RWA and it can be found in \cite{Xiao2020} but we here treat it as a phenomenological correction. 

%The fit free Kerr evolution provides also an independent verification of squeezing amplitude calibration. It is found to be in good agreement with the primary calibration made by studying the Rabi frequency of the continuous quantum-Zeno evolution in the cat manifold as explained in \cite{zenocat, Grimm2020}.

\begin{figure}[t]
    \centering
    \includegraphics[width=0.8\textwidth]{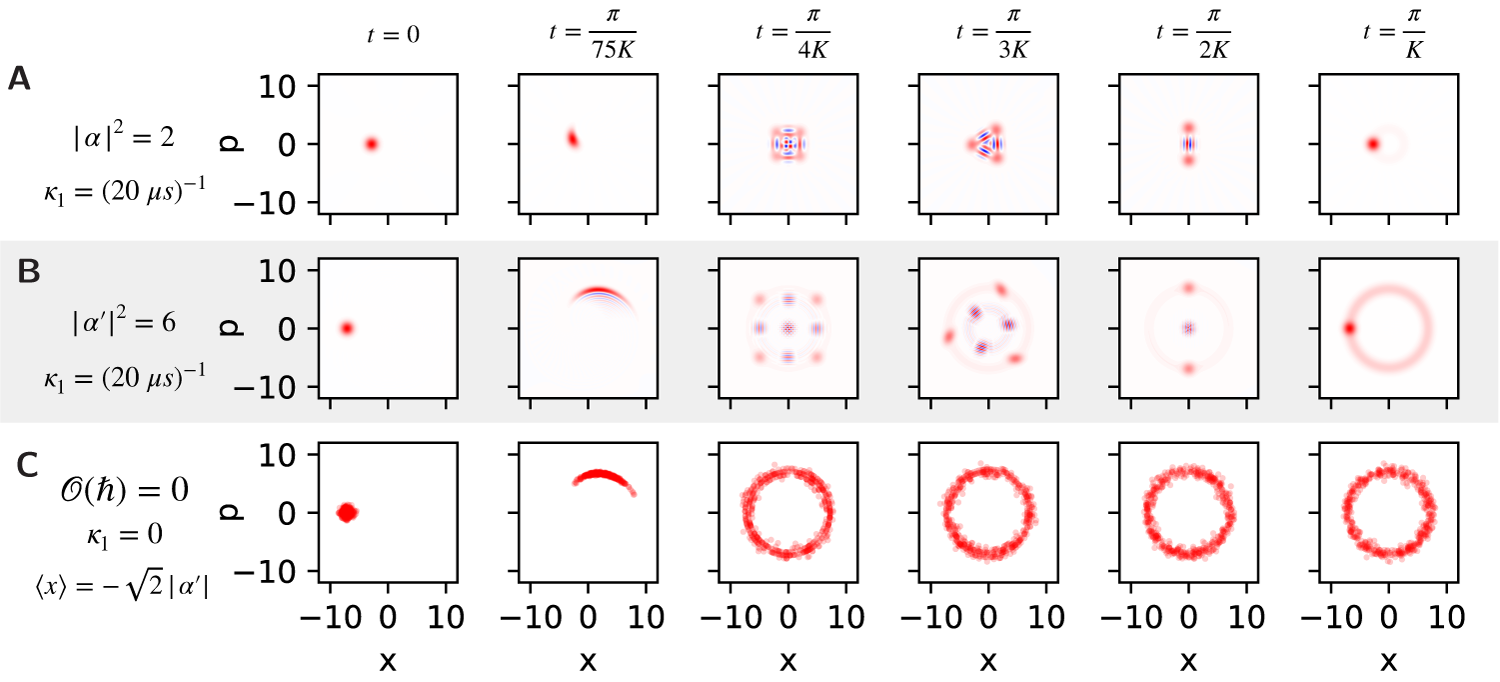}
    \caption{\textbf{The Kerr oscillator in the quantum, quantum dissipative, and classical regimes}. \textbf{A}, Evolution for a Kerr oscillator initialized in a coherent state under mundane dissipation (single photon-loss). The dissipation rate used for the numerical simulation was set to $\kappa_1$=(20~\textmu s)$^{-1}$ and the mean photon number in the initial state was set to $|\alpha|^2= 2$. The formation of $q$-legged cats is an outstanding prediction of quantum theory \cite{yurke1988,haroche2006,kirchmair2013}. \textbf{B}, The same system initialized now with a mean photon number of $|\alpha'|^2=  6$. We see that the purity receives a harder hit from dissipation, which reveals in the enhancement of the background doughnut-shaped halo. The phase of the state vanishes as the distribution occupies the full $360^{\circ}$ around the origine of coordinates. \textbf{C}, The purely reversible ($\kappa_1 =0$) classical evolution generated by the Hamilton's equation via the Poisson bracket. This can be seen as the quantum to classical limit taking $\mathcal{O}(\hbar)\rightarrow 0$ in Moyal's equation of motion. The system is initialized in a Gaussian probability distribution, here represented by a sampling of Dirac deltas around $(x,p)=(-\sqrt{2}|\alpha'|,0)$. The Liouville evolution deforms the state into a ``banana'', making the ``external'' particles go at a higher angular velocity in phase space. After an initial spiral motion, the distribution losses its structure and becomes a circularly-symmetric doughnut. The motion of a continuous classical distribution is non-periodic due to the continuous spectrum of frequencies in its Fourier decomposition. }
    \label{fig:quantumtoclassical}
    %Figure generated with: 
\end{figure}

To clarify the quantum-to-classical transition of our qubit during the gate we show in Figure \ref{fig:quantumtoclassical} \textbf{A}  and \textbf{B}  simulations of the Kerr evolution initialized in coherent states with $\alpha =-\sqrt{2}$, and $\alpha' =-\sqrt{6}$.
For this illustration we set the simulation parameters to $\kappa_1 =1/20$~\textmu s and $K/2\pi = 320$~kHz.
We see that for the smaller photon number, the phase-refocusing is comparatively efficient \cite{kirchmair2013}, while for the larger cat the enhanced diffusion degrades the contrast of the phase space interference.
The destructive interference expected everywhere but at the initial phase is inefficient and a background halo appears. This doughnut-shaped distribution corresponds to that of the classical Kerr oscillator.
In Figure \ref{fig:quantumtoclassical} \textbf{C} we show also the fully classical evolution ($\{\!\!\{H_{\textrm{SK}},W\}\!\!\}\rightarrow \{H_{\textrm{SK}},W\}$) for a Gaussian distribution of Dirac deltas centered at $\langle x \rangle = \alpha'\sqrt{2}$ and $\langle p\rangle =0$.
We additionally set the system to undergo a pure Liouvillian evolution by setting the dissipation $\kappa_{1}$ to zero in Eq.(\ref{eq:wigner_evolution}).
It should be clear that either by, i) waiting for longer times, ii) increasing the dissipation \textit{or} simply by iii) increasing the cat size, the open quantum system evolution will yield a fully classical distribution (a doughnut). 

In the purely quantum case, the periodic Kerr revivals are a direct consequence of the granularity of the electromagnetic field and the discretization of energies, much like in the collapse and revival of the Rabi oscillations of a two-level atom in a mesoscopic field \cite{haroche2006}.
The latter is the signature characteristic of the quantization of light in the Jaynes-Cummings model \cite{Assemat2019}.
The main difference in between these two analogous processes is that the revival in the Jaynes-Cummings model is only approximate even in absence of decoherence.
In the case of the free Kerr evolution, a perfect refocusing is expected, which allows the construction of a qubit gate out of this physical process.
In the classical case, the continuous energy distribution lacks such a periodic behaviour altogether for fundamental reasons.
In the intermediate case, the quantum dissipative scenario, the periodic behavior is degraded until it completely vanishes.
This is a direct consequence of the broadening of the spectroscopic lines to the point in which they effectively allow the quantum system to have a continuous distribution of energies.

Note that the damping rate (setting the radius of rotation  at time $t$ and in particular the radius of the doughnut in the last panel of the Figure) is given by $\kappa_1$, that the decoherence rate (yielding the erasure of the fringes and the lost of the phase coherence) is given by $2\kappa_1\bar n$, and that the collapse of the phase (the onset of the fringe development) occurs after a characteristic time $T_{\textrm{coll}}=\pi/2\sqrt{\bar n} K$ \cite{haroche2006}.
This observation together with the comparison of the three scenarios in Figure \ref{fig:quantumtoclassical} suggest that the transition to the classical appears in the limit $\kappa_{1}\rightarrow0$, $\bar n\rightarrow \infty$ while $\kappa_{1}\sqrt{\bar n}\gg K$. In this sense macroscopic states turn into a classical distribution almost immediately, at a time scale when the radius of rotation is unchanged and before the fringes develop.
In this limit a \textit{quantum dissipative} system naturally behaves as a \textit{classical conservative} system without the need to enforce Dirac's classical limit ``$\hbar\rightarrow0$'' \cite{Zurek2003}. 
Conversely, the ideal conditions for the Kerr-cat qubit architecture are obtained deep in the quantum regime where $\kappa_{1}\rightarrow0$, $\bar n\rightarrow \infty$ and $\kappa_{1}\sqrt{\bar n}\ll K$.
This is a particularly enlightening illustration of the fact that storing redundantly quantum information in the large quantum systems required for error-correction demands ``fast'' control in order to outpace the accentuated decoherence. %The timescale of gate operation is set for us here to be $\sim1/K$.

\subsection{Decoherence of stabilized coherent states}
The most striking result regarding the interplay of nonliterary and decoherence in our work is the experimental discovery of a stepped increase in the coherent state lifetime versus the mean photon-number in the logical state.
This showcases the interplay in between nonlinearity and dissipation and the relevance of the excited state structure in our system to the error correcting properties of our qubit implementation.
Every time that the quantity $|\alpha|^2 /\pi $ hits an integer value the Kerr-cat qubit earns one more order of error protection.
This is a direct consequence of the vanishing tunnel splitting as excited states sink into the wells of the metapotential and under the barrier.
The qualitative effect over the decoherence in the system is largely independent of the specific error channels considered: a staircase in the lifetime is expected.
On the other hand, quantitative predictions do depend on the dominating error channel at play. We here discus how some of the main features of decoherence of our coherent states can be explained by a simple model keeping RWA terms in the coupling to the environment which contains single photon loss ($\kappa_1 (1+ n_{\mathrm{th}}) \mathcal D[\hat a]$), thermal excitations ($\kappa_1 n_{\mathrm{th}}\mathcal D[\hat a^{\dagger}]$), and detuning type noise ($\kappa_{\phi} \mathcal D[\hat a^{\dagger}\hat a]$). A non-Markovian contribution motivated by highly correlated magnetic flux noise threading the superconducting SNAIL-loops is also included.

We first provide a numerical model including single-photon loss, thermal excitations, dephasing, and non-Markovian low-frequency detuning noise.
We then provide an analytical model to showcase the importance of the excited level spectrum, which we restrict to single-photon loss and thermal excitations for brevity.

A beyond-RWA Lindbladian model that explains quantitatively the data by deriving from first principle a set of exotic dissipators created by the interplay of nonlinearity and dissipation will be presented elsewhere.

\subsubsection{RWA Lindbladian evolution}
\begin{figure}[t]
    \centering
    \includegraphics{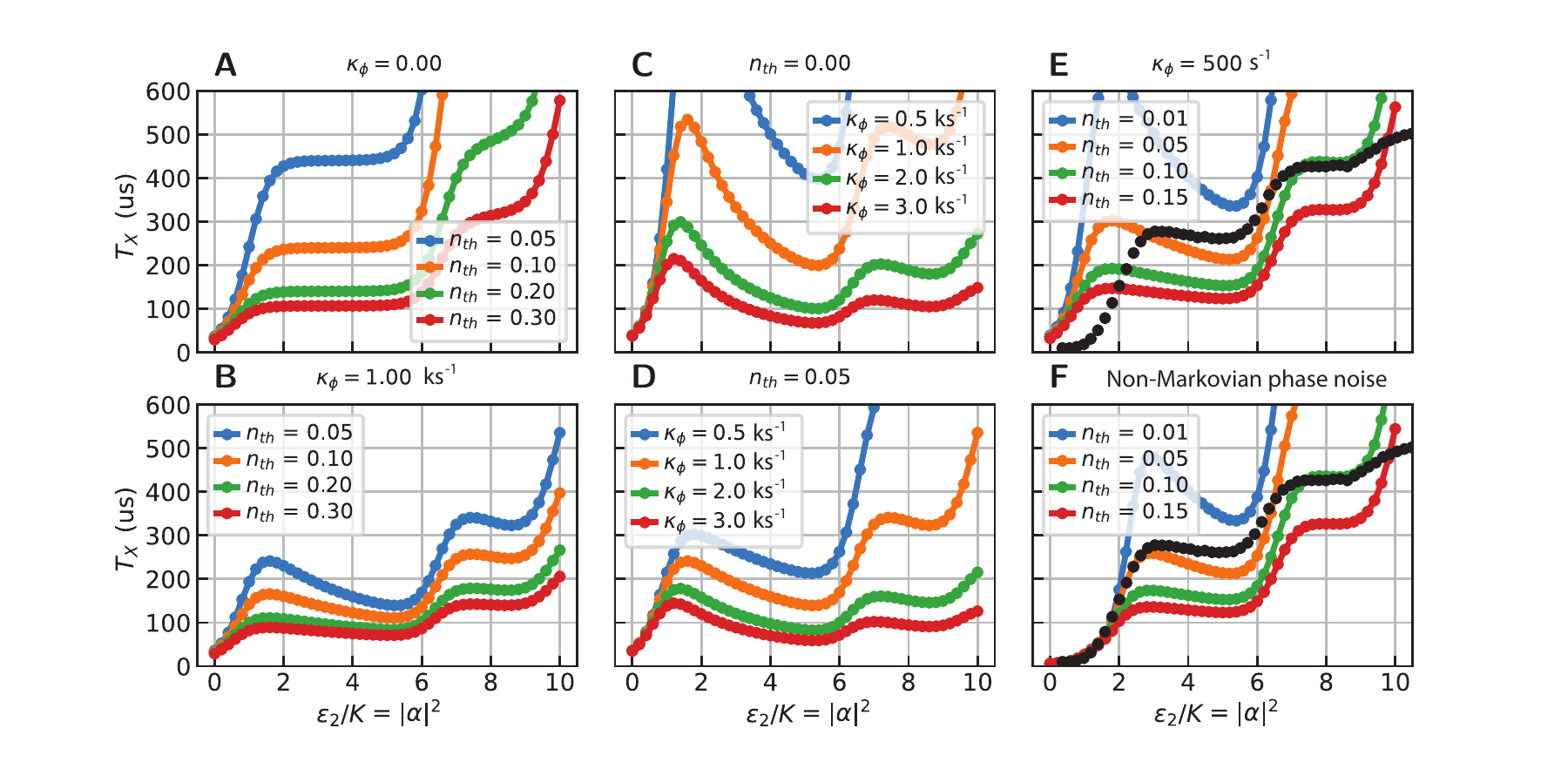}
    \caption{\textbf{A}, Numerical simulations in absence of phase noise ($\kappa_{\phi} = 0$) for a set of different thermal populations. The steps in the coherence state lifetime originated by the discretization of phase space orbits are apparent. \textbf{B}, The lifetime study is performed in presence of white phase-noise for a range of temperatures keeping the coherent state lifetime in the region of interest. \textbf{C} The lifetime study is performed in absence of thermal photons for a set of white phase-noise amplitudes. \textbf{D}, For a given temperature, the effect of different phase-noise amplitude is compared. \textbf{E}, Experimental data compared with a sample of Markovian noise models. \textbf{F}, The low $|\alpha|^2$ behaviour of the data is reproduced by a non-Markovian toy model including low frequency components in the fluctuations of the Hamiltonian term $\Delta \hat a^{\dagger} \hat a$ (see Eq.(1) in the main text).}
    \label{fig:Staircase_Sim}
   % Figure generated with 
%Volumes/qulab-CC1558-FASAPP/shared/Papers/KerrCat_2021/KerrCat_2022/KerrCatMetapot/SupMat/Python/n_bar_cal_9.py'
%/Volumes/qulab-CC1558-FASAPP/shared/Papers/KerrCat_2021/KerrCat_2022/KerrCatMetapot/SupMat/Python/StairCase/FigMaker.py'
\end{figure}

In Figure \ref{fig:Staircase_Sim} \textbf{A} and \textbf{B} we show Markovian master equation simulations considering single-photon loss and a bath at constant temperature for two different magnitudes of \textit{white} dephasing noise ($\kappa_{\phi}$).
Here the colored dots are numerical simulation and the lines are a guide to the eye.
The base temperature of our fridge is close to 30~mK but the radiation environment is believed to be at a higher temperature \cite{grimm2020}.
A thermal population of $n_{\mathrm{th}}\sim0.05$ was used as a lower bound since it corresponds to a black-body temperature at $\sim 6$~GHz of 100~mK, which is pessimistic for superconducting circuits ($n_{\mathrm{th}}\sim0.1$ corresponds to $\sim120$~mK).
In this oversimplified model, the temperature is scanned maintaining the single photon-lifetime fixed at $\sim 20$~\textmu s.
In view of the expected $1/f$ dependence of the flux noise, the computations in absence of dephasing noise is only a heuristic. We represent high-frequency flux noise that causes leakage with an effective \textit{white} noise strength $\kappa_\phi$.

In all our simulations we observe a marked inflection point at $|\alpha|^2\sim 2\pi$ and $|\alpha|^2\sim 3\pi$ as expected form Bohr's phase space quantization (see Section \ref{sec:supMatBohrQuant}).
The first order protection happening at $|\alpha|^2\sim \pi$ presents itself as a subtle inflection point.
This is due to the fact that for small values of  $|\alpha|^2$ ($\lesssim 1$) the limiting factor is not thermal excitation to the first excited state.
Instead, the limiting factor originates from the fact that the Hamiltonian wells are not sufficiently distant in phase space for the states $\ket{\pm X}$ to be well approximated by coherent states; the metapotential wells do not yet contain enough area to host two nonoverlapping states.
As such, single-photon loss and dephasing still dominate the decoherence for $|\alpha|^2 \lesssim \pi$.
Additionally, due to this lack of separation, coherent and incoherent tunneling represented by residual no-jump Hamiltonian terms $-(\Delta + i\kappa_1/2)\adag \ah$ is not yet sufficiently suppressed.
%The limiting factor is an effective incoherent tunneling in between the wells caused by either the effective imaginary detuning term present in the no-jump Hamiltonian describing photon-loss ($i\kappa_1/2\hat a^{\dagger}\hat a$) as well as the effective detuning noise represented here by $\kappa_{\phi}$.
In this regime decoherence is also facilitated by errors in the calibration of the drive frequencies, slow fluctuations of the resonance condition in the setup, and by deterministic AC Stark-shifts over the SNAIL's oscillator frequency due to the stabilization drive. These types of effect for non-zero effective detuning have been studied theoretically in \cite{Marthaler2007} an explained via the WKB approximation as the opening up of phase space tunneling channels in between the wells.
An in-depth experimental study of these physics in our system is left for a future work.
We here restrict to the observation that these tunneling effects are suppressed in the Kerr-cat regime entered as $|\alpha|^2>1$ \cite{puri2017}.

We see that in absence of Markovian dephasing noise ($\kappa_{\phi} = 0$) the plateaus are rather flat, while for $\kappa_{\phi}>0$ they contain a downward slope that we recognize in our data.
This downward trend implies a degradation of the coherence during the intervals where the mean photon number grows but the increase in area of the limiting orbit is not sufficient to host a new excited pair of levels.
A steep increase is then observed when the Bohr's quantization condition is fulfilled.
This behaviour is expected analytically from the photon-number dependence of the dissipators associated with photon-loss (or gain) and the one causing detuning errors.
The imaginary operator contribution of each of the involved dissipators to the no-jump non-Hermitian Hamiltonian is either $\propto |\alpha|^2$ or $\propto |\alpha|^4$ respectively (see Table I).
%Note that, in the interval in between the capture of the second and third excited orbits, the trend in the coherent state lifetime is upwards in absence for $\kappa_{\phi} = 0$.
%This is caused by the remaining exponential kissing of the energy levels.

In Figure \ref{fig:Staircase_Sim} \textbf{C} and \textbf{D}, we plot the behaviour for a few values of $\kappa_{\phi}$ at given sample temperatures.
From Figure \ref{fig:Staircase_Sim} \textbf{C},  we learn that phase-noise by itself cannot explain the qualitative behaviour of our data, even if we allow for a monotonic parameterization of $\kappa_{\phi}$ as a function of $|\alpha|^2$.
This would represent the approximation of in-band white noise at a single relevant frequency.
From Figure \ref{fig:Staircase_Sim} \textbf{D}, we see that a finite temperature limits the coherent state lifetime before the first-order protection peak is achieved, but the value consistent with this saturation fails to control the lifetime at larger values of $|\alpha|^2$.
Larger values of $\kappa_{\phi}$ that control the lifetime at $|\alpha|^2\sim10$ fail instead to reproduce the experimentally observed value of saturation at $|\alpha|^2\sim \pi$.

In Figure \ref{fig:Staircase_Sim} \textbf{E}  and \textbf{F}  we reproduce the experimental data as black dots on top of numerical computations for the expected lifetime under different decoherence models.
In Figure \ref{fig:Staircase_Sim} \textbf{E}, we reproduce a choice of Markovian models including single-photon gain, single-photon loss and a $\kappa_{\phi} = 500$~s$^{-1}$.
We see that, in the center region of the plot, it is possible to assign a temperature to the measurement but the choice becomes inconsistent at either low or high values of $|\alpha|^2$.
At the high end of $|\alpha|^2$, we see what may be interpreted as an increased effective heating as the data traverses different isotherms.
This is consistent with the observations made during in the discussion of the free Kerr evolution and suggest that an effective heating is present at larger cat sizes. %Note that the conditions of the coherent state lifetime measurement are extremely different from those in the free Kerr evolution. The main points of contrast are that i) the coherent state lifetime requires the stabilization drive to remain on during the whole experiment (which is up to two-thousand times longer than in the free Kerr experiment), and thus that ii) the duty cycles of the pumps are orders of magnitude different. Also iii) the state remains in a smooth coherent state during the lifetime measurement instead of visiting complex phase space distributions. The effective temperatures increases only by $\sim 30\%$ for the coherent state lifetime experiment instead of the $\sim 200\%$ increase seen for the free Kerr evolution. This seems to suggest that the effective heating is not caused by a temperature increase in the attenuators and filters of our microwave lines due to the strong squeezing pump. A likely (but unexplored) alternative explanation could be that highly non-linear resonances fed by the drive cause leakage to out-of-manifold states. The matrix elements are expected to depend strongly on the initial states and the abrupt switching of the squeezing pump contains a large frequency spectrum that could favour processes of the sort. Theoretical efforts in our group are addressing these physics \cite{venkatraman2021,Xiao2022}.

The delayed growth of the measured coherent state lifetime for $|\alpha|^2<2$ in comparison with the Lindbladian simulations suggest noise sources beyond the Markov approximation.
This is confirmed experimentally by applying a spin-Echo sequences over the Fock-qubit that produced an increase from $T^*_2 \sim$2~\textmu s to $T_{2E} \sim$13~\textmu s and reveals correlated noise in the experiment.
A model for the noise relying on the assumption that the tunneling oscillations causes by $\Delta< 0$ are close to critically damped seems to explain well our observations (see Figure \ref{fig:Staircase_Sim} \textbf{F}).
To mimic this behaviour, we take the oscillator frequency from a random normal distribution and average over many realizations of the noise.
The frequency fluctuations required to match the experiment ($\sim 10~$kHz) are below the spectroscopic linewidth resolution of the fundamental frequency of our SNAIL-transmon.
As expected \cite{puri2017, grimm2020, Frattini2021}, this effects becomes unimportant for $|\alpha|^2\gtrsim 1$ which is the Kerr-cat operation regime.

%model for this type of noise suggests a coherence damping of the form $e^{-\frac{1}{2}\sigma^2_{\phi}(t)}$, where $\sigma^2_{\phi}(t)$ is the variance of Normally a distributed phase-noise per photon produced by $\phi_j(t) = \int_0^t\Delta_j(\tau)d\tau$ (see Eq.(XXXXXXX) in the main text). Here the index $j$ refers to a given stochastic realization of the noise. Analytical expressions can be derived for a few simple cases \cite{Michel, Girvin, Oliver, Anderson, Cortinas}. We will here limit ourselves to the

% if we take the stochastic detuning variable to have an average of $\langle \Delta_j\rangle_j\sim 2\pi10$~kHz.

It is of notice that in our set-up, measurements can be reliably made for $|\alpha|^2\gtrsim20$, and a decrease in the of coherent state lifetime is observed beyond $|\alpha|^2\gtrsim10$.
For illustrative purposes, we mention that the RWA Lindbladian models would suggest that for $|\alpha|^2\sim 25$ the lifetime of the coherent state in our system should be $\sim$50~minutes in realistic conditions.
This in absence of dissipative two-photon stabilization \cite{lescanne2020} which would, theoretically, provide yet another large improvement factor.
%As mentioned before the inadequacy of the simple models to explain experiments is probably originated in highly non-linear resonances that appear well beyond the rotating wave approximation \cite{venkatraman2021,Xiao2022} and related to chaotic dynamics that remain to be understood \cite{Shillito2022}. The illusion of understanding generated by the laborious fitting of the data up to $|\alpha|^2\sim10$ would be most likely a conceptual artifact. Extending this \textit{decoherence} studies will likely require further theoretical efforts on driven dissipative non-linear superconducting circuits and a new generation of the experimental devices critical for the development of circuit based quantum computing.

\subsubsection{RWA Lindbladian spectrum}
We here elaborate a simplified ($\kappa_{\phi} = 0$) treatment with the objective of developing further the relationship between the nonlinear spectrum of the Lindbladian and the decoherence properties of the system.
This study is based on the fact that the lifetimes of excitations above the equilibrium states are obtained from the real part of the corresponding eigenvalue of the Lindbladian super-operator.
This identification provides a fully quantum interpretation of the experimentally observed steps in the coherent state lifetime of our system, if still only qualitatively.

\begin{figure}[ht]
    \centering
    \includegraphics{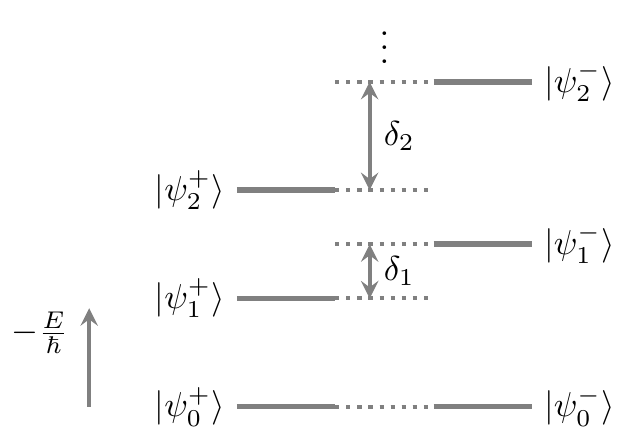}
    \caption{\textbf{Tunnel-splitting for $\hat H_{\textrm{SK}}$.} $\ket{\psi_0^+}$ and $\ket{\psi_0^-}$ are exactly degenerate. The $n{\mathrm{th}}$ excited state manifold has tunnel-splitting $\delta_n \equiv \delta_{nn}^{+-} = \omega_n^+ - \omega_n^-<\delta_{n+1}$.}
    \label{fig:spectrum}
\end{figure}

\textit{Diagonalization of the Lindbladian super-operator}.
We begin this analysis by introducing the notation for the spectrum of the resonantly squeezed ($\Delta=0$) Kerr Hamiltonian Eq. (2): $\hat H_{\textrm{SK}}\ket{\psi_n^{\pm}} = \hbar\omega_n^{\pm}\ket{\psi_n^{\pm}}$.
We choose the eigenstates $\ket{\psi_n^\pm}$ to have overall real expansion coefficients in the Fock basis.
For $n = 0$ we have the degenerate cat state manifold $\ket{\psi_0^{\pm}} = \ket{\mathcal{C}_{\alpha}^{\pm}}$. We denote the energy differences by $\delta_{nm}^{jk} = \omega_n^j-\omega_m^k$, ($j,k=\pm$).
Here $\delta_n = \delta_{nn}^{+-}$ is the tunnel-splitting in between the even and odd parity states in the $n{\mathrm{th}}$ excited manifold in the well (see Figure \ref{fig:spectrum}) and decreases exponentially with $|\alpha|^2$ after the kissing point, whereas the out-of-manifold energy gap $\delta_{n,n+1}^{jk} \approx 4K|\alpha|^2$ increases linearly. 

\begin{figure}[ht]
    \centering
    \includegraphics{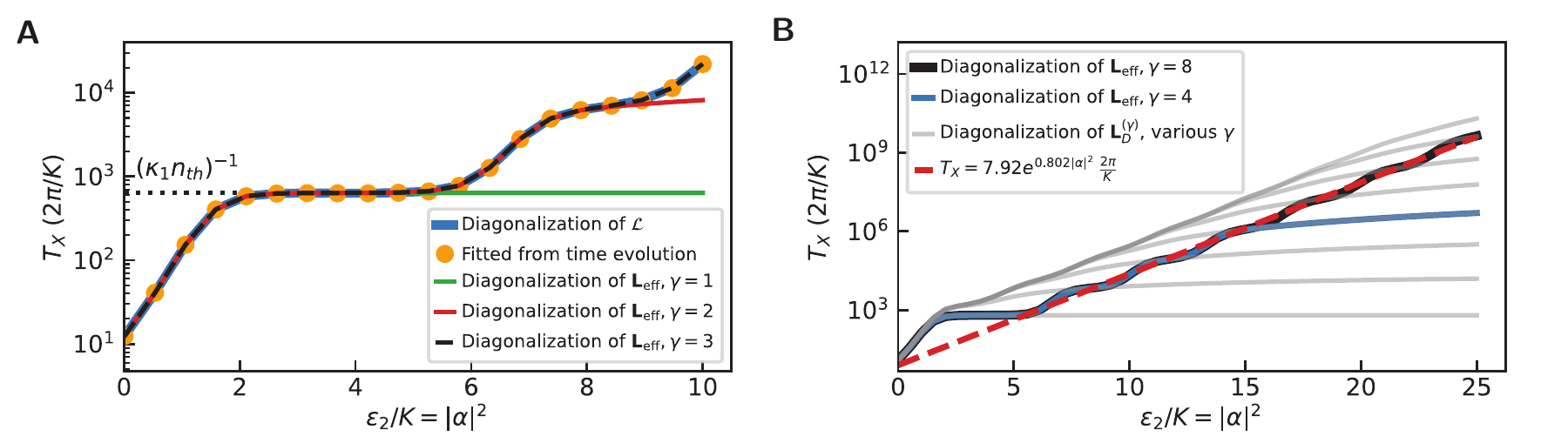}
    \caption{\textbf{Spectral fingerprint of the RWA Lindbladian on the coherent state lifetime.} \textbf{A}, Comparison between exact and approximate coherence time of $\ket{X}$ using $\mathcal{L}$ and $\mathbf{L}_{\text{eff}}^{(\gamma)}$. Orange dots: coherence time extracted from Lindbladian time evolution initialized in state $\ket{X}\bra{X}$. Solid blue line: $[-\mathrm{Re}(\lambda)]^{-1}$ where $\lambda$ is the eigenvalue of $\mathcal{L}$ with the smallest non-zero real part. Solid green line: $[-\mathrm{Re}(\tilde\lambda)]^{-1}$ where $\tilde\lambda$ is the eigenvalue of of the $2\gamma \times 2\gamma$ matrix $\mathbf{L}_{\text{eff}}^{(\gamma)}$ with the smallest non-zero real part, and $\gamma = 1$. Solid red line: $\gamma = 2$. Dashed black line: $\gamma = 3$. \textbf{B}, Comparison of $T_X^{(\gamma)}$ in presences and absences of tunnel splittings. The lifetime in the absences of tunnel splittings is derived from $\mathbf{L}_D^{(\gamma)}$. Black line: extracted from $\mathbf{L}_{\text{eff}}^{(\gamma)}$ with $\gamma = 8$. Blue line: with $\gamma = 4$.  Grey lines: extracted from $\mathbf{L}_D^{(\gamma)}$ with $\gamma = 1, 2, 3, \cdots, 8$ from bottom to top. In both figures, $\kappa_1 / K= 0.025$ and $n_{\mathrm{th}} = 0.01$. Red dashed line: Fit of the black line restricted to $|\alpha|^2 > 6$.}
    \label{fig:results}
\end{figure}

Now we provide a physical interpretation of the eigenvalues of the Lindbladian super-operator.
Under single photon loss and single photon gain, the Lindbladian that governs the evolution of the density operator is 
\begin{align}
    \mathcal{L} = \mathcal{L}_H + \mathcal{L}_D,\;\;\; \mathcal{L}_H \cdot = \frac{1}{i\hbar}[\hat H_{\textrm{SK}}, \ \cdot\ ],\;\;\; \mathcal{L}_D \cdot = \kappa_1(1 + n_{\mathrm{th}})\mathcal{D}[\hat a] \cdot+ \kappa_1 n_{\mathrm{th}}\mathcal{D}[\hat a^\dagger]\cdot .
\end{align}
Here $\kappa_1$ is the rate of single-photon loss to the environment and $n_{\mathrm{th}}$ is the average number of thermal photons.
Then the lifetime of the longest-lived excitation is $T_{X} = [-\mathrm{Re}(\lambda)]^{-1}$, where $\lambda$ is the eigenvalue of $\mathcal{L}$ with the smallest non-zero real component.
In Figure \ref{fig:results} \textbf{A}, we show the lifetime obtained from the eigenvalue of $\mathcal{L}$ is in agreement with that extracted from exponential fits (dots) of the time evolution, obtained as explained in the previous section from numerical time evolution of the master equation (see Figure~\ref{fig:Staircase_Sim}).

\textit{First order perturbation theory}.
We carry the analysis further by exploiting the condition $\kappa_1/K, n_{\mathrm{th}} \ll 1$.
In absence of dissipation, the Lindbladian reduces to $\mathcal{L}_H$, whose eigenoperators are $\ket{\psi_n^{j}}\bra{\psi_m^{k}}$ with eigenvalues $-i\delta_{nm}^{jk}$ directly determined by the energy differences.
The differences are zero for all population eigenoperators $\ket{\psi_n^\pm}\bra{\psi_n^\pm}$ and cat state manifold coherence eigenoperators $\ket{\psi_0^\pm}\bra{\psi_0^\mp}$.
We now treat dissipation as a perturbation to the unitary dynamics generated by $\mathcal{L}_H$.
Moreover, we observe that a truncation of the Lindbladian can be performed because the eigenoperators corresponding to $\delta_{nm}^{jk} \gtrsim \kappa_1 $ do not contribute to the first order perturbative expansion.
That is to say that i) we exclude out-of-manifold coherence eigenoperators $\ket{\psi_n^{j}}\langle\psi_{m\neq n}^{k}|$ where $\delta_{n,m\neq n}^{jk} \sim 4K(n-m)|\alpha|^2 > \kappa_1$, ii) we exclude the nondegenerate in-manifold coherence eigenoperators $\ket{\psi_n^{j}}\bra{\psi_{n}^{k\neq j}}$ when the their tunnel-splittings have not reached the kissing point and $\delta_{nn}^{+-} \equiv \delta_n > \kappa_1$,  whereas iii) we keep the population eigenoperators $\ket{\psi_n^\pm}\bra{\psi_n^\pm}$ because they correspond to zero energy difference, and iv) we keep the pairs of sufficiently degenerate in-manifold coherence eigenoperators $\ket{\psi_n^\pm}\bra{\psi_n^{\mp}}$ corresponding to $\delta_n < \kappa_1$.
In summary, at a fixed value of $|\alpha|^2$ the number of tunnel-splittings that satisfy $\delta_n < \kappa_1$ is $\gamma$, which is a function of both $|\alpha|^2$ and $\kappa_1$. We then restrict the treatment to the quasi-degenerate subspace $\mathcal{B} = \mathcal{B}_{\text{pop}} \cup \mathcal{B}_{\text{coh}}^{(\gamma)}$, where
\begin{align}
    \mathcal{B}_{\text{pop}} &= \left\{\ket{\psi_n^+}\bra{\psi_n^+}, \ket{\psi_n^-}\bra{\psi_n^-}, \cdots,\ \text{where }n \geq 0\right\}, \\
    \mathcal{B}_{\text{coh}}^{(\gamma)} &= \left\{\ket{\psi_n^+}\bra{\psi_n^-}, \ket{\psi_n^-}\bra{\psi_n^+}, \cdots,\ \text{where }0 \leq n < \gamma\right\}.
\end{align}
To further reduce the problem, we consider the initial state $\rho(0) = \ket{X}\bra{X}$ (where we remind $\ket{\pm X} \equiv \frac{1}{\sqrt{2}}(\ket{\mathcal{C}_{\alpha}^+} \pm \ket{\mathcal{C}_{\alpha}^-})$), which can be written as
\begin{align}
    \rho(0) &= \frac{1}{2}\left(\ket{\psi_0^+}\bra{\psi_0^+} + \ket{\psi_0^-}\bra{\psi_0^-}\right) + \frac{1}{2}\left(\ket{\psi_0^+}\bra{\psi_0^-} + \ket{\psi_0^-}\bra{\psi_0^+}\right).
\end{align}
The first term belongs to the subspace $\mathcal{B}_{\text{pop}}$ and closely approximates the sole steady state of the Lindbladian. Thus we expect any change in $\rho$ to result from variations in the second term which belongs to $\mathcal{B}_{\text{coh}}^{(\gamma)}$.
In addition, under perturbations proportional to $\mathcal{D}[\hat{a}]$ and $\mathcal{D}[\hat{a}^\dagger]$ population and coherence operator subspaces are decoupled under the Lindbladian, due to parity. These two observations allow us to restrict to the $2\gamma$-dimensional subspace $\mathcal{B}_{\text{coh}}^{(\gamma)}$, which we rewrite in the following order,
\begin{align}
    \mathcal{B}_{\text{coh}}^{(\gamma)} = \left\{\ket{\psi_0^+}\bra{\psi_0^-}, \ket{\psi_1^+}\bra{\psi_1^-}, \cdots, \ket{\psi_{\gamma-1}^+}\bra{\psi_{\gamma-1}^-};\ \ket{\psi_0^-}\bra{\psi_0^+}, \ket{\psi_1^-}\bra{\psi_1^+},  \cdots, \ket{\psi_{\gamma-1}^-}\bra{\psi_{\gamma-1}^+}\right\} \label{eq:basis}.
\end{align}
We thus construct the effective Lindbladian matrix $\mathbf{L}_{\text{eff}}^{(\gamma)} = \mathbf{L}_H^{(\gamma)} + \mathbf{L}_D^{(\gamma)}$ with the elements of $\mathcal{L}$ in $\mathcal{B}_{\text{coh}}^{(\gamma)}$ as
\begin{align}
    \mathbf{L}_H^{(\gamma)} &=
    \begin{pmatrix}
    -\boldsymbol{\Delta} & 0\\
    0 & \boldsymbol{\Delta}
    \end{pmatrix}, \\
    \mathbf{L}_D^{(\gamma)} &= \kappa_1(1 + n_{\mathrm{th}})
    \begin{pmatrix}
    -\mathbf{A} & \mathbf{B} \\
    \mathbf{B} & -\mathbf{A}
    \end{pmatrix} + \kappa_1n_{\mathrm{th}}
    \begin{pmatrix}
    -\mathbf{A} - \mathbf{I} & \mathbf{B}^T \\
    \mathbf{B}^T & -\mathbf{A} - \mathbf{I}
    \end{pmatrix},
\end{align}
where $\boldsymbol{\Delta}$, $\mathbf{A}$, and $\mathbf{B}$ are $\gamma \times \gamma$ matrices whose entries depend on $|\alpha|^2$ only,
\begin{align}
    &\boldsymbol{\Delta} =
    \begin{pmatrix}
    0 & \\
     & i\delta_1 \\
     & & i\delta_2 \\
     & & & \ddots
    \end{pmatrix},\quad \mathbf{A} = 
    \begin{pmatrix}
    A_0 & \\
     & A_1 \\
     & & A_2 \\
     & & & \ddots
    \end{pmatrix},\quad \mathbf{B} =
    \begin{pmatrix}
    B_{00} & B_{01} & B_{02} & \cdots\\
    B_{10} & B_{11} & B_{12} & \cdots\\
    B_{20} & B_{21} & B_{22} & \cdots\\
    \vdots & \vdots & \vdots & \ddots\\
    \end{pmatrix}, \\
    &\delta_n = \omega_n^+ - \omega_n^-,\quad A_n = \frac{1}{2}\left(\bra{\psi_n^+}\hat a^\dagger \hat a\ket{\psi_n^+} + \bra{\psi_n^-}\hat a^\dagger \hat a\ket{\psi_n^-}\right),\quad B_{mn} = \bra{\psi_m^-}\hat a\ket{\psi_n^+}\bra{\psi_n^-}\hat a^\dagger\ket{\psi_m^+}.
\end{align}
The first order approximation to the coherent state lifetime is then $T_X \approx T_X^{(\gamma)} = [-\mathrm{Re}(\lambda^{(\gamma)})]^{-1}$, where $\lambda^{(\gamma)}$ is the eigenvalue of $\mathbf{L}_{\text{eff}}^{(\gamma)}$ with the smallest real component.
In Figure \ref{fig:results} \textbf{A} we plot $T_X^{(\gamma)}$ as a function of $|\alpha|^2$ for $\kappa_1 / K= 0.025$ and $n_{\mathrm{th}} = 0.01$. While the level of truncation $\gamma$, as defined, should vary with $|\alpha|^2$, we here show the full span for different fixed values of $\gamma$.
We see, then, that each stepped increase in the coherent state lifetime $T_X$ can be associated to an excited state manifold; truncating at a particular value of $\gamma$ amounts to the assumption that $\delta_{n>\gamma}\gg\kappa_1$.
Thus, the tunneling associated to states with $n>\gamma$ is considered immediate, and $T_X^{(\gamma)}$ only approximates well $T_X$ if the number of degenerate levels ($N\approx|\alpha|^2/\pi$) is $\lesssim(\gamma+1)$. If instead $|\alpha|^2>\pi(\gamma+1)$, $T_X^{(\gamma)}$ plateaus and informs the rate of excitation outside of the first $2\gamma$ states considered in the truncation.

%The reason for this attribution is as follows. Our perturbation theory implies that the effect of truncating at a particular order $\gamma$ can be emulated by a fictional \textcolor{red}{Kerr-cat} Hamiltonian whose spectrum (at a given $|\alpha|^2$) is modified to satisfy the condition $\delta_{n \geq \gamma} \gg \kappa_1$ on the tunnel-splittings, thereby causing tunneling as soon as any manifold $n \geq \gamma$ is populated. This is a faithful simplification if $|\alpha|^2$ is small enough such that the real number of sufficiently degenerate manifolds satisfies $\gamma(\alpha, \kappa_1) \leq \bar\gamma$, hence the agreement between $T_X^{(\gamma)}$ and $T_X$ for $|\alpha|^2$ bounded above by the corresponding kissing point. Otherwise, the incorrectly-assumed tunneling rate in the manifold $n = \gamma$ becomes the limiting factor of $T_X^{(\gamma)}$, which plateaus at a value that informs us of the rate of leakage out of the first $\gamma$ manifolds.

%by the agreement between the coherence time obtained from time evolution $e^{t\mathcal{L}}\ket{\tilde\alpha}\bra{\tilde\alpha}$ and that from the identified eigenvalue of $\mathcal{L}$. The perturbation theory analysis is justified by the agreement between time evolution and $\mathcal{L}_{\text{eff}}$.

To further illustrate the role of the tunnel-splittings in $T_X$ enhancement we set $\mathbf{L}^{(\gamma)}_H = 0$ and plot the lifetime extracted from $\mathcal{L}_{D}$ as grey curves in Figure \ref{fig:results} \textbf{B}.
The double well structure is thus only encoded in the unperturbed eigenbases.
The first observation we make is that stepped increase in lifetime vanishes in absence of the kissing spectrum.
Secondly, we observe that for $|\alpha|^2$ smaller than the kissing point of the manifold $n = \gamma$, ($|\alpha|^2 < \pi(\gamma+1)$), the tunneling due to finite splitting in the manifold $n = \gamma$ is a limiting factor for the lifetime \cite{gautier_combined_2022, putterman_stabilizing_2022}.
Finally we observe that the exponential trend shown by the lifetime as a function of $|\alpha|^2$ is largely independent of the particular Hamiltonian spectrum involved, provided the tunnel splitting vanishes for excited states progressively. This features defines ``double-well" potentials.
This is a property strictly related to the ``non-local'' character of the phase space double-well encoding in the classical limit $\epsilon_2\gg K $.

\bibliography{bib}